\documentclass{bmcart}

\usepackage{amsthm,amsmath}
\usepackage[utf8]{inputenc} 

\usepackage{graphicx}

\startlocaldefs
\endlocaldefs

\begin{document}

\begin{frontmatter}

\begin{fmbox}
\dochead{Research}


\title{A survey of results on  mobile phone datasets analysis}


\author[
   addressref={aff1},                   
   corref={aff1},                       
   email={vincent.blondel@uclouvain.be}   
]{\inits{VDB}\fnm{Vincent D} \snm{Blondel}}
\author[
   addressref={aff1},
   email={adeline.decuyper@uclouvain.be}
]{\inits{AD}\fnm{Adeline} \snm{Decuyper}}
\author[
   addressref={aff1,aff2},
   email={gautier.krings@realimpactanalytics.com}
]{\inits{GK}\fnm{Gautier} \snm{Krings}}


\address[id=aff1]{
  \orgname{Department of Applied Mathematics, Universit\'e{} catholique de Louvain}, 
  \street{Avenue Georges Lemaitre, 4},                     %
  \postcode{1348}                                
  \city{Louvain-La-Neuve},                              
  \cny{Belgium}                                    
}
\address[id=aff2]{%
  \orgname{Real Impact Analytics},
  \street{Place Flagey, 7},
  \postcode{1050}
  \city{Brussels},
  \cny{Belgium}
}



\end{fmbox}


\begin{abstractbox}

\begin{abstract} 
In this paper, we review some advances made recently in the study of \textbf{mobile phone datasets}. This area of research has emerged a decade ago, with the increasing availability of large-scale anonymized datasets, and has grown into a stand-alone topic. We will survey the contributions made so far on the \textbf{social networks} that can be constructed with such data, the study of \textbf{personal mobility}, \textbf{geographical partitioning}, \textbf{urban planning}, and \textbf{help towards development} as well as \textbf{security and privacy issues}.
\end{abstract}




\end{abstractbox}
%

\end{frontmatter}




\section{Introduction}
As the Internet has been the technological breakthrough of the '90s, mobile phones have changed our communication habits in the first decade of the twenty-first century. In a few years, the world coverage of mobile phone subscriptions has raised from 12\% of the world population in 2000 up to 96\% in 2014 -- 6.8 billion subscribers -- corresponding to a penetration of 128\% in the developed world and 90\% in developing countries \cite{ITU}.  
Mobile communication has initiated the decline of landline use -- decreasing both in developing and developed world since 2005 -- and allows people to be connected even in the most remote places of the world.\\
In short, mobile phones are \textit{ubiquitous}. In most countries of the developed world, the coverage reaches 100\% of the population, and even in remote villages of developing countries, it is not unusual to cross paths with someone in the street talking on a mobile phone. Due to their ubiquity, mobile phones have stimulated the creativity of scientists to use them as millions of potential sensors of their environment. Mobile phones have been used, as distributed seismographs, as motorway traffic sensors, as transmitters of medical imagery or as communication hubs for high-level data such as the reporting of invading species \cite{kwok2009personal} to only cite a few of their many side-uses. \\

Besides these applications of voluntary reporting, where users install applications on their mobile phones in the aim to serve as sensor, the essence of mobile phones have revealed them to be a source of even much richer data.
The call data records (CDRs), needed by the mobile phone operators for billing purposes, contain an enormous amount of information on how, when, and with whom we communicate.\\
In the past, research on social interactions between individuals were mostly done by surveys, for which the number of participants ranges typically around 1000 people, and for which the results were biased by the subjectivity of the participants' answers. Mobile phone CDRs, instead, contain the information on communications between millions of people at a time, and contain real observations of communications between them rather than self-reported information.\\

In addition, CDRs also contain location data and may be coupled to external data on customers such as age or gender. Such a combination of personal data makes of mobile phone CDRs a extremely rich and informative source of data for scientists. The past few years have seen the rise of research based on the analysis of CDRs. First presented as a side-topic in network theory, it has now become a whole field of research in itself, and has been for a few years the leading topic of NetMob, an international conference on the analysis of mobile phone datasets, of which the fourth edition is in preparation for 2015. Closely related to this conference, a side-topic has now risen, namely the analysis of mobile phone datasets for the purpose of development. The telecom company Orange has, to this end, proposed a challenge named D4D, which concept is to give access to a large number of research teams throughout the world to the same dataset from an African country. Their purpose is to make suggestions for development, on the basis of the observations extracted from the mobile phone dataset. The first challenge, conducted in 2013 was such a success that the results of a second challenge will be presented at the NetMob conference in April 2015. \\

Of course, there are restrictions on the availability of some types of data and on the projected applications. First, the content of communications (SMS or phone discussions) is not recorded by the operator, and thus inaccessible to any third party --  exception made of cases of phone tapping, which are not part of this subject. Secondly, while mobile phone operators have access to all the information filed by their customers and the CDRs, they may not give the same access to all the information to a third party (such as researchers), depending on their own privacy policies and the laws on protection of privacy that apply in the country of application. For example, names and phone numbers are never transmitted to external parties. In some countries, location data, i.e., the base stations at which each call is made, have to remain confidential --  some operators are even not allowed to use their own data for private research.\\
Finally, when a company transmits data to a third party, it goes along with non-disclosure agreements (NDA's) and contracts that strongly regulate the authorised research directions, in order to protect the users' privacy.\\

Yet, even the smallest bit of information is enough for triggering bursts of new applications, and day after day researchers discover new purposes one can get from CDRs. The first application of a study of phone logs (not mobile, though) appeared in 1949, with the seminal paper by George Zipf modeling the influence of distance on communication \cite{zipf1949human}. Since then, phone logs have been studied in order to infer relationships between the volume of communication and other parameters (see e.g. \cite{cortes2001communities}), but the apparition of mobile phone data in massive quantities, and of computers and methods that are able to handle those data efficiently, has definitely made a breakthrough in that domain. Being personal objects, mobile phones enabled to infer real social networks from their CDRs, while fixed phones are shared by users of one same geographical space (a house, an office). The communications recorded on a mobile phone are thus representative of a part of the social network of one single person, where the records of a fixed phone show a superposition of several social actors. By being mobile, a mobile phone has two additional advantages: first, its owner has almost always the possibility to pick up a call, thus the communications are reflecting the temporal patterns of communications in great detail, and second, the positioning data of a mobile phone allows to track the displacements of its owner.\\

Given the large amount of research related to mobile phones, we will focus in this paper on contributions related to the analysis of massive CDR datasets. A chapter of the (unpublished) PhD thesis of Gautier Krings \cite{Krings2012thesis} gives an overview of the litterature on mobile phone datasets analysis. This research area is growing fast and this survey is a significantly expanded version of that chapter, with additional sections and figures and an updated list of references. The paper is organized following the different types of data that may be used in related research. In Section \ref{sec:sn} we will survey the contributions studying the topological properties of the static social network constructed from the calls between users. When information on the position of each node is available, such a network becomes a geographical network, and the relationship between distance and the structure of the network can be analyzed. This will be addressed in Section \ref{sec:gn}. Phone calls are always localized in time, and some of them might represent transient relationships while others rather long-lasting interactions. This has led researchers to study these networks as temporal networks, which will be presented in Section \ref{sec:dn}. In Section \ref{sec:hm}, we will focus on the abundant literature that has been produced on human mobility, made possible by the spatio-temporal information contained in CDR data. As mobile phone networks represent in their essence the transmission of information or more recently data between users, we will cover this topic in Section \ref{sec:dyn}, with contributions on information diffusion and the spread of mobile phone viruses. Some contributions combine many of these different approaches to use mobile phone data towards many different applications, which will be the object of Section \ref{sec:apps}. Finally, in Section \ref{sec:pi} we will consider privacy issues raised by the availability and use of personal data.

\section{Social networks} \label{sec:sn}
\begin{figure}[!t]
\begin{center}
\includegraphics[width=\textwidth]{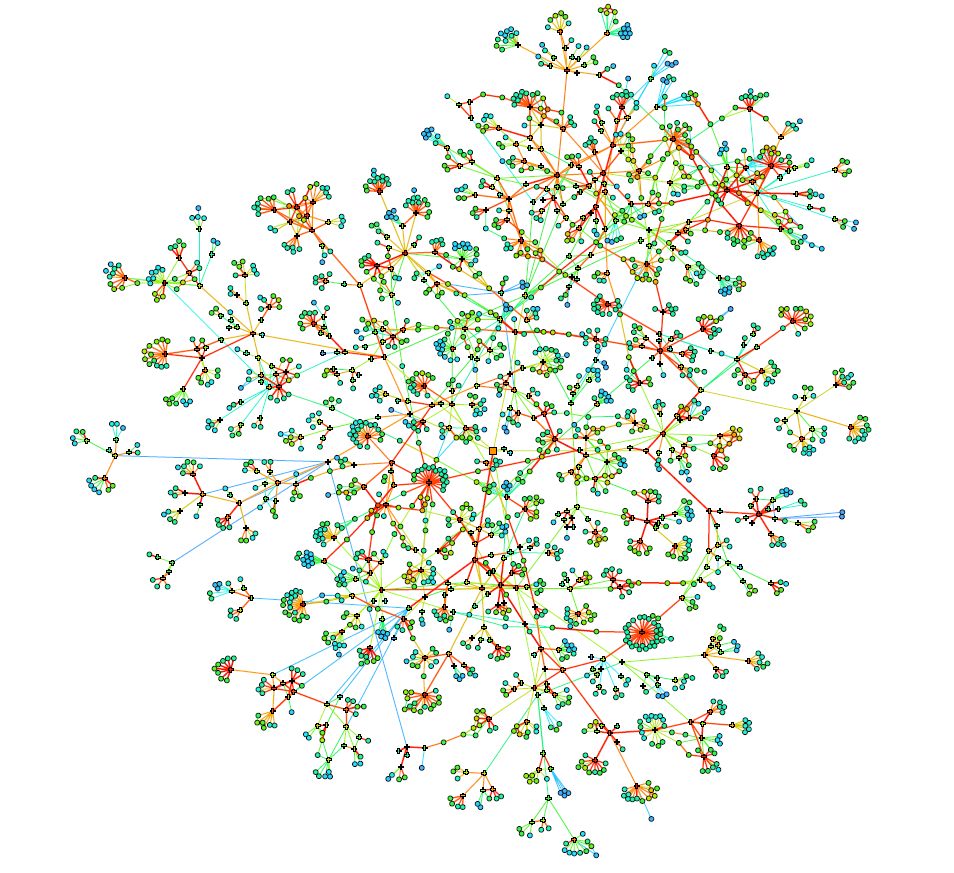}
\end{center}
\caption{\csentence{Sample of a mobile phone network, obtained with a snowball sampling.} The source node is represented by a square, bulk nodes by a + sign and surface nodes by an empty circle. Figure reproduced from \cite{onnela2007analysis}.}
\label{fig:survey_snowball}
\end{figure}

In its simplest representation, a dataset of people making phone calls to each other is represented by a network where nodes are people and links are drawn between two nodes who call each other. 
In the first publications related to telecommunications datasets, the datasets were rather used as an example for demonstration of the potential applications of an algorithm \cite{abello1999maximum} or model \cite{aiello2000random} rather than for a purpose of analysis. However, it quickly appeared that the so-called \textit{mobile call graphs} (MCG) were structurally different from other complex networks, such as the web and internet, and deserved a particular attention, see Figure \ref{fig:survey_snowball} for an example of snowball sampling of a mobile phone network.
We will review here the different contributions on network analysis. We will address the construction of a social network from CDR data, which is not a trivial exercise, simple statistical properties of such networks and models that manage to reproduce them, more complex organizing principles, and community structure, and finally we will discuss the relevance of the analysis of mobile phone networks.

\subsection*{Construction}
While the network construction scheme mentioned above seems relatively simple, there exist many possible interpretations on how to define a link of the network, given a dataset.  \\
The primary aim of social network analysis is to observe social interactions, but not every phone call is made with the same social purpose. Some calls might be for business purposes, some might be accidental calls, some nodes may be call centers that call a large number of people, and all such interactions are present in CDRs. In short, CDRs are noisy datasets. ``Cleaning'' operations are usually needed to eliminate some of the accidental edges. For example, Lambiotte \textit{et al.}\ \cite{lambiotte2008gdm} imposed as condition for a link that at least one call is made in both directions (reciprocity) and that at least 6 calls are made in total over 6 months of the dataset. This filtering operation appeared to remove a large fraction of the links of the network, but at the same time, the total weight (the total number of calls passed in by all users) was reduced by only a small fraction. The threshold of 6 calls in 6 months may be questionable, but a stability analysis around this value can comfort that the exact choice of the threshold is not crucial. Similarly, Onnela \textit{et al.}\ \cite{onnela2007structure} analyzed the differences between the degree distribution of two versions of the same dataset, one containing all calls of the dataset, and the other containing only calls that are reciprocated. Some nodes in the complete network have up to 30,000 different neighbors, while in the reciprocated network, the maximal degree is close to 150. Clearly, in the first case it is hard to imagine that node representing a single person, while the latter is a much more realistic bound. However, even if calls have been reciprocated, the question of setting a meaningful weight on each link is far from easy. Li \textit{et al.}\ suggest another more statistical approach in \cite{li2014statistically}, and use multiple hypothesis testing to filter out the links that appeared randomly in the network and that are therefore not the mirror of a true social relationship. It is sometimes convenient to represent a mobile call network by an undirected network, arguing that communication during a single phone call goes both ways, and set the weight of the link as the sum of the weights from both directions. However, who initiates the call might be important in other contexts than the passing of information, depending on the aim of the research, and Kovanen \textit{et al.}\ have showed that reciprocal calls are often strongly imbalanced \cite{kovanen2011reciprocity}. In the interacting pair, one user is often initiating most of the calls, so how can this be represented in an undirected network by a representative link weight? In a closely related question, most CDRs contain both information on voice calls and text messages, but so far it is not clear how to incorporate both pieces of information into one simple measure. Moreover, there seems to be a generational difference in the use of text messages or preference between texts and voice calls which may introduce a bias in measures that only take one type of communication into account \cite{ling2012sociodemographics}. \\

Besides these considerations on the treatment of noise, the way to represent social ties may vary as well: they may be binary, weighted, symmetric or directed. Different answers to such decisions lead to different network characteristics, and result in diverse possible interpretations of the same dataset. For example, Nanavati \textit{et al.}\ \cite{nanavati2006structural} keep their network as a directed network, in order to obtain information on the strongly connected component of the network, while Onnela \textit{et al.}\ \cite{onnela2007structure} rather focus on an undirected network, weighted by the sum of calls going in both directions.\\

\subsection*{Topological properties}
\begin{figure}[t!]
\begin{center}
\includegraphics[width=\textwidth]{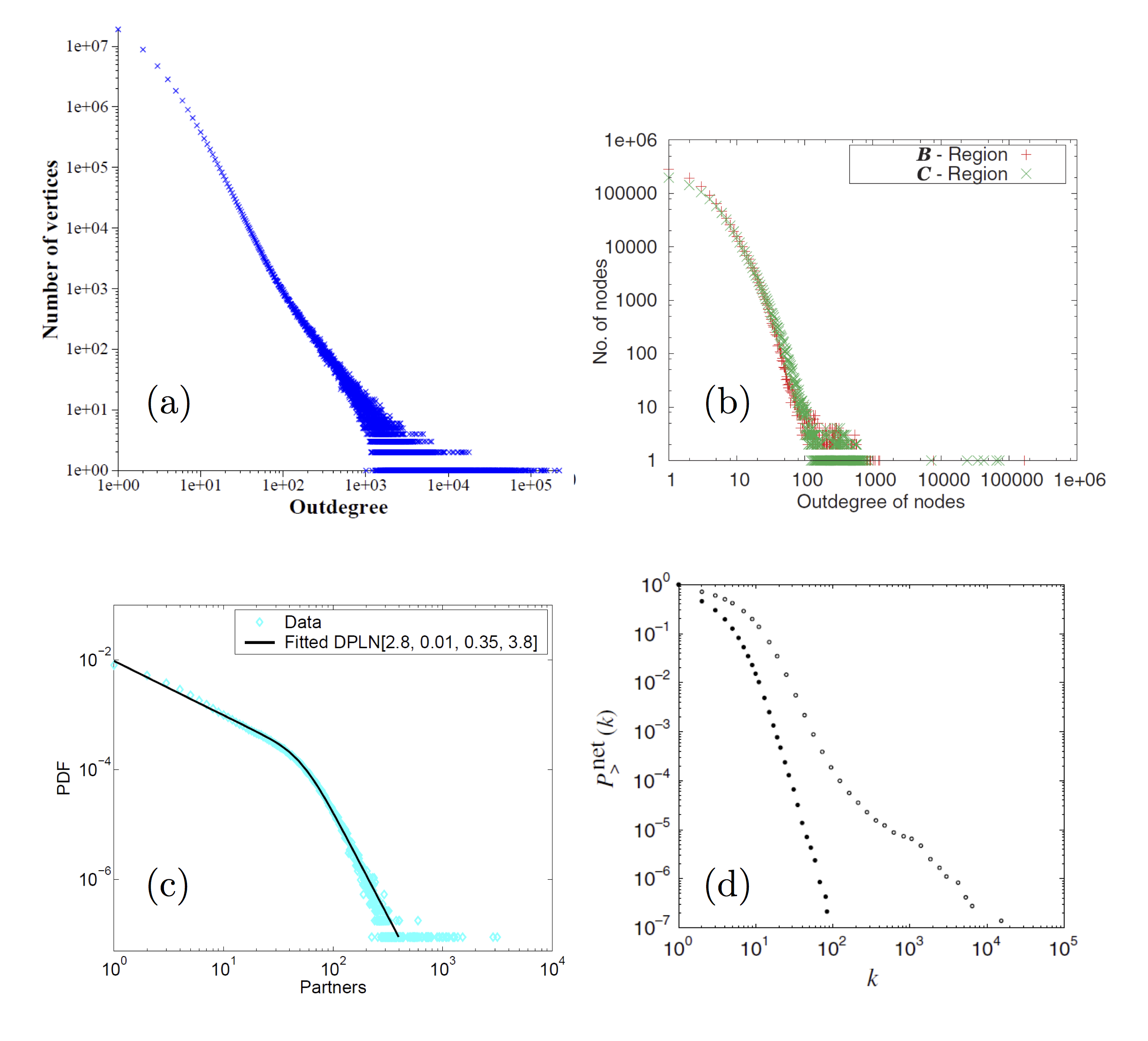}
\end{center}
 \caption{\csentence{Degree distributions in mobile phone networks.} The degree distributions of several datasets have comparable features, but differences in the construction, the time range of the dataset and the size of the system lead to different shapes. Note the bump in (d), when non-reciprocal links are taken into account. 
(a) Aiello, W. \textit{et al.}, ``A random graph model for massive graphs", in Proceedings of the thirty-second annual ACM symposium on Theory of computing, pages 171--180 \cite{aiello2000random}  \copyright 2000 Association for Computing Machinery, Inc. Reprinted by permission. \url{http://doi.acm.org/10.1145/335305.335326} 
(b) Nanavati, A.A. \textit{et al.}, ``On the structural properties of massive telecom call graphs: findings and implications.", in Proceedings of the 15th ACM international conference on Information and knowledge management, pages 435--444 \cite{nanavati2006structural}  \copyright 2006 Association for Computing Machinery, Inc. Reprinted by permission. \url{http://doi.acm.org/10.1145/1183614.1183678}
(c) Seshadri, M. \textit{et al.}, ``Mobile call graphs: beyond power-law and lognormal distributions." in Proceedings of the 14th ACM SIGKDD international conference on Knowledge discovery and data mining, pages 596--604 \cite{seshadri2008mobile} \copyright 2008 Association for Computing Machinery, Inc. Reprinted by permission.  \url{http://doi.acm.org/10.1145/1401890.1401963}
(d) Figure reproduced from \cite{onnela2007analysis}.}
 \label{fig:survey_degdists}
\end{figure}

The simplest information one can get out of CDRs is statistical information on the number of acquaintances of a node, on the local density of the network or on its connectivity. Like social networks, mobile call graphs differ from random networks and lattices by their broad degree distribution \cite{barabasi2009scale}, their small diameter and their high clustering \cite{watts1998collective}.\\
While all analyzed datasets present similar general shapes for those distributions, their fine shape and their range differ due to differences between the datasets, the construction scheme, the size, or the time span of the collection period.\\
In one of the first studies involving CDR data Aiello \textit{et al.}\ \cite{aiello2000random} observed a power law degree distribution, which was well explained by a massive random graph model $P(\alpha,\beta)$ described by its power-law degree distribution $p(d=x) = e^{\alpha}x^{-\beta}$.\\
Random graph models have often been used in order to model networks, and manage to reproduce some observations from real-world networks, such as the small diameter and the presence of a giant component, such as observed on mobile datasets. However, they fail to uncover more complex features, such as degree-degree correlations. Nanavati \textit{et al.}\ \cite{nanavati2006structural} observed in the study of 4 mobile datasets that besides the power-law tail of the degree distribution, the degree of a node is strongly correlated with the degree of its neighbors.\\
Characterizing the exact shape of the degree distribution is not an easy task, which has been the focus of a study by Seshradi \textit{et al.}\ \cite{seshadri2008mobile}. They observed that the degree distribution of their data can be fitted with a Double Pareto Log Normal (DPLN) distribution, two power-laws joined by a hyperbolic segment -- which can be related to a model of social wealth acquisition ruled by a lognormal multiplicative process. Those different degree distributions are depicted on Figure \ref{fig:survey_degdists}. Interestingly, let us note that the time span of the three aforementioned datasets are different, Aiello \textit{et al.}\ have data over one day, Nanavati \textit{et al.}\ over one week, and Seshadri \textit{et al.}\ over one month. \\

Krings \textit{et. al.} dig a bit deeper into this topic, and investigated the effect of placement and size of the aggregation time window \cite{krings2012effects}. They showed that the size of the time window of aggregation can have a significant influence on the distributions of degrees and weights in the network. The authors also observed that the degree and weight distributions become stationary after a few days and a few weeks respectively. 
The effect of the placement of the time window has most influence for short time windows, and depends mostly on whether it contains holiday periods or weekends, during which the behavioral patterns have been shown to be significantly different than during normal weekdays. 

What information do we get from these distributions? They mostly reflect the heterogeneity of communication behaviors, a common feature for complex networks \cite{barabasi2009scale}. The fat tail of the degree distribution is responsible for large statistical fluctuations around the average, indication that there is no particular scale representative of the system. The majority of users have a small number of contacts, while a tiny fraction of nodes are hubs, or super-connectors. However, it is not clear whether these hubs represent true popular users or are artefacts of noise in the data, as was observed by Onnela \textit{et al.}\ \cite{onnela2007analysis} in their comparison of the reciprocated and non-reciprocated network.\\
The heterogeneity of degrees is also observed on node strengths and link weight, which is also to be expected for social networks. All studies also mention high clustering coefficient, which indicates that the nodes arrange themselves locally in well-organized structures. We will address this topic in more detail further.

\subsection*{Advanced network characteristics}
\begin{figure}[!t]
\begin{center}
\includegraphics[width=\textwidth]{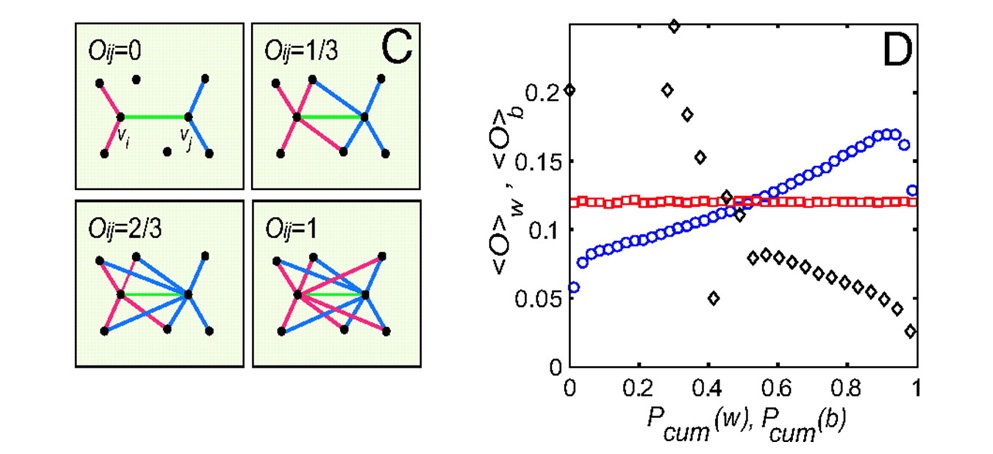}
\end{center}
 \caption{\csentence{Overlap of a link in a network.} (Left) The overlap of a link is defined as the ratio between the common neighbors of both nodes and the maximum possible common neighbors. Here, the overlap is given for the green link. (Right) The average overlap increases with the cumulative weight in the real network (blue circles) and is constant in the random reference where link weights are shuffled (red squares). The overlap also decreases with the cumulative betweenness centrality $P_{cum}(b)$ (black diamonds). Figure reproduced from \cite{onnela2007structure}.}
 \label{fig:survey_overlap}
\end{figure}

Beyond statistical distributions, more complex analyses provide a better understanding of the structure of our communication networks. The heterogeneity of link weights deserves particular attention. Strong links represent intense relationships, hence the correlation between weight and topology is of primary interest. Recalling that mobile call graphs show high clustering coefficient, and thus are locally dense, one can differentiate links based on their position in the network.\\
The overlap of a link, introduced in \cite{onnela2007structure} (and illustrated on Figure \ref{fig:survey_overlap}), is an appropriate measure which characterizes the position of a link as the ratio of observed common neighbors $n_{ij}$ over the maximal possible, depending on the degrees $k_i$ and $k_j$ of the nodes and defined as:
\begin{equation}
O_{ij} = \frac{n_{ij}}{(d_i - 1) + (d_j - 1) - n_{ij}}
\end{equation}
The authors show that link weight and topology are strongly correlated, the strongest links lying inside dense structures of the network, while weaker links act as connectors between these densely organized groups. This finding has an important consequence on processes such as link percolation or the spread of information on networks, since the weak ties act as bridges between disconnected dense parts of the network, illustrating Granovetter's hypothesis on the strength of weak ties \cite{granovetter1973strength}.\\

The structure of the dense subparts of the network provides essential information on the self-organizing principles lying behind communication behaviors. Before moving to the analysis of communities, we will focus on properties of cliques. The structure of cliques is reflected by how weights are distributed among their links. In a group where everyone talks to everyone, is communication balanced? Or are small subgroups observable? A simple measure to analyze the balance of weights is the measure of coherence $q(g)$. This measure was introduced in \cite{onnela2005intensity} before its application to mobile phone data in \cite{onnela2007analysis}, and is calculated as the ratio between the geometric mean of the link weights and the arithmetic mean, 
\begin{equation}
q(g) = \frac{\left(\prod_{ij \in l_g}w_{ij}\right)^{1/|l_g|}}{\frac{\sum_{ij\in l_g}w_{ij}}{|l_{g}|}}
\end{equation}
where $g$ is a subgraph of the network and $l_g$ is its set of links. This measure takes values in the range $]0,1]$, 1 corresponding to equilibrium. On average, cliques appear to be more coherent than what would be expected in the random case, in particular for triangles, which show high coherence values.\\

On a related topic, Du \textit{et al.}\ \cite{du2009large} focused instead on the propensity of nodes to participate to cliques, and in particular on the balance of link weights inside triangles. Their observations differ slightly from Onnela \textit{et al.}\ : on average, the weights of links in triangles can be expressed as powers of one another. The authors managed to reproduce this singular situation with a utility-driven model, where users try to maximize their return from contacts.\\

\subsection*{Communities}
The previous analysis of cliques and triangles opens the way for an analysis of more complex structures, such as communities in mobile phone networks. The analysis of communities provides information on how communication networks are organized at large scale. In conjunction with external data, such as age, gender or cultural differences, it provides sociological information on how acquaintances are distributed over the population. From a corporate point of view, the knowledge of well-connected structures is of primary importance for marketing purposes. In this paragraph, we will only address simple results on community analysis, but this topic will be addressed again further in the document, when it relates to geographic dispersal of networks or dynamical networks. \\
At small scale, traditional clustering techniques may be applied, see \cite{kianmehr2009calling} and \cite{zhang2008discovery} for examples of applications on small datasets. However, on large mobile call graphs involving millions of users, such clustering techniques are outplayed by community detection algorithms.\\

\begin{figure}[t!]
\begin{center}
\includegraphics[width=\textwidth]{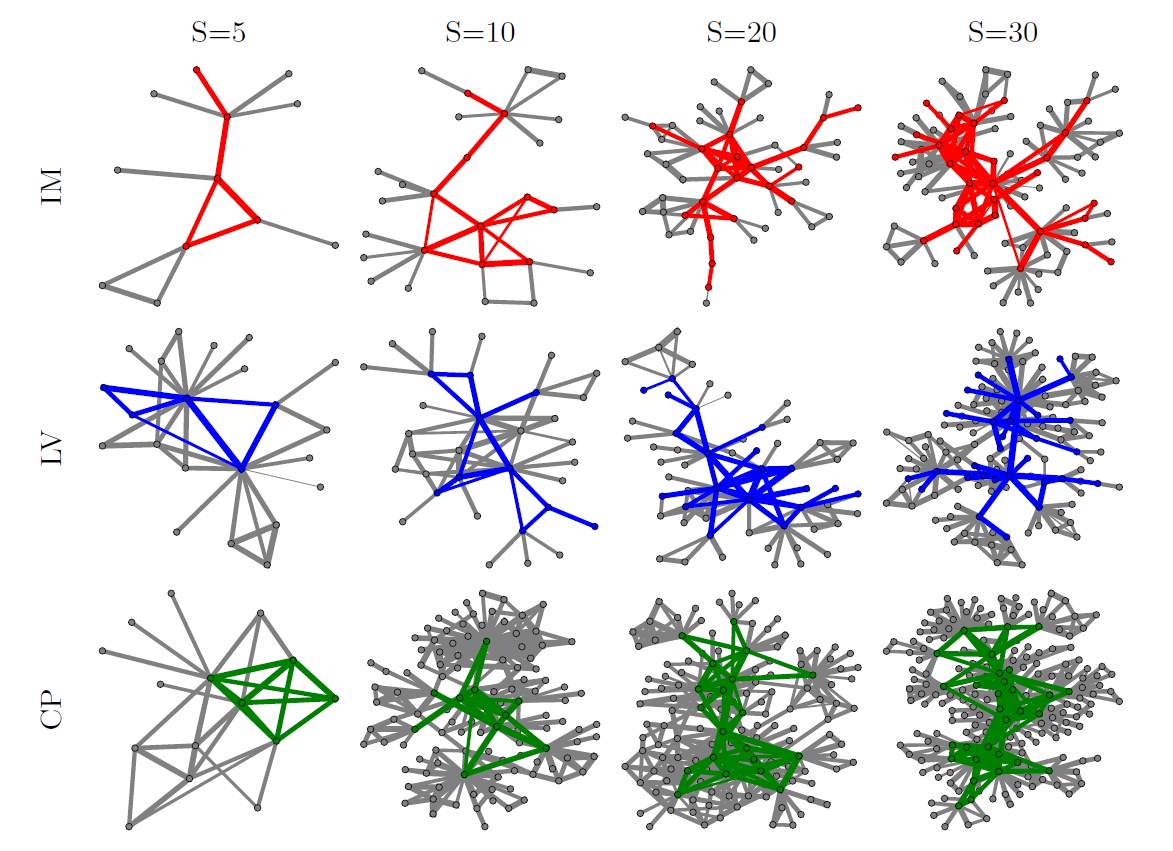}
\end{center}
 \caption{\csentence{Examples of communities detected with different methods.} The different methods are the InfoMap method (IM, red), Louvain method (LV, blue) and Clique percolation method (CP, green). For each method, four examples are shown, with 5, 10, 20 and 30 nodes. The coloured links are part of the community, the grey nodes are the neighbors of the represented community. While IM and LV find almost tree-like structures, CP finds dense communities \cite{tibely2011communities}. Reproduced figure with permission from Tib\'ely, G. \emph{et al.}, Physical Review E. 83(5):056125, 2011. Copyright (2011) by the American Physical Society. \url{http://dx.doi.org/10.1103/PhysRevE.83.056125} }
 \label{fig:survey_communities}
\end{figure}

Uncovering the community structure in a mobile phone network is highly dependent on the used definition of communities and detection method. One could argue that there exist as many plausible analyses as there are community detection methods. Moreover, the particular structure of mobile call graphs induces some issues for traditional community detection methods. Tibely \textit{et al.}\ \cite{tibely2011communities} show that even though some community detection methods perform well on benchmark networks, they do not produce clear community structures on mobile call graphs. Mobile call graphs contain many small tree-like structures, which are badly handled by most community detection methods. The comparison of three well-known methods: the Louvain method \cite{blondel2008fuc}, Infomap \cite{rosvall2008maps} and the Clique Percolation method \cite{palla2007quantifying} produce different results on mobile call graphs. The Louvain method and Infomap both build a partition of the nodes of the network, so that every node belongs to exactly one community. In contrast Clique Percolation only keeps as community dense subparts of the network (see Figure \ref{fig:survey_communities}). 

\begin{figure}[t!]
\begin{center}
\includegraphics[width=\textwidth]{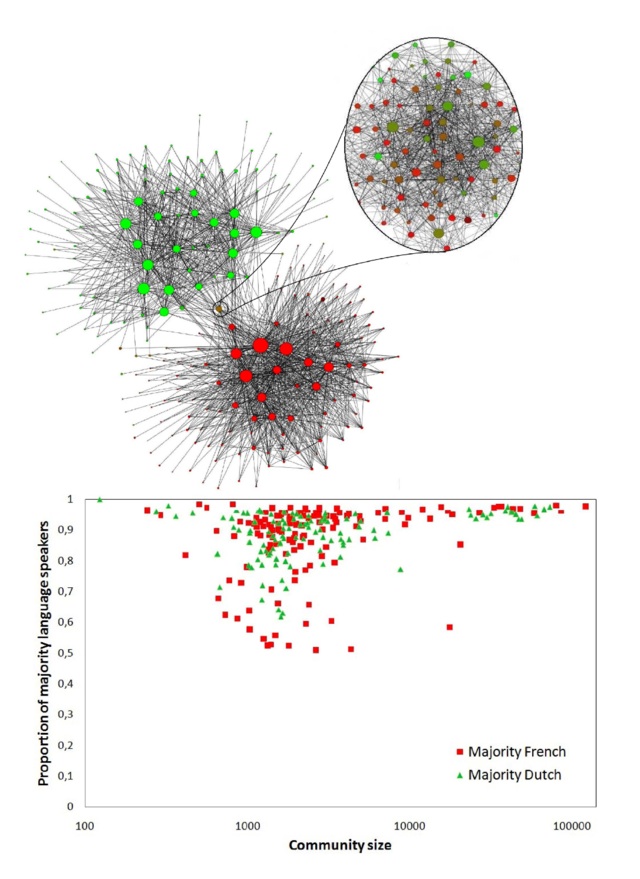}
\end{center}
 \caption{\csentence{Community detection in Belgium} (top) The communities of the Belgian network are colored based on their linguistic composition: green for Flemish, red for French. Communities having a mixed composition are colored with a mixed color, based on the proportion of each language. (bottom) Most communities are almost monolingual. Figures reproduced from \cite{blondel2008fuc}.}
 \label{fig:survey_combelg}
\end{figure}

As observed in Tibely \textit{et al.}\ the small tree-like structures are often considered as communities, although their structure is sparse. Such a result is counter-intuitive given the intrinsic meaning of communities and raises the question: is community detection hence unusable on mobile call graphs? The results have probably to be considered with caution, but as this is always the case for community detection methods, whatever network is used, this special character of communities in mobile call graphs appears rather as a particularity than a problem. Although they might have singular shapes, communities can provide significant information, when usefully combined with external information. Proof is made by the study of the linguistic distribution of communities in a Belgian mobile call graph \cite{blondel2008fuc}, where the communities returned by the Louvain method strikingly show a well-known linguistic split, as illustrated on Figure \ref{fig:survey_combelg}.\\

The notion of communities in social networks, such as rendered by mobile phone networks, has raised a debate on the exact vision one has of what a community is and what it is not. In particular, several authors have favored the idea of overlapping communities, such that one node may belong to several communities, in opposition with the classical vision that communities are a partition of the nodes of a network. An argument in favor of this vision is that one is most often part of several groups of acquaintances who do not share common interests, such as family, work and sports activities. In \cite{ahn2010link}, Ahn \textit{et al.}\ show how overlapping communities can be detected by partitioning edges rather than nodes, and illustrated their methods with a mobile phone dataset. For each node, they had additional information about its center of activities, with which they showed that communities were geographically consistent.

\subsection*{Social analysis}
The use of mobile call data in the purpose of analysis of social relationships raises two questions. First, how faithful is such a dataset of real interactions? Second, can we extract information on the users themselves from their calling behavior?\\

It has often been claimed that mobile phone data analysis is a significant advance for social sciences, since it allowed scientists to use massive datasets containing the activity of entire populations. The study of mobile phone datasets is part of an emerging field known as computational social science \cite{Lazer2009computational}. These massive datasets, it is said, are free from the bias of self-reporting, which is that the answers to a survey are usually biased by the own perception of the subject, who is not objective. Still, the question remains: how much does self-reporting differ from our real behavior, what is the exact added value of having location data? This has been studied by Eagle \textit{et al.}\ \cite{eagle2009inferring} in the well-known Reality Mining project. By studying the behavior of about 100 persons both by recording their movements and encounters using GSM and Bluetooth technology and with the use of surveys, they managed to quantify the difference between self-reported behavior and what could be observed. It appears that observed behavior strongly differs from what has been self-reported, confirming that the subjectivity of the subjects' own perception produces a significant bias in surveys. In contrast, collected data allows to reduce this bias significantly. However, mobile phone data introduce a different bias, namely, that they only contain social contacts that were expressed through phone calls, thus missing all other types of social interactions out \cite{wiese2015you}. 

While most studies use external data as validation tool to confirm the validity of results, Blumenstock \textit{et al.}\ shortly addressed a different question, namely if it was possible to infer information on people's social class based on their communication behavior. Apparently, this task is hard to perform, even if significant differences appear in calling behavior between different classes of the population \cite{blumenstock2010who}. 
While inferring information about users from their calling activity still seems difficult, many studies show strong correlations between calling behavior and other information included in some datasets, such as gender or age. In a study on landline use, Smoreda \textit{et al.}\ highlight the differences in the use of the domestic telephone based on the genders of both the caller and the callee \cite{smoreda2000gender}, and show not only that women call more often than men but also that the gender of the callee has more influence than the gender of the caller on the duration of the call. Those same trends have also been observed in later studies of mobile phone datasets \cite{kovanen2013temporal}. Further than just observing the gender differences in mobile phone use, Frias-Martinez \textit{et al.}\ propose a method to infer the gender of a user based on several variables extracted from mobile phone activity \cite{frias2010gender}, and achieve a success rate of prediction between 70\% and 80\% on a dataset of a developing economy. In a later study on data from Rwanda, Blumenstock  \textit{et al.}\ show that differences of social class induce more striking differences in mobile phone use than differences of gender \cite{blumenstock2012divided}. 

Further than analyzing the nodes of a network, Chawla \textit{et. al.} take a closer look at the links of the network, and introduce a measure of reciprocity to quantify how balanced the relationship between two users is \cite{chawla2011reciprocity}: 
\begin{equation}
R_{ij} = | ln(p_{ij}) - ln(p_{ji}) |
\end{equation}
where $p_{ij}$ is the probability that if $i$ makes a communication, it will be directed towards $j$. 
They also test this measure on a mobile communications dataset, and show that there are very large degrees of non-reciprocity, far above what could be expected if only balanced relationships were kept. 

Going one step further, instead of inferring information on the nodes of the mobile calling graph, Motahari \textit{et al.}\ study the difference in calling behavior depending on the relationship between two subscribers, characterizing different types of links. They show that the links within a family generate the highest number of calls, and that the network topology around those links looks significantly different from the topology of a network of utility communications \cite{motahari2012affinity}.

\section{Adding space -- Geographical networks}\label{sec:gn}
\begin{figure}[!t]
\begin{center}
\includegraphics[width=0.9\textwidth]{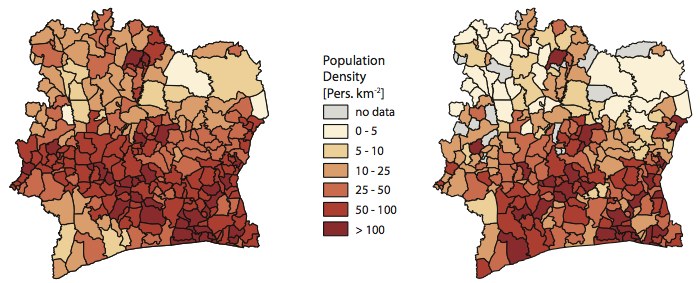}
\end{center}
 \caption{\csentence{Population density estimates.} (left) population density estimates from the Afripop project \cite{afripop}. (right) Population density estimates from mobile phone data. Figure reproduced from \cite{sterlyD4Ddensity}.}
 \label{fig:survey_CIVdensity}
\end{figure}

Besides basic CDR data, it happens that geographic information is available about the nodes, such as the home location (available for billing purposes) or the most often used antenna. This allows then to assign each node to one geographic point, and to study the interplay between geography and mobile phone usage. Studies on geographical networks have already been performed on a range of different types of networks \cite{barthelemy2011spatial}. One of the very basic applications is to use mobile phone data to estimate the density of population in the different regions covered by the dataset. Deville \textit{et al.}\ explored this idea \cite{deville2014dynamic}, using the number of people who are calling from each antenna, they are able to produce timely estimates of the population density in France and Portugal. In the developing world, census data is often very costly or even impossible to obtain, and existing data is often very old and outdated. Using CDRs can then provide very useful and updated information on the actual density of population in remote parts of the world. Another example is given by Sterly \textit{et al.}\ who mapped an estimate of the density of population of Ivory Coast using a mobile phone dataset \cite{sterlyD4Ddensity}, as illustrated on Figure \ref{fig:survey_CIVdensity}.  \\

\subsection*{Relationship space-communication}
\begin{figure}[!t]
\begin{center}
\includegraphics[width=\textwidth]{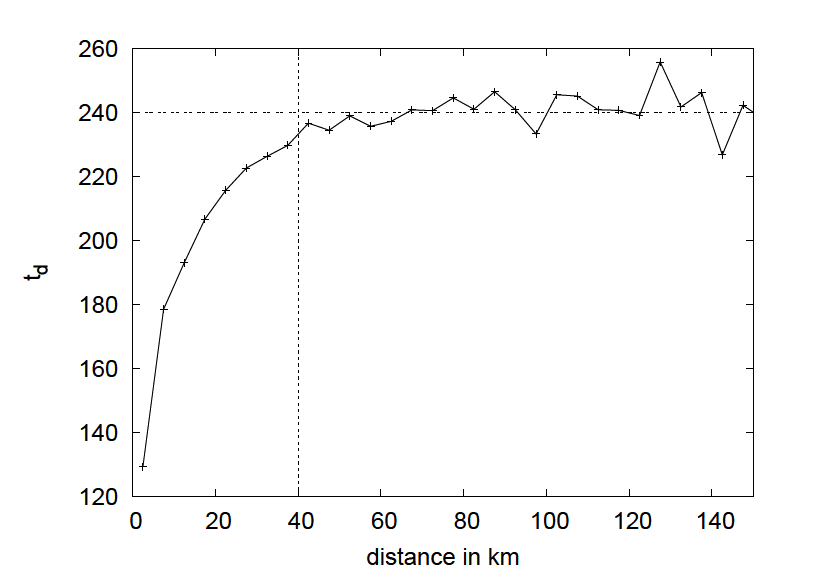}
\end{center}
 \caption{\csentence{Average duration of a call depending on the distance between the callers.} A saturation point is observed at 40 km. Figure reproduced from \cite{lambiotte2008gdm}.}
 \label{fig:survey_duration}
\end{figure}

Lambiotte \textit{et al.}\ \cite{lambiotte2008gdm}, investigated the interplay between geography and communications, and assigned each of the 2.5 million users from a Belgian mobile phone operator to the ZIP code location where they were billed. By approximating the position of the users to the center of each ZIP code area, they showed that the probability of two users to be connected decreases with the distance $r$ separating them, following a power law of exponent $-2$. The probability of a link to be part of a triangle decreases with distance, until a threshold distance of 40 km, after which the probability is constant. 
Interestingly, this threshold of 40 km is also a saturation point for the average duration of a call (see Figure \ref{fig:survey_duration}). A different study on the same dataset also showed that total communication duration between communes in Belgium was well fitted by a gravity law, showing positive linear contribution of the number of users in each commune and negative quadratic influence of distance \cite{krings2009urban,krings2009scaling}:
\begin{equation}
l_{ab} = \frac{c_ac_b}{r_{ab}^2}
\end{equation}
where $l_{ab}$ represents the total communication between communes $a$ and $b$, $c_a$ and $c_b$ the number of customers in each commune and $r_{ab}$ the distance that separates them.\\

While it seems sure that distance has a negative impact on communication, its exact influence is not unique. Onnela \textit{et al.}\ \cite{onnela2011geographic} observed in a different dataset a probability of connection decreasing as $r^{-1.5}$ rather than the gravity model observed by Lambiotte \textit{et al}, and a later study on Ivory Coast by Bucicovschi \textit{et al.}\ \cite{bucicovschiD4Dsocdiv} observe that the total duration of communication between two cities decays with $r^{-1/3}$. However, these differences might be explained by the differences that exist between the studied countries, such as the distribution of the population density. A different study on mobility data from the location-based service Foursquare \cite{noulas2012tale} levelled those variations using a rank-based distance \cite{liben2005geographic}, which could also be helpful in this case.  Another comparison is presented by Carolan \textit{et al.}\  \cite{carolan2012issc} who compare two different types of distance, namely the spatial travel distance and the travel time taken to link two cities. Interestingly, it appears that the use of the spatial distance rather than the time taken gives a better fit of the number of communications between two cities with the gravity model. Their observations also show that the gravity model fits the data better when data is collected during the daytime on weekdays than during evenings and weekends.\\
Instead of studying the communication between cities, Schl\"apfer \textit{et al.}\ looked at the relationship between city size and the structure of local networks of people living in those cities \cite{schlapfer2014scaling}. They show that the number of contacts and communication activity both grow with city size, but that the probability of being friends with a friend's friend remains the same independently of the city size. Jo \textit{et al.}\ propose another approach and study the evolution with age of the distance between a person and the person with whom they have the most contacts \cite{jo2014spatial}. They thus show that young couples tend to live within longer distances than old couples. \\

Instead of only taking into account the distance between two places to predict the number links between them, Herrera-Yag{\"u}e \textit{et al.}\ make another hypothesis, namely that the probability of someone living in a location $i$ has contacts with a person living in another location $j$ is inversely proportional to the total population within an ellipse \cite{herrera2014elliptic}. The ellipse is defined as the one whose foci are $i$ and $j$, and whose surface is the smallest such that both circles of radius $r_{ij}$ centred around $i$ and $j$ are contained in the ellipse. If we name $e_{ij}$ the total population within the ellipse, the number of contacts between locations $i$ and $j$ is thus described by: 
\begin{equation}
T_{ij} = K\frac{n_in_j}{e_{ij}}
\end{equation}
where $K$ is a normalisation parameter depending on the total number of relationships to predict, and $n_i$ and $n_j$ are the populations of locations $i$ and $j$ respectively. \\

\begin{figure}[!t]
\begin{center}
\includegraphics[width=0.9\textwidth]{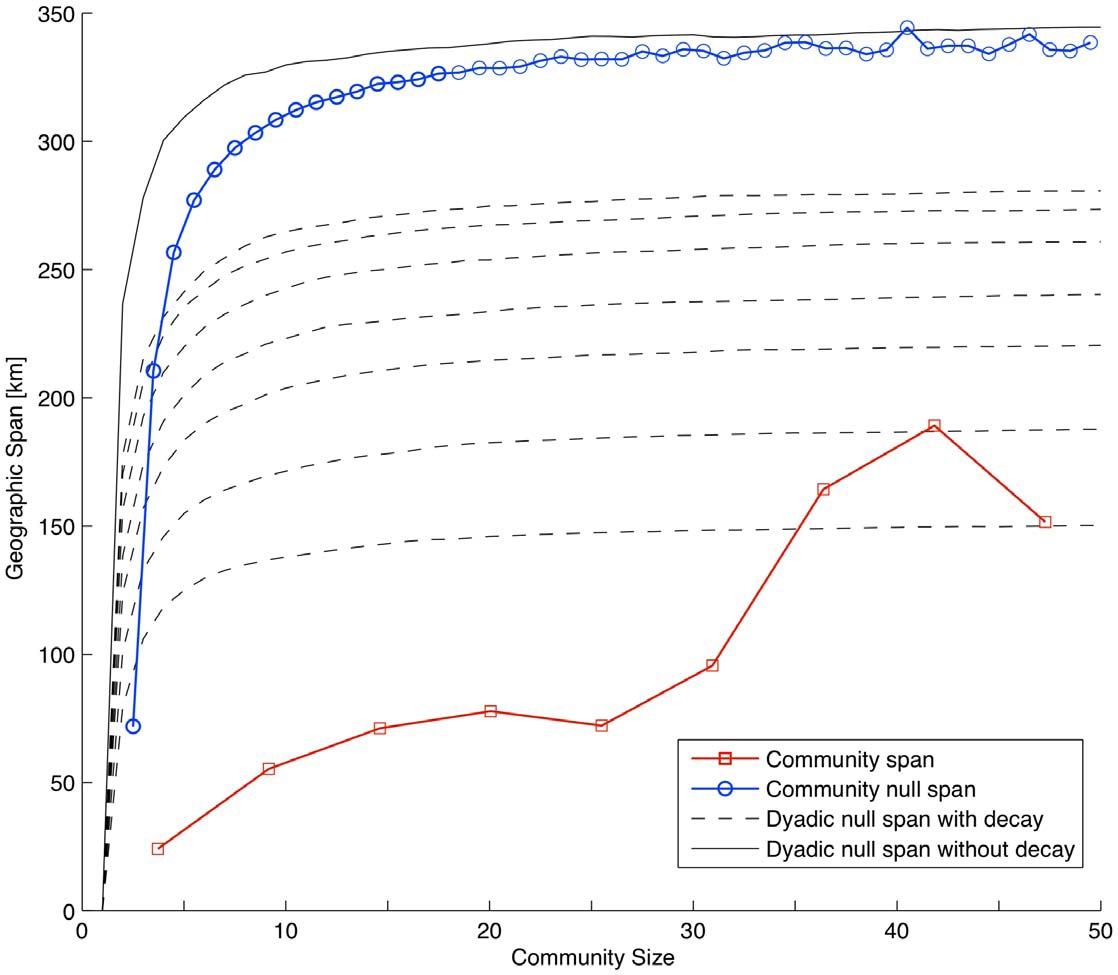}
\end{center}
 \caption{\csentence{Average geographic span (red) for communities and average geographic span for the null model (blue)}. A bump is observed for communities of size 30 and more, which could not be reproduced by the different null models. Figure reproduced from \cite{onnela2011geographic}.}
 \label{fig:survey_geospan}
 \end{figure}

Further, Onnela \textit{et al.}\ also studied the geographic structure of communities, and showed on the one hand that nodes that are topologically central inside a community may not be central from a geographical point of view, and on the other hand that the geographical shape of communities varies with their size. Communities smaller than 30 individuals show a smooth increase of geographical span with size, but bounces suddenly at the size of 30, which could not be clearly explained by the authors, see Figure \ref{fig:survey_geospan}.\\

\subsection*{Geographic partitioning}
The availability to place customers in higher level entities, such as communes or counties, gave researchers the idea of drawing the ``social borders'' inside a country based on the interactions between those entities \cite{grady2012modularity}. Individual call patterns of users are aggregated at a higher level to a network of entities, which can in turn be partitioned into a set of communities based on the intensities of calls between the nodes of this \textit{macroscopic} network. It is important to notice that, in contrast with the \textit{microscopic} network (the network of users), the macroscopic network is not a sparse network at all. Since the nodes represent the aggregated behavior of many users, there is a high chance of having a link between most pairs of communes or counties. Hence, the weights on the links of the macroscopic network are of crucial importance, since they define the complete structure of the network. Such a partition exercise using CDR datasets has been applied, among others, on Belgium, or Ivory Coast \cite{blondel2011voice} \cite{bucicovschiD4Dsocdiv}. An initial study of the communities in Belgium \cite{blondel2010regions} used the Louvain method optimizing modularity for weighted directed networks to partition the Belgian communes based on two link weights: the frequency of calls between two communes and the average duration of a call. The obtained partitions were geographically connected, with the influence of distance, of influential cities, and the cultural barrier of language being observable in the optimal partitions.\\

Given that the intensity of communication between two cities can be well-modeled by a gravity law, Expert \textit{et al.}\ \cite{expert2011uncovering} proposed to replace Newman's modularity by a more appropriate null model, given that geographic information was available. The spatial modularity (SPA) compares the intensity between communes to a null model influenced both by the sizes $c_a$ and $c_b$ of the communes and the distance that separates them
\begin{equation}
p_{ab}^{Spa} = c_ac_bf(r_{ab}).
\end{equation}
The influence of distance is estimated from the data by a function $f$, which is calculated for distance bins $[r-\epsilon, r+\epsilon]$ as
\begin{equation}
f(r) = \frac{\sum\limits_{a,b|r_{ab}\in [r-\epsilon,r+\epsilon]}A_{ab}}{\sum\limits_{a,b|r_{ab}\in [r-\epsilon,r+\epsilon]}c_ac_b}.
\end{equation}
Using their null model, the authors obtained an almost perfect bipartition of the Belgian communes which renders the Belgian linguistic border. Moreover, they showed with a simple example that such a null model allows to remove the influence of geography and obtain communities showing geography-independent features.\\

\begin{figure}[!t]
\begin{center}
\includegraphics[height=10cm]{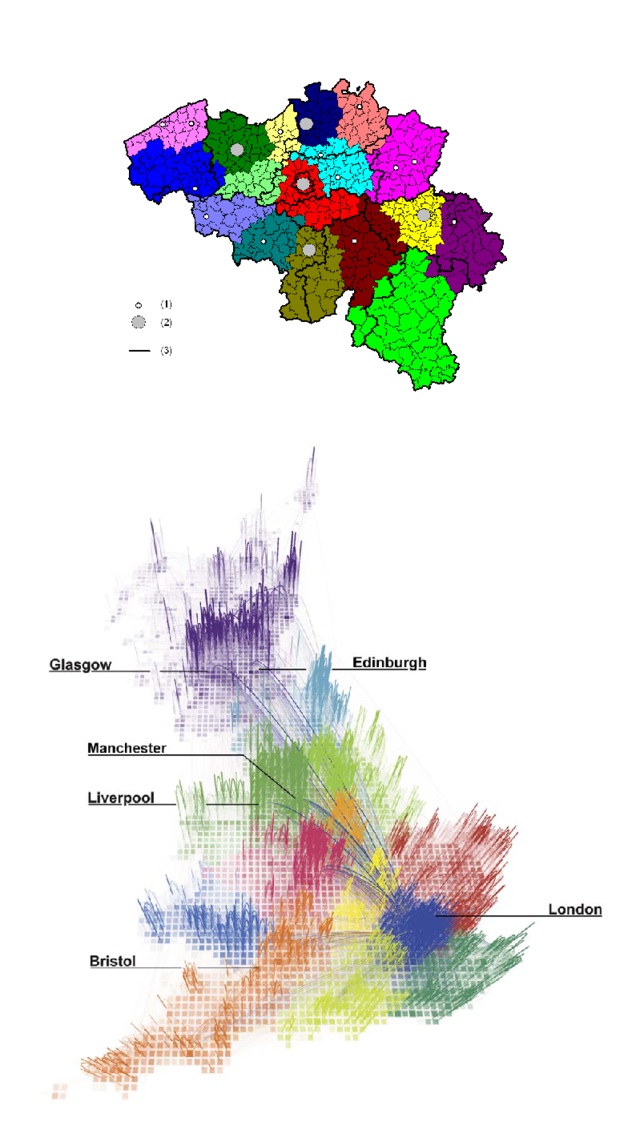}
\end{center}
 \caption{\csentence{Geographic partitioning of countries} (top) Communities in Belgium, obtained through modularity optimization. Communities are geographically well-balanced and are centred around important cities (gray dots). Figure reproduced from \cite{blondel2010regions}. (bottom) Communication network in Great Britain (80\% of strongest links). The colors correspond to the communities found by spectral modularity optimization. Figure reproduced from \cite{ratti2010redrawing}.}
 \label{fig:survey_geopart}
\end{figure}

On an identical topic, Ratti \textit{et al.}\ used an algorithm of spectral modularity optimization, to partition the map of Great Britain \cite{ratti2010redrawing} based on phone calls between geographic locations. Similarly to results obtained on Belgium, they obtained spatially connected communities after a fine-grain tuning of their algorithm, which correspond to meaningful areas, such as Scotland or Greater London, see Figure \ref{fig:survey_geopart}. A stability analysis of the obtained partition showed that while some variation appears on the boundary of communities, the obtained communities are geographically centered at the same place. The intersections between several results of the same algorithm showed 11 spatially well-defined ``cores'' corresponding to densely populated areas of Great Britain. Interestingly, the map of the cores loosely corresponds to the historical British regions. \\

A later study using the data of antenna to antenna volumes of communications in Ivory Coast confirmed the very strong influence of language on the formation of communities in a large country. Using the same method as was used by Blondel \textit{et al.}\  for the Belgian dataset, they show that the borders of the communities formed in Ivory Coast strongly correlate with the language borders, even in the presence of much more than two language groups \cite{bucicovschiD4Dsocdiv}. \\

Going a bit further, Blumenstock \textit{et al.}\ introduce a measure of the social and spatial \textit{segregation} that can be observed through mobile phone communication records \cite{blumenstock2013social}. They define the \textit{spatial segregation} as the proportion of people from ethnicity $t$ in a region $r$ as :
\begin{equation}
w_{tr} = \frac{N_{tr}}{N_r}
\end{equation}  
where $N_r$ is the total population of region $r$. They also define \textit{social segregation} of ethnicity $t$ as the fraction of contacts that individuals of ethnicity $t$ form with the same type of people: 
\begin{equation}
H_{tr} = \frac{s_t}{s_t + d_t}
\end{equation} 
where $s_t$ is the number of contacts that a person of type $t$ has with people from the same ethnicity, and $d_t$ is the number of contacts that people of type $t$ have with people from other ethnicities. With these measures, it is then possible to map the more or less segregated parts of a city, see which ethnicities occupy which regions, and show how strong or weak the links between these ethnicities are. \\

\subsection*{Communications reveal regional economy}
Lately, with the growth of mobile phone coverage even in the most remote regions of the developing world, a new question has risen, namely: is it possible to use CDR data to evaluate the socio-economic state of the different regions of a country? Being able to estimate and update poverty rates in different regions of a country could help governments make informed political decisions knowing how their country is developing economically. \\

\begin{figure}[!t]
\begin{center}
\includegraphics[width=\textwidth]{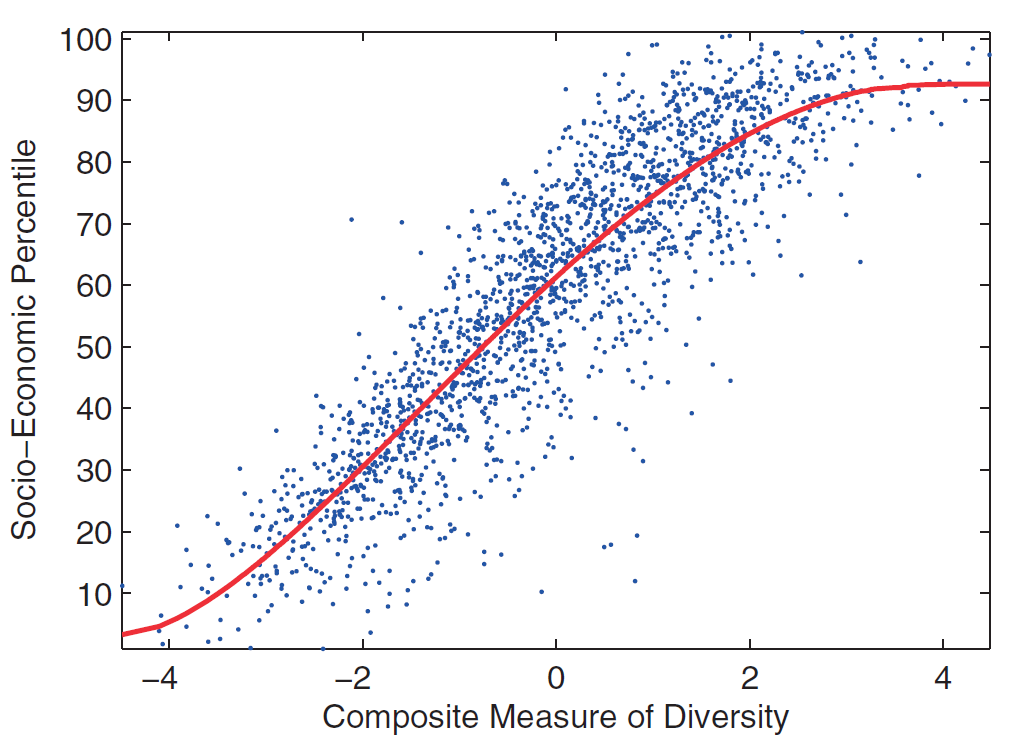}
\end{center}
 \caption{\csentence{Average social wealth as a function of social and geographic diversity.} From Eagle \emph{et al.}, \href{http://dx.doi.org/10.1126/science.1186605}{Network diversity and economic development}, Science 328(5981):1029 (2010)  \cite{eagle2010network}. Reprinted with permission from AAAS. }
 \label{fig:survey_socwealth}
\end{figure}

A first step in that direction was explored by Eagle \textit{et al.}\ in a study using data from the UK \cite{eagle2010network}. The authors investigated if some relationship could be found between the structure of a user's social network and the type of environment in which they live. Using both CDRs of fixed landline (99\% coverage) and mobile phones (90\% coverage), they showed that the social and geographical diversity of nodes' contacts, measured using the entropy of contact frequencies, correlates positively with a socio-economic factor of the neighborhood. Given a node $i$, calling each of his $d_i$ neighbors $j$ at frequency $p_{ij}$, and calling each of the $A$ locations $a$ at frequency $p_{ia}$, his social and spatial diversity are given by
\begin{equation}
D_{social}(i) = \frac{-\sum\limits_{j}p_{ij}\log p_{ij}}{\log d} \qquad D_{spatial}(i) = \frac{-\sum\limits_{a} p_{ia}\log p_{ia}}{\log A},
\end{equation}
which is 1 if the node has diversified contacts. On Figure \ref{fig:survey_socwealth}, the authors compare a composite measure of both diversities with the socio-economic factor of the neighborhood.\\

In a recent study, this time with data from Africa, Mao \textit{et al.}\ tried to determine which characteristics of the mobile phone network could best describe the socio-economic status of a developing region \cite{maoD4Dregeco}. They introduce an indicator named \textit{CallRank}, obtained by running the weighted PageRank algorithm on an aggregated mobile calling graph of Ivory Coast, where nodes are the antennas and the weight of the links are the number of calls between each pair of antenna. They observe that a high CallRank index seems to correspond well to a region that is important for the national economy. However, lacking accurate data to validate the results, they only conclude that this measure is probably a good indicator, without being able to evaluate its accuracy quantitatively. Another analysis of the same dataset was proposed by Smith-Clarke \textit{et al.}\, who extracted a series of features to see which ones showed the best correlation with poverty levels \cite{smithclarke2014poverty}. The authors show that besides the total volume of calls, poverty levels are also linked to deviations from the expected flow of communications: if the amount of communications is significantly lower than expected from and to a certain area, then higher poverty levels are to be expected in that area. Another indicator of poverty was also explored by Frias-Martinez \textit{et al.}\, who analyzed the link between the mobility of people and socio-economic levels of a city in Latin-America \cite{friasmartinez2010socio}. The authors propose several measures to quantify the mobility of users, and show that socio-economic levels present a linear correspondence with three indicators of mobility, namely the number of different antennas used, the radius of gyration and diameter of the area of often visited locations, indicating that the more mobile people are, the less poor the area in which they live seems to be. In a further study by the same research group, Frias-Martinez \textit{et al.}\ go one step further, and propose a method not only to estimate, but also to forecast future socio-economic levels, based on time series of different variables gathered from mobile phone data \cite{frias2013forecasting}. They show preliminary evidence that the socio-economic levels could follow a pattern, allowing for prediction with mobile phone data. \\

\begin{figure}[!t]
\begin{center}
\includegraphics[height=9cm]{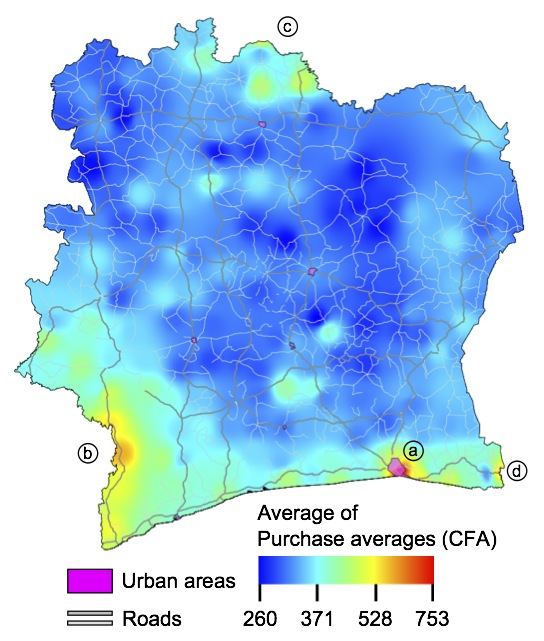}
\end{center}
 \caption{\csentence{Average purchases of airtime credit in Ivory Coast.} (a) Abidjan, (b) Liberian border, (c ) Roads to Mali and Burkina Faso, (d) Road to Ghana. Figure reproduced from \cite{gutierrez2013evaluating}.}
 \label{fig:survey_gutierrez_CIV}
\end{figure}

Another valuable, and rather new, source of data extracted from mobile phone activity is the history of airtime purchases of each user. Using this data on the network of Ivory Coast, Gutierrez \textit{et al.}\ propose another approach to infer the socio-economic state of the different regions of a developing country \cite{gutierrez2013evaluating}. The authors make the hypothesis that people who make many small purchases are probably less wealthy than those who make fewer larger purchases, supposing that the poorer will not have enough cash flow to buy large amounts at the same time. Figure \ref{fig:survey_gutierrez_CIV} shows the map of average purchases throughout the country. Here again, lacking external reliable data to validate those results and compare them with socio-economic data, the authors provide an interpretation of the differences observed between the different regions, and show that the hypothesis they make seems plausible. \\

\section{Adding time -- Dynamical networks} \label{sec:dn}
A particularity of a mobile call graph is that the links are very precisely located in time. Although each call has a precise time stamp and duration, the previously presented studies consider mobile call graphs as static networks, where edges are aggregated over time. This aggregation leads to a loss of information on the one hand about the dynamics of the links (some may appear or disappear during the collection period) but on the other hand about the dynamics \textit{on} the links. Recently, some authors have attempted to avoid this issue by taking the dynamical component of links into account in the definition of such networks. The topic of dynamical -- or temporal--  networks has been studied broadly regarding several types of networks \cite{holme2012temporal}, but the study of mobile phone graphs as evolving ones is rather recent, and given their inherent dynamical nature, mobile call graphs are excellent sources of information for such studies. \\

\subsection*{Dynamics of structural properties}
\begin{figure}[!t]
\begin{center}
\includegraphics[width=\textwidth]{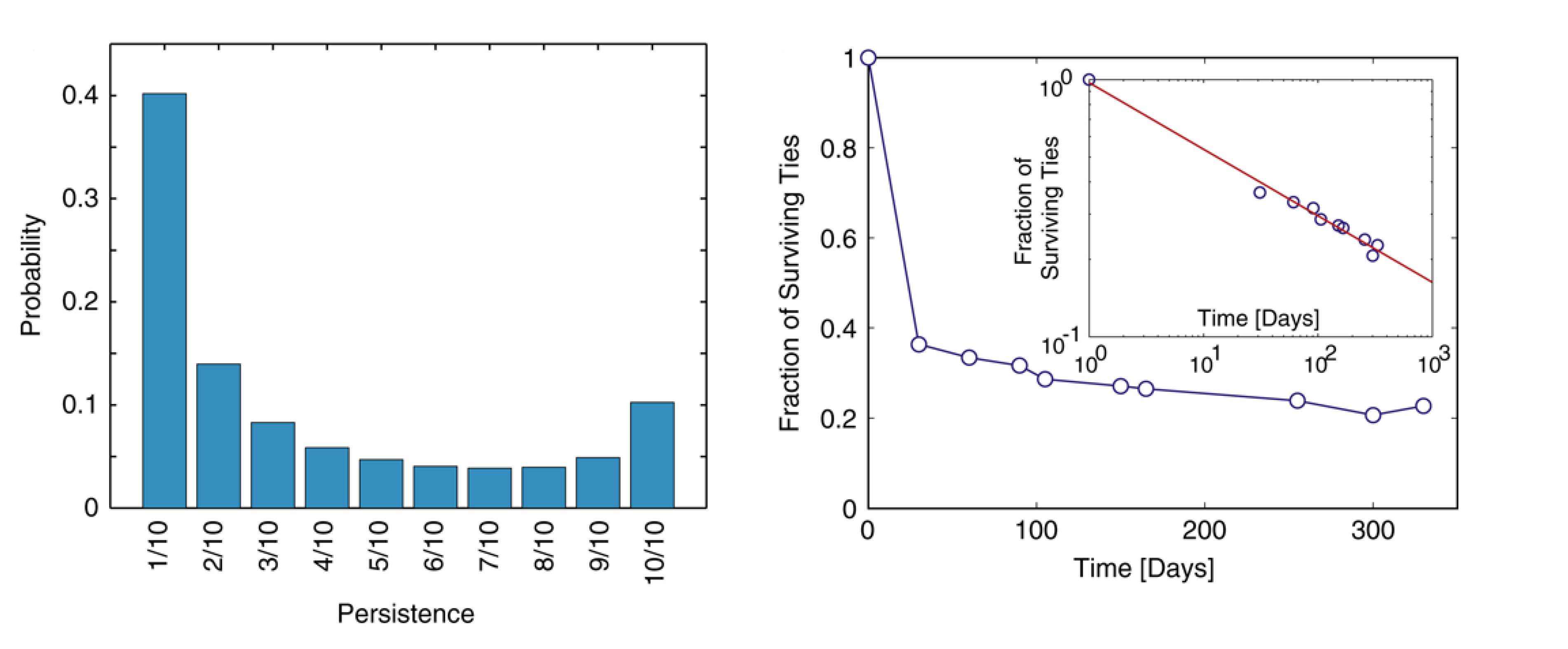}
\end{center}
  \caption{\csentence{Measures of the strength of links over time.} (left) Distribution of the persistence of links. (right) The fraction of surviving links as a function of time follows a power-law like decrease \cite{hidalgo2008dynamics}. Figures reprinted from Physica A: Statistical Mechanics and its Applications, 387(12), Hidalgo, C.A. and Rodriguez-Sickert, C. The dynamics of a mobile phone network, 3017-3024, Copyright (2008), with permission from Elsevier.  }
  \label{fig:survey_linkstrength}
\end{figure}

One such question regards the persistence of links in a mobile phone network. How long does a link last in a network? By analyzing slices of 2 weeks of a mobile phone network, Hidalgo and Rodriguez-Sickert observed that the frequency of presence of links in the different slices, the \textit{persistence}, followed a bimodal distribution \cite{hidalgo2008dynamics}, as illustrated on Figure \ref{fig:survey_linkstrength}. The persistence of link $(i,j)$ is defined as:
\begin{equation}
p_{ij} = \frac{\sum\limits_{T} A_{ij}(T)}{M},
\end{equation}
where $A_{ij}(T)$ is 1 if the link $(i,j)$ is in slice $T$ and 0 otherwise, $M$ being the number of slices.
Most links in the network are only present in one window, and the probability of a link to be observed in several windows decreases with the number of windows, but there is an unexpectedly large number of links that are present in all windows. These highly recurrent links represent thus strong temporally consistent relationships, in contrast with the large number of volatile connections appearing in only one of the slices. A deeper analysis of correlations between the persistence and static measures further shows that clustering, reciprocity and high topological overlap are usually associated with a strong persistence. \\

Raeder \textit{et al.}\ \cite{raeder2011predictors} dig a bit deeper into that last topic, by attempting to predict which link will decay and which will persist, based on several local indicators. They quantify the information provided by each indicator with the decrease of entropy on the probability of an edge to persist, and obtain that the most informative indicators are the number of calls passed between both nodes as well as its scaled version. By trying both a decision-tree classifier and a logistic regression classifier, they manage to predict correctly about 70\% of the persistent edges and decays.\\

\begin{figure}[!t]
\begin{center}
\includegraphics[width=\textwidth]{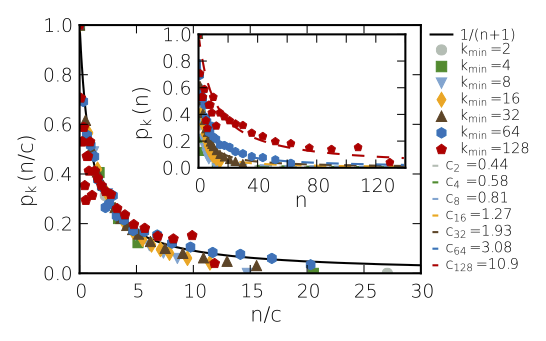}
\end{center}
 \caption{\csentence{Probability of a new communication to form a new tie.} Probability functions $p_k(n)$ calculated for different degree groups. In the inset, symbols show the averaged $p_k(n)$ values for groups of nodes with degrees between the corresponding $k_{min}$ values. Figure reprinted by permission from Macmillan Publishers Ltd: \href{http://www.nature.com/srep/index.html}{Scientific Reports} \cite{karsai2014time}, copyright (2014).}
 \label{fig:survey_karsaidegdist}
\end{figure}

On a very close topic, Karsai \textit{et al.}\ studied how the weights of the links in a network vary with time, how strong ties form, and how this process is related to the formation of new ties \cite{karsai2014time}. They start by measuring the probability $p_k(n)$ that the next communication of an individual that has degree $n$ will occur with the formation of a new $(n+1)^{th}$ tie. This probability depends on the parameter $k$ that corresponds to the final degree of the individual at the end of the observation period. They find that the process of the formation of new ties follows a very consistent pattern, namely 
\begin{equation}
p_k(n) = \frac{c(k)}{n+c(k)}
\end{equation}
where $c(k)$ is an offset constant that depends on the degree $k$ considered. Using the measured $c$ for each degree class, the authors then show that rescaling the distributions $p_k(n)$ allows to collapse all curves into one (see Figure \ref{fig:survey_karsaidegdist}), suggesting that the evolution of the ego-network of each individual is governed by roughly the same mechanism. \\

The reasons for the decay and persistence of links remain various and unknown. However, Miritello \textit{et al.}\ addressed a related question, namely how many links can a person maintain active in time \cite{miritello2013limited}? By looking at a large time-window (around 19 months of data), they evaluate how many contacts are new acquaintances, and how many ties are de-activated during a smaller time-window. It appears that individuals show a finite communication capacity, limiting the number of ties that they are able to maintain active in time:  in the network of a single user, the number of active ties remains approximately constant on the long term. From a social point of view, apart from the balanced social strategy between a user's communication capacity and activity, the authors discern between two kinds of rather extreme behavior that they name \textit{social explorer} and \textit{social keeper}. While the social explorer shows a very high turnover in his social contacts and has a very high activity compared to his capacity, keeping only a very little stable network, the social keeper has a very stable social circle, and only has a very small pace of activating and deactivating ties. The authors further show that the social strategy of an individual can be linked to the topology of its local network. In a related paper, Miritello \textit{et al.}\ \cite{miritello2013time} further show that even though people who have a large network tend to spend more time on the phone than those who have few contacts, the total communication time seems to reach a maximum, and the strength of ties starts decaying for people who have more than 40 contacts. \\

Despite this turnover in links and the fact that links appear and disappear, there seems to be some consistency in a person's network of contacts. In a related study, Saram\"aki \textit{et al.}\ showed how a turnover in contacts did not imply a change in the structure of the local network around a person\cite{saramaki2014persistence}. They study a network of students who, during the time window covered by the dataset, move from high school to college. Despite the very high turnover in a user's contacts, the distribution of the weights on the links around the user, that the authors call the \textit{social signature} of this user, stays very similar through time. \\

From an evolving network perspective, the question of stability and survival of communities is closely linked to the previous questions. Palla \textit{et al.}\ studied the temporal stability of a mobile phone network \cite{palla2007quantifying}, analyzing communities detected on slices of two weeks. They observed that communities have different conditions to survive, depending on their size; small communities require to be stable, while large groups require to be highly dynamic and often change their composition.\\

On a shorter time scale, Kovanen \textit{et al.} identified temporal motifs of sequences of adjacent events involving a small number of nodes (typically 3 or 4) \cite{kovanen2011temporal}. Events are said to be $\Delta t$-adjacent if they have at least one node in common, and the timing between the two events is less than $\Delta t$ (typically of the order of minutes). The authors analyze the most common motifs present in a mobile phone database and find that the most common temporal motifs of three events involve only two nodes, and motifs that allow a causal hypothesis are more frequent than those that do not.\\

The availability of timestamps in datasets allows to segment the calls between office hours and home hours. By supposing that calls made during office hours are for a purpose of business, while private calls are made early morning, in the evening or over the weekend, Cebrian \textit{et al.}\ managed to build two separate networks based on a mobile and landline dataset from the UK \cite{cebrian2010disentangling}. The degree and clustering coefficient distributions of both networks are mostly similar, but a deeper analysis of the network structure shows that some important differences exist between them. By decomposing the network into k-cores and monitoring the speed of information diffusion, they observe that the work network is much more connected than the leisure network, and that information diffuses almost twice as fast. \\

\subsection*{Burstiness}
The dynamics of many random systems are modeled by a Poisson process, where the average interval between two events is distributed following an exponential, well-characterized by its average. However, it has appeared that human interactions show a different temporal pattern, with many interactions happening in very short times, separated by less frequent long waiting times \cite{barabasi2005origin}.

The same holds for mobile phone calls. Karsai \textit{et al.}\ studied the implications of the bursty patterns on the links of a mobile call graph \cite{karsai2011small}. They observed that indeed, the inter-event time ranges over a multiple orders of magnitude, and in particular, the burstiness of human communication induce long waiting times, which slows down the spreading of information over the network (see Section \ref{sec:dyn} for more results on spreading processes). In a further paper \cite{karsai2012universal}, Karsai \textit{et al.}\ also analyzed the distribution of numbers of events in bursty cascades, thus better explaining the correlations and heterogeneities in temporal sequences that arise from the effects of memory in the timing of events. In another study, Wu \textit{et al.}\ find that the distribution of times between two consecutive events is neither a power-law nor exponential, but rather a bimodal distribution represented by a power-law with an exponential tail \cite{wu2010evidence}. 

It is interesting to note that in the previous papers, the authors observed the inter-event time on links, by sorting links by weight. In \cite{candia2008uncovering}, Candia \textit{et al.}\ perform a similar task but for nodes, and measure the inter-event time for nodes, by grouping them based on the number of calls they made. Similarly to Karsai \emph{et al.}'s observations, the inter-event times range over several orders of magnitude, and the distribution is shifted to higher inter-event times for nodes of lower activity. By rescaling with the average of each distribution, the inter-event time distributions collapse into a single curve fitted by a power law with exponent 0.9 followed by an exponential cutoff at 48 days.
\begin{equation}
p(\Delta T) = (\Delta T)^{-\alpha}exp(\Delta T/\tau_c).
\end{equation}

The origin of this burstiness in human behavior has been discussed in several papers in the last few years. It is expected, for example, that people will have more activity during the daytime than at night, and that some times of the day will represent peaks of activity. Therefore, could the burstiness of phone calls only be due to the daily patterns present in our lives? Jo \textit{et al.}\ studied this question and looked at how much of the burstiness of events still remained if they removed the circadian and weekly patterns that appear in a mobile phone dataset\cite{jo2012burstiness}. They dilated (contracted) the time of their dataset at times of high (low) activity. They observed that much of the burstiness remained after removing the circadian and weekly patterns, indicating that there is probably another cause of burstiness coming from the mechanisms of correlated patterns of human behavior. \\

Mobile phone networks are composed of complex patterns and interactions, but still only little work has been done yet in order to characterize these interactions. The temporal arrival and disappearance of more complex structures than simple edges and the timescales of human communication are only two examples of the wide possible research that still needs to be explored in this matter.

\section{Combining space and time -- Mobility} \label{sec:hm}
Given their portability, mobile phones are trusty devices to record mobility traces of users. The availability of spatio-temporal information of mobile phone users has already led to a tremendous number of research projects, and potential applications (see Section \ref{sec:apps}) which would be too large to review exhaustively here. The increasing number of smartphone applications that offer services based on the geolocation of the user are a proof that this information still has a lot of potential uses that are yet to be discovered. In this section, we concentrate on the contributions that present new observations or methods for analyzing and modeling human mobility, while the contributions that propose new applications or uses of these methods are presented in Section \ref{sec:apps}. \\

\subsection*{Individual mobility is far from random}
\begin{figure}[!t]
\begin{center}
\includegraphics[width=\textwidth]{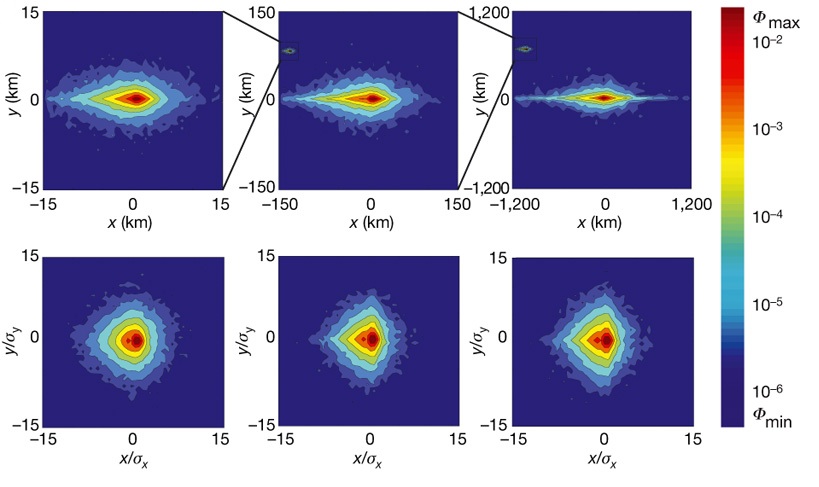}
\end{center}
 \caption{\csentence{Probability of finding a mobile phone user in a specific location.} Probability density function $\Phi(x,y)$ of finding a mobile phone user in location $(x,y)$. The plots, from left to right, were generated for users having a different radius of gyration. After rescaling based on the variance of each distribution, the resulting distribution show approximately the same shape. Figure reprinted by permission from Macmillan Publishers Ltd: \href{http://www.nature.com}{Nature} \cite{gonzalez2008understanding}, copyright (2008)}
 \label{fig:survey_probalocation}
\end{figure}

A mobility trace is represented as a sequence of cell phone towers at which a specific user has been recorded while making a phone call. By studying the traces of 100,000 mobile phone users over 6 months, Gonz\'a{}lez \textit{et al.}\ found that human trajectories show a high degree of temporal and spatial regularity \cite{gonzalez2008understanding}, as illustrated on Figure \ref{fig:survey_probalocation}. This result contrasts with usual approximations of human motions by random walks or L\'e{}vy flights. Their main results showed that all users show very similar patterns of motion, up to a parameter defining their radius of gyration. The regularity is mainly due to the fact that users spend most of their time in a small number of locations. If rescaled and oriented following its principal axis, the mobility of all users can then be described by a single function.
These findings are supported by an additional work produced by Song \textit{et al.}\ \cite{song2010modelling}, who identify significant differences between observational data and two typical models of human displacement: the continuous time random walk and the L\'e{}vy flight. Instead, the authors show that a model mixing the propensity of users to return to previously visited locations and a drift for exploration manages to reproduce characteristics present in their data but absent from traditional models. In their model, each time a user decides to change location, they can either choose a new location with a probability that decreases with the number of already visited locations ($p_{new} \propto S^{-\gamma}$, where $S$ is the number of visited locations, and $\gamma$ a constant), or they can return to a previously visited location. Despite the simplicity of this model they manage to explain the temporal growth of the number of distinct locations, the shape of the probability distribution of presence in each location, and the slowness of diffusion.\\

In another approach, Cs\'{a}ji \textit{et al.}\ show how small the number of frequently visited locations is \cite{csaji2013exploring}. They define a frequently visited location of a user as a place where more than $5\%$ of phone calls were initiated. Using a sample of 100,000 users randomly chosen in a dataset of communications of Portugal, the authors find that the average number of frequently visited locations is only 2.14, and that $95\%$ of the users visit frequently less than 4 locations. Instead of making a list of frequently visited locations, Bagrow \textit{et al.}\ propose another method to group frequently visited locations representing recurrent mobility into one ``habitat" \cite{bagrow2012mesoscopic}. The primary ``habitats" will therefore capture the typical daily mobility, and subsidiary ``habitats" will represent occasional travel. Interestingly, they show that the mobility within each habitat presents universal scaling patterns and that the radius of gyration of motion within a habitat is usually an order of magnitude smaller than that of the total mobility. \\

However synchronized and predictable the mobility of most countries presented here seem to be, most of these studies are based on data from developed countries, where the cultural and lingual diversity do not play as big a role as in the developing world. Amini \textit{et. al.} analyze and quantify the differences between mobility patterns in Portugal and Ivory Coast, and show that models that perform well for developed countries can be challenged by the cultural and lingual diversity of Ivory Coast, that counts 60 distinct tribes \cite{amini2013differing}. They show, for example, that commuters in Ivory Coast tend to travel much longer distances than their counterparts in Portugal, and that mobility patterns vary much more across the country in Ivory Coast than in Portugal. \\

\begin{figure}[!t]
\begin{center}
\includegraphics[width=\textwidth]{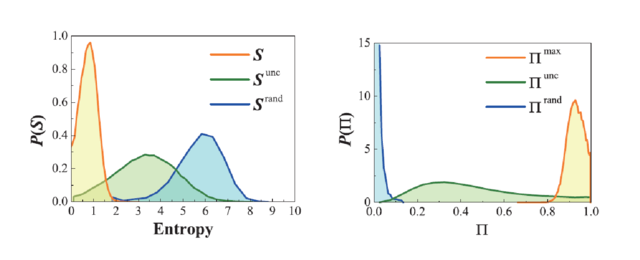}
\end{center}
 \caption{\csentence{Entropy and predictability of the location of users.} (left) Entropy rate of the location of users, for the real, uncorrelated and random data. (right) Maximal predictability of the location of users, for the real, uncorrelated and random data. From Song \emph{et al.}, \href{http://dx.doi.org/10.1126/science.1177170}{Limits of predictability in human mobility}, Science 327(5968):1018 (2010)  \cite{song2010limits}. Reprinted with permission from AAAS.}
 \label{fig:survey_entropylocation}
\end{figure}

If mobility traces are not random, and if users often return to their previous visited locations, could one state that human mobility could be predicted? Song \textit{et al.}\ \cite{song2010limits} addressed this question and investigated to what extent one could predict the subsequent location of a user based on the sequence of his previous visited locations. This predictability is given by the entropy rate of the sequence of locations at which the user is observed. Importantly, one has to point out that not only the frequency of visits at each location is taken into account, but also the temporal correlations between those visits. Their results show that the temporal correlations of the users' displacements reduces drastically the uncertainty on the presence of a mobile phone user, see Figure \ref{fig:survey_entropylocation}. Using Fano's inequality, they deduce that an appropriate algorithm could predict up to 93\% of a user's location on average. The most surprising finding is that not only users are highly predictable on average, but this predictability remains constant across the whole population, whatever distance users are used to travel. While one would expect that people traveling often and far would be less predictable than those who stay in their neighborhood, Song's results seem to point out that there is no variation in predictability in the population.\\

While the aim of the previous work was to show how predictable human motion could be, the authors did not provide any prediction algorithm, keeping their contribution on the theoretical side. Calabrese \textit{et al.}\ went a step further and proposed in \cite{calabrese2010human} a predictive model for the location of people. Their algorithm is both based on the past trajectory of the targeted user and on a general drift of the collectivity, imposed by geographical features and points of interest. The prediction is then a weighted average between an individual behavior and a collective behavior. The individual behavior is modeled as a first-order approximation of the concept proposed by Song \cite{song2010limits}, building a Markov chain where states are locations visited by the user and the probability of moving from state $i$ to state $j$ is proportional to the number of times it has been observed in the data. The collective behavior is then modeled as a weighted average between the influence of distance, points of interest and land use. The predictions of their model on a sample of a dataset containing the records of 1 million people on 4 months shows that in 60\% of their predictions, they manage to predict correctly the next location of a user.  \\
The Markov chain approach used by Calabrese \textit{et al.}\ for modeling the individual behavior is also at the base of a study proposed by Park \textit{et al.}\ \cite{park2010eigenmode}. They showed how the temporal evolution of the radius of gyration of a user can be explained by the eigenmode analysis of the transition matrix of the Markov chain. More precisely, the eigenvectors of the transition matrix provide fine-grain information on the traces of individuals.\\

Instead of looking at the general mobility of people, Simini \textit{et al.}\ focused on the modeling the commuting fluxes between cities, and introduced the \emph{radiation model} \cite{simini2012universal}, overcoming some of the limitations of the gravity model (recall Section \ref{sec:gn}). The radiation model is a stochastic model, assigning a person from a county $i$ to a job of another county $j$ with a probability depending on the estimated number of job opportunities close to the county of origin $i$. The estimated number of job opportunities in a given county is also a stochastic variable proportional to the total population of the county. If we name $d_{ij}$ the distance between counties $i$ and $j$, the average number of commuters between the two counties depends on the population of both counties ($m_i$ and $n_j$, respectively), and of $s_{ij}$, representing the total population in a circle of radius $d_{ij}$: 
\begin{equation}
\langle T_{ij} \rangle = T_i \frac{m_in_j}{(m_i + s_{ij})(m_i+n_j+s_{ij})}
\end{equation}
where $T_i$ is the total number of commuters from county $i$. The radiation model, however efficient, still relies on the knowledge of the distribution of the population, which may be difficult to get in some areas such as the developing world. Overcoming this limitation, Palchykov \textit{et al.}\ suggest a new model using only communication patterns \cite{palchykov2014inferring}. The \emph{communication model} supposes that the mobility between two places $i$ and $j$ is a function of the distance $d_{ij}$ separating the two locations, and of the intensity of communication between these two locations, $c_{ij}$ :  
\begin{equation}
T_{ij} = k \frac{c_{ij}}{d_{ij}^{\beta}},
\end{equation}
where $k$ is a normalization constant. The authors find fitting values for the parameter $\beta$ around 0.98 or 1.08 depending on whether they consider the mobility at intra- or inter-city level, respectively. \\

As it appears, the massive amount of mobility data, which would on first view be considered as random motion, respects a strict routine. Mathematical models, prediction algorithms and visualization tools (see for example Martino's work \cite{martino2010ocean}) have recently shed light on this routine, allowing to construct better human displacement models which can be used to predict epidemics outbreaks. At individual level, this routine appears to be strictly ruling our daily behavior, as Eagle and Pentland \cite{eagle2009eigenbehaviors} show that six eigenvectors of the mobility patterns of users are sufficient to reconstruct 90\% of the variance observed. They also observed that individuals tend to have synchronized behaviors, which will be described in the next paragraph. \\

\subsection*{Aggregate mobility reveal synchronized behavior of populations}
At a higher level, those datasets allow to consider whole populations from a God-eye point of view. More practically, the availability of such massive data allows us first to observe and quantify the interaction of people with their environment, and second to quantify the synchronicity of those interactions.\\

\begin{figure}[!t]
\begin{center}
\includegraphics[width=\textwidth]{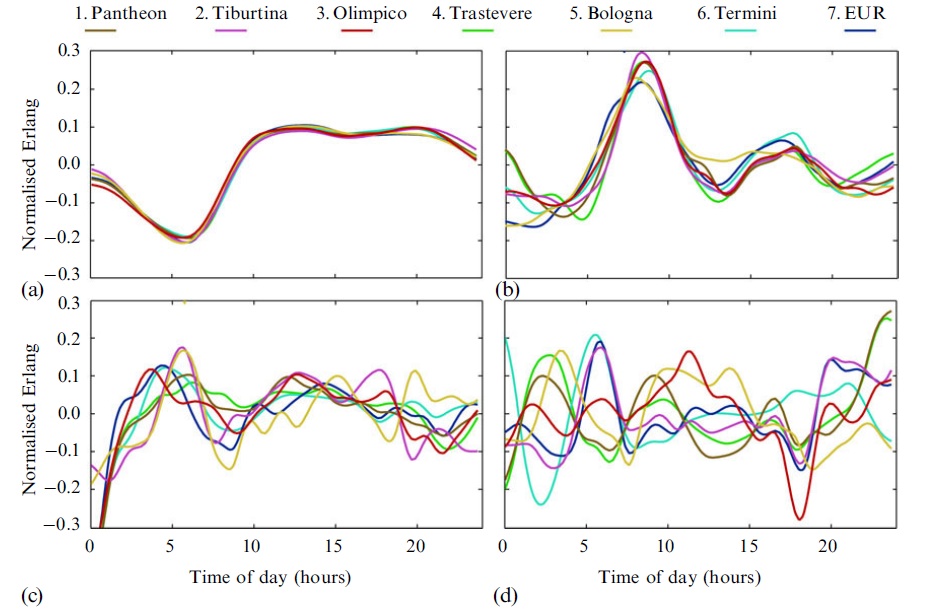}
\end{center}
 \caption{\csentence{Eigenvectors of the Erlang signature of a weekday.} Four principal eigenvectors of the Erlang signature for a weekday of 7 places in Rome. While most of the variance is dominated by the principal eigenvector, representing the normal daily activity, the differences between other eigenvectors indicate differences in space usage. Figure reproduced from \cite{reades2009eigenplaces}.}
 \label{fig:survey_eigenvectors}
\end{figure}

\begin{figure}[!t]
\begin{center}
\includegraphics[width=0.9\textwidth]{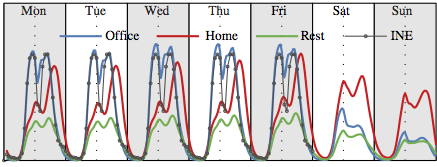}
\end{center}
 \caption{\csentence{Weekly pattern of clusters.} We observe clear differences between calling behavior of work and home locations. Figure reproduced from \cite{csaji2013exploring}.}
 \label{fig:survey_csaji_clusters}
\end{figure}

Initial projects, such as the Mobile Landscapes \cite{ratti2006mobile} project and Real Time Rome \cite{calabrese2011real} have shed light on the potential of such an approach, contributions being essentially visual. However, the next step has been made by Reades \textit{et al.}\ \cite{reades2007cellular}, who used tower signals as a digital signature of the neighborhood. They showed how similar locations presented similar signatures, which implies that a clustering of the urban space is possible, based on the phone usage recorded by its antennas. In particular, the obtained clusters reveal known segmentations of the town, such as residential areas, commercial areas, bars or parks. In short, such a technique may be used as a cheap census method on area usage, which could be of great interest to local authorities. Going a bit further, the same team showed how using an eigendecomposition \cite{reades2009eigenplaces} of the signatures of different locations in town it is possible to extract significant information on differences and similarities in space usage, see Figure \ref{fig:survey_eigenvectors} for the four principal eigenvectors of the signature of a weekday. 
With the same goal in mind, Cs\'{a}ji \textit{et al.}\ \cite{csaji2013exploring} used a k-means clustering algorithm on the activity patterns of different areas to detect which places show the same weekly calling patterns, and thus identify which places typically correspond to work or home calling patterns (see Figure \ref{fig:survey_csaji_clusters}). \\

Beyond the analysis of a single city, Isaacman \textit{et al.}\ explored behavioral differences between inhabitants of different cities \cite{isaacman2010tale}. By analyzing the mobility of hundreds of thousands of inhabitants of Los Angeles and New York City, they showed that Angelenos travel on average twice as far as New Yorkers. Finding an explanation for such a significant difference seems possible, if the inhomogeneities of population density and city surfaces are taken into account. See, for example the work of Noulas \textit{et al.}\ \cite{noulas2012tale}, who show using Foursquare location data that using a rank-based distance, the differences between cities are leveled. A rank-based distance measures the distance between two places $i$ and $j$ as the number of potential opportunities (people, places of interest) being closer to $i$ than $j$. Given the geographic distance $r_{ij}$ and the density of opportunities expressed in radial coordinates and centered in $i$, $p_{i}(r,\theta)$, such a distance reads
\begin{equation}
rank(i,j) = \int_{0}^{2\pi} \int_{0}^{r_{ij}} p_i(r,\theta) r dr d\theta .
\end{equation}
In a city of large population density, there will be more opportunities at short geographical distance than in a city with low population density. Hence, users are likely to travel over shorter distances in city of large population density. These distortions of the use of geographical distance are here leveled by the rank-based distance.
In a recent study, Louail \textit{et al.}\ suggest another way to formalize these differences and analyze the spatial structure of cities by detecting hot-spots or points of interest in 31 spanish metropolitan areas  \cite{louail2014mobile}. The authors show that the average distance between individuals evolves during the day, highlighting the spatial structure of the hot spots and the differences and similarities between different types of cities. They distinguish between cities that are \textit{monocentric} where the spatial distribution is dependent on land use, and \textit{polycentric} cities where spatial mixing between land uses is more important.  
In a similar approach, Trasarti \textit{et al.}\ also analyze the correlations that arise in terms of co-variations of the local density of people, and uncover highly correlated temporal variations of population, at the city level but also at the country level \cite{trasarti2013discovering}. \\

If the detection of the hot-spots and places of interest in a city is possible, then is it possible to go one step further and infer the type of activity that people engage in, from looking at their mobility patterns ? Jiang \textit{et. al.} present a first approach to achieve this in \cite{jiang2013review}, by first extracting and characterizing areas where people will stay or only pass-by, and then infer the type of activity that they engage in depending on the timing of their visit to certain specific locations. In many cases, modeling the mobility of users starts by creating an Origin-Destination matrix that represents how many people will travel between a specific pair of (origin, destination) locations within a given time frame \cite{berlingerio2013allaboard, nanni2013mp4a, angelakis2013mobility}. After extracting which places and times of the day correspond to which activities, Alexander \textit{et al.} propose a method to estimate OD-matrices depending on the time of the day and on the purpose of the trip. The authors' results extracted from data in the area of Boston, are surprisingly consistent with several travel survey sources.  \\

\subsection*{Extreme situation monitoring}
\begin{figure}[!t]
\begin{center}
\includegraphics[width=0.9\textwidth]{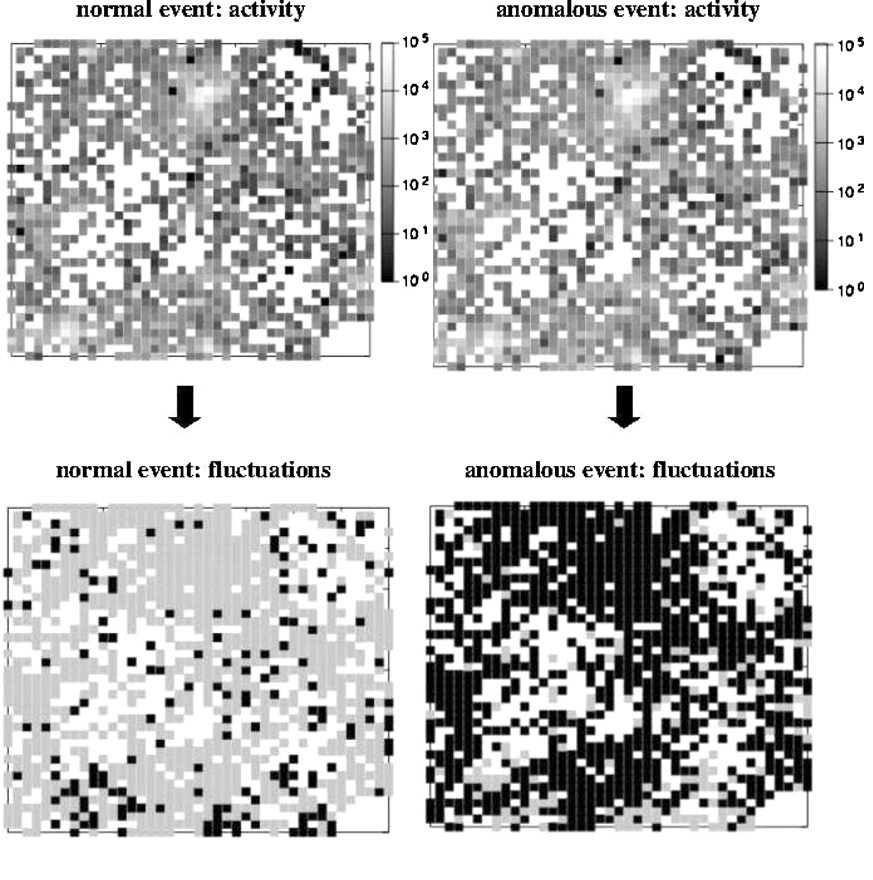}
\end{center}
  \caption{\csentence{Activity and fluctuations during anomalous events.} Activity (top) and fluctuations (bottom) for a normal day (left) and an anomalous event (right). Note that even if no difference is observed on activity, fluctuations are significantly different. Figure reproduced from \cite{candia2008uncovering}. \copyright IOP Publishing.  Reproduced by permission of IOP Publishing.  All rights reserved.}
  \label{fig:survey_candia}
\end{figure}

If the availability of data containing the time-stamped activity of a large population allows to perform monitoring of routine in population activities, it also enables to observe the population's collective response to emergencies. Many recent papers addressed this interesting question. Candia \emph{et al.}, for first, focused on the temporal activity of users at antennas \cite{candia2008uncovering}. They propose a method that is based on the study of the statistical fluctuations of individual users behaviors with respect to their average behavior. As shown on Figure \ref{fig:survey_candia}, in an anomalous case, users show many high fluctuations from their average, while the overall average is close to that of a normal activity. The variance 
\begin{equation}
\sigma(a,t,T) = \sqrt{\frac{1}{N-1}\sum\limits_{i=1}^{N}\left( n_i(a,t,T)- \left\langle n(a,t,T)\right\rangle \right)^2 }
\end{equation}
is computed for each place $a$, for the time interval $[t,t+T]$ between the different individual behaviors $n_{i}(a,t,T)$ and the average expected behavior. Comparing this variance with the normally expected variance allows to identify locations where users are acting abnormally, and that such locations are, in case of emergencies, spatially clustered. In cases of extreme emergencies, the response of populations can even be monitored as geographically and temporally located spikes of activity. 

\begin{figure}[!t]
 \begin{center}
\includegraphics[width=\textwidth]{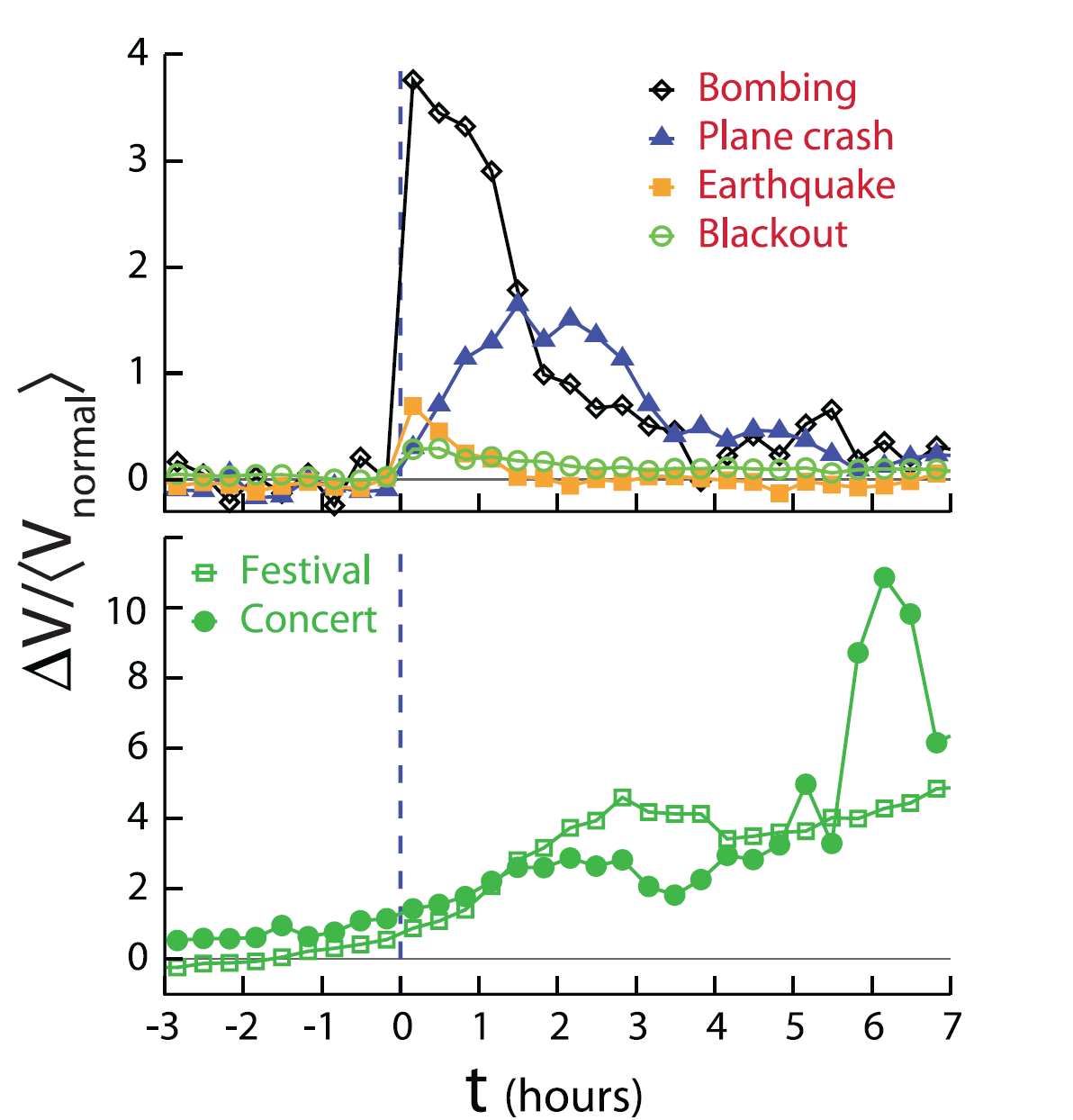}
\end{center}
 \caption{\csentence{Spikes of activity during emergency situations.} The activity has been recorded for users close to the center of activity of several emergency situations, relatively to the normal activity. Figure reproduced from \cite{bagrow2011collective}.}
 \label{fig:survey_bagrow}
 \end{figure}

In a related paper, Bagrow \textit{et al.}\ \cite{bagrow2011collective} analyzed the reaction of populations to different emergency situations, such as a bombing, a plane crash or an earthquake (Figure \ref{fig:survey_bagrow}). They observed such spikes of information when eye witnesses and their neighbors reacted almost directly after the event. The reaction was mostly driven by calls made by nodes who don't usually call at that time, rather than an increase of call rate of usually active nodes. A detailed study of the paths followed by the information during its propagation shows the efficiency of the collective response, with 3 to 4 degrees from eye witnesses being contacted within minutes after the situation. Gao \textit{et al.}\ further analyzed these dynamics in \cite{gao2014quantifying}, and observed that the reciprocity of calls, i.e., ``call-back" actions, showed a sharp increase in emergency cases, such as a bombing or plane crash. The same kind of spikes of behavior, though with different characteristics, are also known to appear at large-scale events, such as concerts or demonstrations \cite{xavier2013understanding, gao2014quantifying}.

Altshuler \textit{et al.}\ have recently also introduced another method they call the \textit{social amplifier} to detect anomalous behavior and thus detect emergencies \cite{altshuler2013social}. Hubs of the network are nodes that have a very high degree, and are thus very well connected to the rest of the network, enabling them to amplify the diffusion of information through the social graph. Using those particular nodes as social amplifiers, the authors show that only analyzing the local behavior of nodes that are close to the hubs of the network can be efficient to detect anomalies of the whole network, and thus detect emergencies. This approach has the advantage that only keeping an eye on a limited fraction of the network is computationally much easier than monitoring and keeping updates on the whole network activity.  \\
Further than \textit{detecting} emergencies, Lu \textit{et al.}\ studied whether the mobility of populations after a disaster could be predicted, analyzing as case study the mobility of populations before and after the 2010 Haiti earthquake \cite{lu2012predictability}. Interestingly, the predictability of people's trajectories remained high and even increased in the three months following the earthquake. The authors also show that the destinations of people who left the capital were highly correlated with their previous mobility patterns, and thus that, with further research,  mobile phone data could be used in the future to monitor extreme situations and predict the movements of populations after natural disasters. These results are very encouraging for many humanitarian organizations who are now trying to use Big Data to save lives. After the earthquake and the following tsunami that struck Japan in 2011, several research teams started a project together combining several big data sources, such as GPS devices, mobile phones, twitter or Facebook to analyze how the analysis of this data could help save lives in the future, if natural disasters were to strike these regions again. This area of research still needs to be explored, especially as so many data sources are now becoming available, combining datasets could prove very useful, and even life-saving for some people. \\

\subsection*{Mobility and social ties}
The common availability of mobility traces and social interactions in the same dataset allows to address causality questions on the creation of social links. From the work of Calabrese \textit{et al.}\, it appears that users who call each other have almost always physically met at least once over a one year interval \cite{calabrese2011interplay}. Users call each other mostly right before or after physical co-location, and interestingly, the frequency of meetings between users is highly correlated with their frequency of calls as well as with the distance separating them.\\

Going a step further, one may wonder if social ties could be predicted using mobility data. Wang \textit{et al.}\ \cite{wang2011human} showed that indeed, nodes that are not connected in the network, but topologically close, and who show similar mobility patterns are likely to create a link. By combining the mobility similarity and the topological distances in a decision-tree classifier, they manage to improve significantly classical link prediction algorithms, yielding in an average precision of 75\% and a recall of 66\%.
Closely related, Eagle \textit{et al.}\ showed on 4 years of data how the social network of people changes drastically when moving from one geographical environment to another \cite{eagle2009community}.\\

\section{Dynamics on mobile phone networks} \label{sec:dyn}
Many networks represent a transport between nodes via their links. In mobile phone networks, the links transport either information (exchanged during phone calls or contained in messages) or non-voice exchanges (SMS, MMS). Information diffusion has opened questions on the speed of the diffusion or on the presence of super-spreaders, with applications in viral marketing or crowd management. The transmission of data has been at the centre of attention only recently, with the rise of new types of computer viruses running on smartphones. \\

\subsection*{Information diffusion}
A phone call is associated to the transfer of information between caller and callee. However, as paradoxical as it may sound, mobile phone datasets are not appropriate to \textit{observe} real propagations of information. The content of phone calls or text messages is, for evident privacy reasons, unknown. Yet, without having access to the content, it is impossible to decide for sure if an observed pattern of calls reflects the transmission of information or if it happens by chance. One can imagine a network with a number of indistinguishable balls circulating between the nodes. Each time a node receives a ball from one of its neighbors, it decides to keep it for a random time interval and after that to transmit it to one of its neighbors. Suppose now that one decides to track the movement of one specific ball. If the number of balls is small compared to the number of nodes, this can still be doable, as long as each node has maximum one ball in its possession. However, if the number of balls increases to become equivalent to the number of nodes, there is a high probability to confuse the paths of several balls. Add to this that balls might be added, removed or duplicated during the process, and one gets a similar situation as trying to track a piece of information in a mobile phone network.\\
This artificial example reflects well the issue of tracking information. Peruani and Tabourier addressed this issue and showed that cascades of information, such as observed in mobile call graphs are statistically irrelevant, and correspond thus probably not to real propagations \cite{peruani2011directedness}. Tabourier \textit{et al.}\ show in a further paper \cite{tabourier2012how} that even though large cascades of information spreading don't seem to happen in mobile call graphs, local short chain-like patterns and closed loops seem to be the effects of some causality and could very well be related to information spreading.\\
In a small number of cases, however, the actual observation of large diffusion of information might be possible. Studying the case of emergencies, such as a plane crash or a bombing, Bagrow \textit{et al.}\ \cite{bagrow2011collective} observed an unusual activity in the geographical neighborhood of the catastrophe. In this case, the knowledge of both the temporal and spatial localization of an unexpected event that is likely to generate a cascade of information allows to assume that the observed sequences of calls are correlated for a specific reason.\\

If, in most cases, the observation of real propagations seems an unreachable objective, a more complete research has been driven in the simulation of propagation of information on complex networks, which results have been extended to questions related to mobile phone networks. There are several ways of modeling information diffusion on networks. A simple way is used in \cite{onnela2007structure} with an SI or SIR model where at each time step, infectious nodes try to infect their neighbors with a probability proportional to the link weight, which corresponds to a sequence of percolation processes on the network. However, mobile phone networks are known to have very particular dynamics (recall Section \ref{sec:dn}), which are not taken into account here. Miritello et al. \cite{miritello2011dynamical} used a formalism similar to the one presented by Newman \cite{newman2002spread} for epidemics, to characterize the \textit{dynamical strength} of a link, which can be used as link weight to map the dynamical process onto a static percolation problem. The dynamical strength, given an SIR model of recovery time $T$ and probability of transmission $\lambda$, is given by
\begin{equation}
\mathcal{T}_{ij}[\lambda,T] = \sum\limits_{n=0}^{\infty}P(w_{ij} = n;T)[1-(1-\lambda)^{n}],
\end{equation}
which is the expected probability of having $n$ calls between $i$ and $j$ in a time range of $T$ multiplied by the probability of propagation given these $n$ calls, summed over all possible values for $n$. Using an approximation of this expression, they manage to link the observed outbreaks to classical percolation theory tools.\\

\begin{figure}[!t]
 \begin{center}
\includegraphics[width=\textwidth]{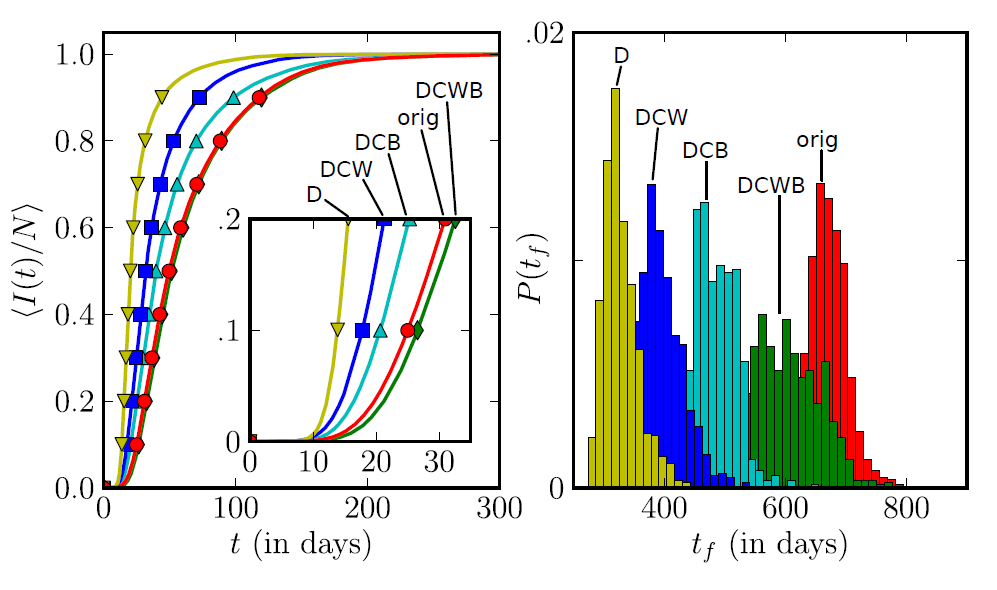}
\end{center}
 \caption{\csentence{Comparison of the speed of spreading processes using different randomization schemes.} (left) Fraction of infected nodes as a function of time for the real (red) data and different randomization schemes. (right) Average prevalence time distribution for nodes. Reprinted figure with permission from Karsai \emph{et al.}, Physical Review E, 83(2):025102, 2011 \cite{karsai2011small}. Copyright (2011) by the American Physical Society. \url{http://dx.doi.org/10.1103/PhysRevE.83.025102}}
 \label{fig:survey_karsaispeed}
\end{figure}

However, such a formalism still neglects the impact of temporal correlations between calls, which significantly slows down the transmission of information over a network. Social networks often exhibit small-world topologies, characterized by average shortest paths between pairs of nodes being very short compared to the size of the network \cite{newman2006structure}.  However, Karsai \textit{et al.}\ \cite{karsai2011small} used different randomization schemes to show that even though social networks have a typical small-world topology, the temporal sequence of events significantly slows down the spreading of information, as illustrated on Figure \ref{fig:survey_karsaispeed}. Kivel{\"a} \textit{et al.}\ \cite{kivela2012multiscale} analyze this topic further, and introduce a measure they call the \textit{relay time}, specific to each link, that represents the time it takes for a newly infected node to spread the information through that link. By analyzing several computations of this relay time, in randomized and empirical networks, they show that the bursty behavior of links and the timings of event sequences are the components that slow down the most the spreading dynamics in mobile phone networks. In another study, Karsai \textit{et al.}\ \cite{karsai2014time} confirm this influence and show that neglecting the time-varying dynamics by aggregating temporal networks into their static counterparts introduces serious biases of several orders of magnitude in the time-scale and size of a spreading process unfolding on the network. \\

From a more theoretical point of view, diffusion processes can be seen as particular cases of dynamical systems. Liu \textit{et al.}\ \cite{liu2011controllability} questioned in this framework the controllability of complex networks. The problem was stated as follows; given a linear dynamical system with time-invariant dynamics
\begin{equation}
\frac{d\mathbf{x}(t)}{dt} = A\mathbf{x}(t) + B\mathbf{u}(t),
\end{equation}
where $\mathbf{x}(t) = (x_{1}(t),\dots,x_{N}(t))^{T}$ defines the state of the nodes of the network at time $t$, $A$ is the (possibly weighted) adjacency matrix of the network, and $B$ an input matrix, what is the minimal number of nodes needed for the input such that the state of each node is controllable, i.e., the system is entirely controllable? From control theory, one knows that a sufficient and necessary condition is that the \textit{reachability matrix} $C=(B,AB,A^2B,\dots,A^{N-1}B)$ is of full rank. From previous work, it is known that the minimal number of nodes required is related to the maximal matching in the network, which can be computed with a reasonable complexity. For example, the authors show that in a mobile phone network, one needs to control about 20\% of the nodes in order to achieve full controllability of the system. Surprisingly, most nodes needed for controlling the network are low-degree nodes, while hubs, that are commonly used as efficient spreaders, are under-represented in the set of input nodes. While the practical interest of this research still needs to be defined, this first result on controllability of networks might open new ideas in the field of information spreading.\\

Finally, one may wonder if the patterns of phone usage are efficient in a collaborative scheme. Cebrian \textit{et al.}\ \cite{cebrian2010measuring} studied this with a small model, where each node of a mobile phone graph is represented as an agent assorted with a state represented by a binary string. The agents are all given the same function $f$, that takes their binary string as input and which is hard to optimize, and which computes their personal score. After each communication, the two communicating agents can modify their state in order to increase their personal score. This modification is done with a simple genetic algorithm, which simulates a cross-over of the states of both agents.\\
Practically, suppose that two agents $i$ and $j$ are respectively in state $\mathbf{x}_{i}^{(t)}$ and $\mathbf{x}_{j}^{(t)}$ at time $t$. These states are both binary strings of length $T$. The agents choose a random integer $c$ in the interval $[1,T]$ and both update their state as
\begin{align}
\mathbf{x}_{i}^{(t+1)} &= \arg\max_{x\in\{\mathbf{x}_{i}^{(t)},\mathbf{y}_1,\mathbf{y}_2\}} f(x)\\
\mathbf{x}_{j}^{(t+1)} &= \arg\max_{x\in\{\mathbf{x}_{j}^{(t)},\mathbf{y}_1,\mathbf{y}_2\}} f(x)
\end{align}
where $\mathbf{y}_1$ is the vector with the $c$ first entries of $\mathbf{x}_{i}^{(t)}$ and the $T-c$ last entries of $\mathbf{x}_{j}^{(t)}$ and $\mathbf{y}_2$ is the vector with the $c$ first entries of $\mathbf{x}_{j}^{(t)}$ and the $T-c$ last entries of $\mathbf{x}_{i}^{(t)}$.\\
The authors observe with this model that the average score on all agents obtained in the real dataset is smaller than for a random topology, which is in line with similar known results from population genetics. Also, perturbation of the time sequence of calls produces a small enhancing of the global fitness.\\

\subsection*{Mobile viruses}
\begin{figure}[!t]
 \begin{center}
\includegraphics[width=\textwidth]{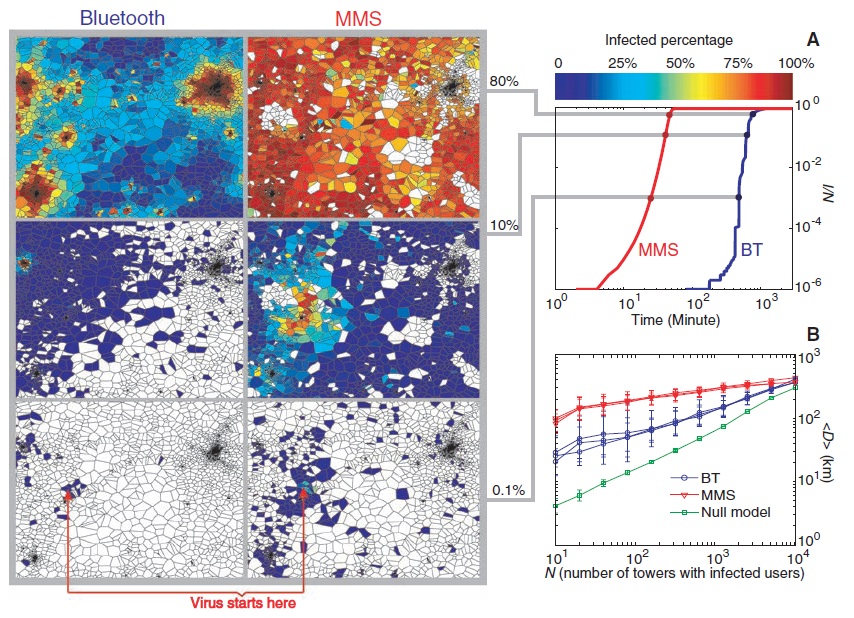}
\end{center}
 \caption{\csentence{Propagation of a mobile virus, either via MMS or Bluetooth service, over the observed area.} From Wang \emph{et al.}, \href{http://dx.doi.org/10.1126/science.1167053}{Understanding the spreading patterns of mobile phone viruses}, Science 324(5930):1071 (2009) \cite{wang2009understanding}. Reprinted with permission from AAAS.}
 \label{fig:survey_mobvirus}
\end{figure}

The study of virus propagations has a long history, may it be biological viruses or more recently computer viruses. Wang \textit{et al.}\ \cite{wang2009understanding} studied a new kind of virus, which spreads over mobile phone networks. Their work is motivated by the increasing number of smartphones, which have high-level operating systems like computers, which leads to a higher risk of an outbreak. So far, despite the large number of known \textit{mobile viruses}, no real outbreak has been noticed. The reason for this is that mobile viruses function only on the operating system for which they are designed for. An infected phone can hence only transfer the virus to its contacts running on the same operating system. As exposed by Wang \textit{et al.}\, this situation corresponds to a site percolation procedure on the network of possible contacts. Given the actual market shares of the main operating systems, the authors showed that those were below the percolation transition of the contact network. The study concerns two types of spread available for viruses: the diffusion via Bluetooth and via Multimedia Messaging System (MMS). Both diffusions show major differences in spreading patterns; Bluetooth viruses spread relatively slow and depend on user mobility. In contrast, MMS epidemics spread extremely fast and can potentially reach the whole network in a short time, see Figure \ref{fig:survey_mobvirus}. However, currently they are contained in small parts of the network, due to the different operating systems. In conclusion, the authors deduce thus that if no outbreak has taken place so far, it is not due to the lack of efficient viruses, but it is rooted in the fragmentation of the call graph. However, the current evolution of the market leads to a situation where some operating systems are gaining a large market share, which could lead to a more risky situation.\\

In a subsequent study, Wang \textit{et al.}\ \cite{wang2010new} show how the scanning technique, where MMS malware generate random phone numbers to which they try to propagate instead of using the address book of their host, increases the probability of a major outbreak, even when the market share of operating systems are too low for having a giant component. Operators can detect such outbreaks by monitoring the MMS traffic of their network and observe suspicious increases of volume. However, given enough time, viruses can infect a large fraction of the network without being detected by operators. Smart anomaly detection schemes may prevent such outbreaks, as well as a reduction of market shares of operating systems. 
Wang \textit{et al.}\ also compare the last two strategies in a further paper \cite{wang2013understanding}. They study the effectiveness of topological viruses versus viruses that also use a scanning technique. The authors show that topological viruses, i.e., those that spread through the contact network of infected phones, are the most effective for an operating system that has a large market share, whereas the scanning technique will generate a bigger outbreak in the case of a low market share operating system.

\section{Applications in urban sensing, epidemics, development.}\label{sec:apps}
The last few years have seen the rise of Big Data and of its uses, and in many regards, this is rapidly changing our lives and way of thinking. Further than observing those networks of mobile phone calls, or modeling social behavior, many researchers now engage in finding new ways of using mobile phone data in everyday life. 

\subsection*{Urban sensing}
As showed in the previous sections, mobile phone data allows to observe and quantify human behavior as never before. Besides purely sociological questions, this data also opens a number of potential applications, which gives to this data an intrinsic economical value, thinking of geo-localized advertising applications \cite{baccelli2011modeling}. Recalling that an increasing fraction of the available smartphone applications record the user's geolocation -- whether it is necessary for the app to work or not -- it is easy to understand that this information is valuable to target the right users when making advertising campaigns, or simply to understand the profile of the application's users. Mobile phones are more and more becoming a way of taking the pulse of a population, or the pulse of a city, and we expect that in the future, more and more cities will make development plans based on information gathered from mobile phone data. In this framework, recent research has shown that mobile phone data could detect where people are \cite{deville2014dynamic} and where people travel to \cite{csaji2013exploring} including the purpose of their trips \cite{jiang2013review}. If these findings are applied to a whole city and points of interest are uncovered via mobile phone data (recall Section \ref{sec:hm}), then the whole organization of urban places can be influenced by the knowledge gained from this data. Urban sensing is only shortly addressed here, but has been a popular topic in the last few years, and we refer the interested reader to a recent survey of contributions in this specific field \cite{calabrese2014urban}.\\

We have previously addressed the possibility of using mobile phone signatures as a cheap census technique, Isaacman \textit{et al.}\ take this analysis a step further and show how one can derive the carbon footprint emissions \cite{isaacman2011identifying} based on the mobility observed from mobile phone activity.\\

\begin{figure}[!t]
 \begin{center}
\includegraphics[width=0.9\textwidth]{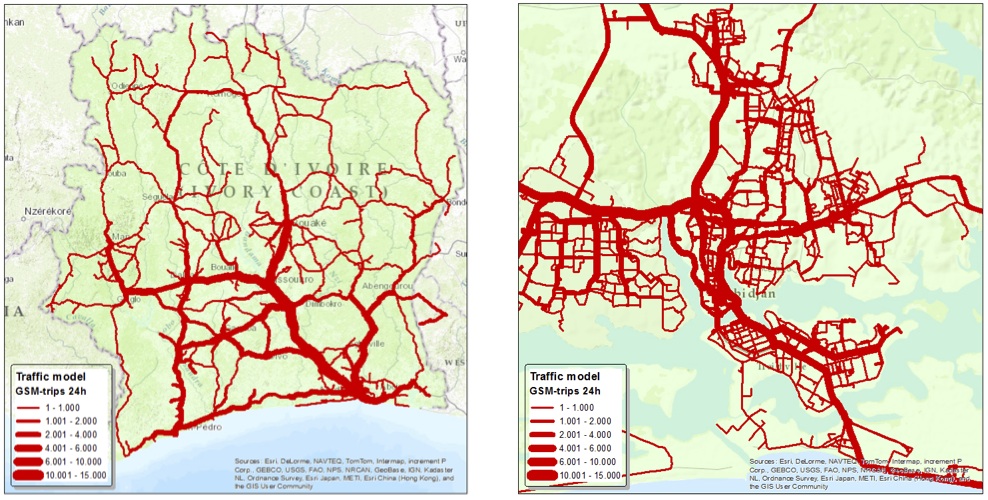}
\end{center}
 \caption{\csentence{Traffic model for 24 hour period for Ivory Coast (left) and Abidjan Area (right).} Figure reproduced from \cite{nanni2013mp4a}.}
 \label{fig:survey_nanniroads}
\end{figure}

Many applications of modeling mobility aim towards transport planning and monitoring traffic with evident applications in accident management and traffic jam prevention. Over the last (almost) 20 years, a large number of attempts have been made to enhance prediction using mobile phone data. This topic is only shortly addressed here with a few recent contributions, but for more information on the research in this field, we will refer the interested reader to a review published in 2011 \cite{steenbruggen2011mobile}. One example of such an application was proposed by Nanni \textit{et al.}, who create the OD-matrix of Ivory Coast and then assign this matrix to the road network \cite{nanni2013mp4a} to produce a map (see Figure \ref{fig:survey_nanniroads}) modeling the traffic of the main roads of the country, showing estimated traffic flows. In a similar approach, Toole \textit{et al.}\ estimate the flow of residents between each pair of intersections of a city's road map \cite{toole2014path}. They show that these estimations, coupled with traffic assignment methods can help estimate congestion and detect local bottlenecks in the city. In a related study, Wang \textit{et al.}\ examine in more details the usage patterns of road segments, and show that a road's usage depends on its topological properties in the road network, and that roads are usually used only by people living a small number of different locations \cite{wang2012understanding}. The authors further show that taking advantage of this observation helps create better strategies for reducing travel time and congestion in the road network of a city. \\

Going one step further, Berlingerio \textit{et. al.} designed an algorithm to detect which means of transport people would chose, including public transportation or private means, to infer how many people used which public transportation routes \cite{berlingerio2013allaboard} throughout the day. The authors then proposed a model of the network of local transportation of Abidjan highlighting the routes that are taken most often. Then, they are able to show how specific little changes to the network could improve the average travel time of commuters by $10\%$.  Among other possible uses of information on commuting flows, McInerney \textit{et. al.} suggested using the regular mobility of people for physical packages delivery to the most rural areas \cite{mcinerney2013crowdsourcing}, showing on the one hand, the feasibility of this method, and on the other hand reducing by $83\%$ the total delivery time for rural areas. Other applications of prediction algorithms for the next journey of users include, for example, a recommender system for bush taxis such as suggested by Gambs \textit{et. al.} \cite{gambs2013towards}, using the predicted next location of users to recommend to pedestrians adapted means of transport that are in their neighborhood.  \\

By monitoring the movements of people towards special planned events, Calabrese \textit{et al.}\ \cite{calabrese2010geography} show that the type of events highly correlates to the neighborhood of origin of the users. Such a cartography of taste can be used by authorities when planning the congestion effects of large events, or for targeted advertising of events (see Quercia \textit{et al.}\ \cite{quercia2010recommending}). In a closely related approach, Cloquet and Blondel use the analysis of anomalous behavior in mobile phone activity to predict the attendance to large-scale events such as demonstrations or concerts. The authors propose, as a first step in that direction, a method to determine the time when no more people will arrive to a certain event \cite{cloquet2014forecasting}. To do this, they propose two methods. The first method uses the mobility of people that are traveling towards the event to model the flux of the arriving or leaving crowd. The second method is based on the recorded interactions between people that are already at the event and other users that are within 20km. The authors show that using these methods, they are able to predict the time when no more people will join the event up to 43 minutes in advance. Another related application was explored by Xavier \textit{et al.}\, who analyzed the workload dynamics of a telecommunication operator before and after an event such as a soccer match \cite{xavier2012analyzing} in order to help the management of mobile phone networks during such events. \\

Finally, mobility traces can also be used to monitor temporal populations \cite{manfredini2011monitoring}, such as tourists. Kuusik \textit{et al.}\ \cite{kuusik2009analysing} studied the mobility of roaming numbers in Estonia for 5 consecutive years, showing the potential for authorities to understand and efficiently target visiting tourists.\\

\subsection*{Infectious Diseases}
In recent years, a lot of research has been done in order to use Big Data to help monitor and prevent epidemics of infectious diseases. If one can model information spreading in mobile phone networks (recall Section \ref{sec:dyn}), then the same theory could also be used to model the spreading of real infectious diseases. As mobile phone data can help follow the movements of people (recall Section \ref{sec:hm}), these movements can also provide information about how a disease could travel and spread across a country. The dynamics at hand usually depend on the type of disease and how it can be transmitted, hence many articles, of which we will review a few here, propose different models based on the mobility of people to predict the spread of an epidemic.  \\

\begin{figure}[!t]
 \begin{center}
\includegraphics[width=\textwidth]{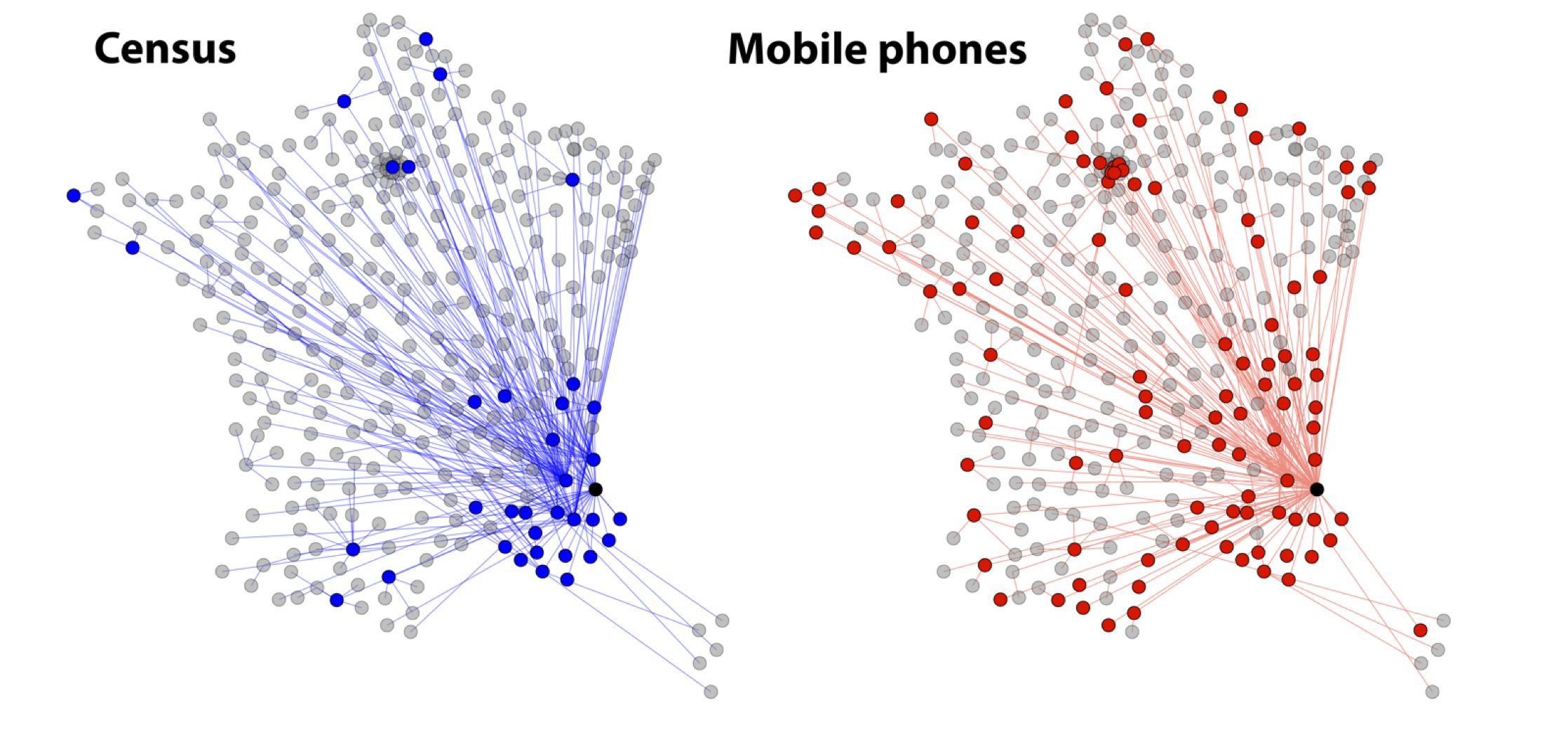}
\end{center}
 \caption{\csentence{Epidemic invasion trees.} Invasion trees observed using the census (left) and the mobile phone network (right), the seed of the simulation is in Barcelonnette (black node). Figure reproduced from \cite{tizzoni2014use}.}
 \label{fig:survey_tizzoniFrance}
\end{figure}

Using mobile phone traces, Wesolowski \textit{et al.}\ measure the impact of human mobility on malaria, comparing the mobility of mobile phone users to the prevalence of malaria in different regions of Kenya, and identify the main importation routes that contribute to the spreading of malaria \cite{wesolowski2012quantifying}. In another study, Tizzoni \textit{et al.}\ \cite{tizzoni2014use} validate the use of mobile phone data as proxy for modeling epidemics. The authors extract a network of commuters in three European countries by detecting home and work locations for each mobile phone user, and compare this network with the numbers of commuters obtained by census. On these networks of commuters, they trace agent-based simulations of epidemics spreading across the country. They show that the invasion trees and spatio-temporal evolution of epidemics are similar in both census and mobile phone extracted networks of commuters (see Figure \ref{fig:survey_tizzoniFrance}). Most models assume, lacking additional information, homogenous mixing between people that are physically within the same region or area. Frias-Martinez \textit{et al.} propose another agent-based model of epidemic spreading, using individual mobility \emph{and} social networks of individuals to build a more realistic model \cite{frias2011agent}. Instead of assuming homogeneous mixing within a given area, an individual will have more probability of meeting an infected agent that is in the same area if they have communicated with each other before. The authors further divide the social network of contacts and the mobility model of an individual between weekday and weekend to achieve better accuracy. \\

Going a step further, a few contributions to the D4D challenge \cite{blondel2012data} investigated which would be the best ways to monitor and \emph{influence} an epidemic rather than just predicting its spread. In this framework, Kafsi \textit{et al.}\ \cite{kafsi2013mitigating} propose a series of measures applicable at the individual level that could help limit the epidemic. They investigate the effect of three different recommendations, namely (1) do not cross community boundaries; (2) stay with your social circle and (3) go/stay home. Considering that either of these three recommendations could be sent via their mobile phone to different users in the network, and that probably only a fraction of the contacted users would participate, the authors evaluate the impact that implementing this system could have on the spreading process. They show that these measures can weaken the epidemic's intensity, delay its peak, and in some regions, even seriously limit the number of infected individuals. Using the same dataset, Lima \textit{et al.}\ proposed a different approach \cite{lima2013exploiting}, namely using the connection between people to launch an information campaign about the epidemic, in the hope to reduce the probability of infection if an individual is better informed about the risks. The authors use an SIR model and the observed mobility of mobile phone users to simulate epidemics unfolding on a population, and evaluate the impact of geographic quarantine on the spreading of the disease, as well as the impact of an information campaign reducing the risks of infection for ``aware" individuals. They show that the quarantine measures don't seem to delay the endemic state, even when almost half the population is limited to their own sub-prefecture, whereas the information campaign, less invasive, seems to limit significantly the final fraction of infected individuals, opening this topic for further research. \\

This field of research has shown again how valuable mobile phone data could be to save lives, and potentially monitor and limit epidemics of infectious diseases. However, most models and studies are limited by the lack of ground-truth data to compare their results with. Indeed, how would you know who an individual got the disease from, and what was its exact route towards each infected person?  Another shortcoming of this area of research comes from the current difficulty of gaining access to those mobile phone datasets, especially to cross-border mobility. If modeling mobility in Africa could be useful to containing the current Ebola outbreak, cross-border mobility would be very valuable data, as discussed in \cite{wesolowski2014containing}. However, gaining access to these data is more difficult as it involves getting the approval from more than one country for a single dataset. In \cite{demontjoye2014enabling}, the authors suggest guidelines to share data for humanitarian use, while preserving the privacy of users. \\

\subsection*{Viral marketing}
In 1970, Katz and Lazarsfeld introduced the breakthrough idea that, more than mass media, the neighborhood of an individual is influencing their decisions \cite{katz1970personal}. This idea has induced the concept of opinion leaders -- persons who have a high influence on their neighborhood --, although some debate exists on the exact role played by opinion leaders \cite{watts2007influentials}, and introduced the concept of viral marketing. In opposition to direct marketing, the principle of viral marketing is that consumers respond better to information accessed from a friend than to information provided through direct means of communication. Viral marketing searches thus for means of making people communicate about a brand, in order to push friends of an early adopter to adopt the product in their turn. In particular, mobile viral marketing has proved to be an effective means of propagation of such marketing campaigns. The influence of one's neighbors can be observed using CDR data coupled to data on product adoption. In a study of the adoption of 4 mobile services, Szab\'o{} and Barab\'a{}si \cite{szabo2006network} showed that the adoption of a product by a user was highly correlated to the adoption of their neighbors for some services only, while other services were not showing any viral attribute. A similar study by Hill \textit{et al.}\ \cite{hill2006network} on the adoption of an undisclosed technological service showed again that neighbors of nodes that had adopted the service were 3 to 5 times more likely to adopt the service than the best-practice selection of the company's marketing service. A related result was also obtained in the FunF project by Aharony \textit{et al.}\ \cite{aharony2010tracing}, who showed that the number of common installed applications was significantly larger for pairs of users having often physical encounters. Risselada \textit{et al.}\ \cite{risselada2014dynamic} further showed that the influence of one's neighbors on the adoption of a product evolved with time, depending on the elapsed time since the introduction of the product on the market.\\

Even though one could use a simple SI or SIR model to characterize viral marketing, it is more likely in this case, that a user will adopt a product if several of its neighbors have already adopted it and the information comes from several different sources. One of the possible ways to model these dynamics is to use a threshold model: each user is assigned a threshold. A node will adopt a product if the proportion of its neighbors that have adopted the product is above the node's threshold. The model can be either deterministic, and decide a priori a same threshold for all nodes, or stochastic and draw thresholds from a probability distribution. To take into account the timing of contacts between people, one can then add to this model the condition that a node will adopt a product if it has enough contacts with different neighbors that have adopted the product within a given time frame. Backlund \textit{et al.}\ have studied the effect of timings of call sequences on those models \cite{backlund2014effects}. Here again, they observe that the burstiness of events tends to hinder propagation of adoption of a product, increasing the waiting times between contacts compared to a randomized sequence of contacts. \\

The identification of ``good'' spreaders for a viral marketing campaign is tough work, especially given the usually very large size of the datasets, which makes it hard to extract informational data in a small time frame. With this in mind, the authors of \cite{blondel2006social} proposed a local definition of \textit{social leaders}, nodes that are expected to play an influential role on their  neighborhood. They defined the social degree of a node as the number of triangles in which the node participates, and social leaders as nodes that have a higher social degree than their neighbors. This definition has its use in marketing campaigns, to identify the customers who should be contacted to start the campaign, which proved to be efficient \cite{deranking}. Moreover, social leaders can also be used to reduce the complexity of a network, by only analyzing the network of social leaders instead of the whole network, with possible uses in visualization and community detection. \\

\subsection*{Data for development}
The last couple of years have seen a spectacular rise of interest for applications of mobile data for the purpose of helping towards development. Many contributions to the ``Data for Development" (or D4D) challenge launched by Orange \cite{blondel2012data} used different bits of information from the data of mobile phone users to help the development of Ivory Coast. Several of these contributions have already been reviewed in the previous paragraphs, for the full set of research projects, see \cite{D4Dbook}. \\

While in the developed world, much information of what can be inferred from mobile phone data is already known (population density, some of the mobility traces,...), this information can be very valuable in the developing world where census data is often unavailable or several years old. Modeling the mobility of people in developing countries can provide very useful information for local governments when making decisions regarding changes in local transportation networks, or urban planning. Indeed, in rural areas of low income countries where the most recent technologies are not always available, up to date information on how many people commute from one place to another can be very useful and help policy makers to decide on the next steps towards development.  Sometimes, very basic information such as drawing the road network can be difficult in remote places. Salnikov \textit{et al.}\ used the D4D challenge dataset to detect high traffic roads by selecting displacements only within a certain range of velocities \cite{salnikov2014geography}. They were able to redraw the main road structure of the country and even identified unknown roads, which they validated a posteriori. Between techniques for cheap census, mobility planning and fighting infectious diseases applications, we expect that in the next few years, the developing world will profit from the availability of such rich databases, and research will provide useful insights into how to better help towards development. 

\subsection*{Data representativity}
Finally, one may raise the question of the significance of the data: given that only a fraction of a country's population is reached by one operator, to which extent may the results on a dataset be generalized to larger populations? Clearly, quantitative results obtained in these studies, such as the degrees of nodes, cannot be taken for granted, but one may expect that as long as the population sample is not biased, qualitative observations such as the broadness of degree distribution or the organization of nodes in communities are significant information on the structure of communication networks. However, the question of knowing whether the sample is biased or not is almost impossible, especially given the lack of information about the users in CDR databases.\\

Frias-Martinez \textit{et al.} raised this question in \cite{frias2012relationship}, regarding e.g. the socio-economic level that could be biased among mobile phone users compared to the whole population. They validate their results by performing a series of statistical tests to compare the population in their sample to the overall population using census data, and show that no significant difference was observed. However, in the general case, data about users in CDR databases is often missing, and census data may not always be available for comparison. Regarding mobility models, one could argue that active mobile phone users are more likely to be on the move than the rest of the population. A mobility model based on mobile phone users is therefore likely to overestimate the number of people within a population that are traveling. Buckee \textit{et al.}\ raised this question regarding those models, further arguing that bias in models of mobility could, in turn, influence the spreading of modeled epidemics \cite{buckee2013mobile}. Onnela \textit{et al.}\ also address this problem studying how paths differ depending how much of the network is observed \cite{onnela2012spreading}. They show that, counterintuitively, paths in partially observed networks may appear shorter than they actually are in the underlying full network. \\

Ranjan \textit{et. al.} studied a related question regarding the mobility of users \cite{ranjan2012are}: given that one only sees data points where and when a user has made a phone call, to which extent are these points representative of a user's mobility?. They found that sampling only voice calls of an individual will most of the time do well to uncover locations such as home and work, but will also, in some cases, incur biases in the spatio-temporal behavior of the user. In a recent study, Stopczyncki \textit{et al.}\ widen their coverage by coupling databases from many sources on the same set of users \cite{stopczynski2014measuring}. While this approach clearly captures more than just studying mobile phone records, its coverage is limited (1,000 subjects) as the users had to give their explicit consent to share their data: facebook interactions, face-to-face encounters, and answers to a survey. The authors are therefore able to analyze a bigger picture than other studies based on only mobile phone data and show that only studying mobile phone data may not be enough to capture a user's comprehensive profile. Learning from these studies, one should therefore be cautious when drawing conclusions from such analyses, and keep in mind that observing the traces left by mobile phones is only observing selected parts of the whole picture.

\section{Privacy issues} \label{sec:pi}
The collection and availability of personal behavioral data such as phone calls or mobility patterns raises evident questions on the security of users'privacy. The content of phone calls or text messages is not recorded, but even the simple knowledge of communication patterns between individuals or their mobility traces contains highly personal information that one typically does not want to be disclosed. During the past decade, a fairly high amount of personal data was made available to researchers via, among others, CDR datasets. The companies sharing their data do not always know how much personal information can be inferred from the analysis of such large datasets, and this has led, so far in other cases than mobile phone data, to a few scandals in the recent years \cite{singel2009netflix, barth2012re}. In turn, these incidents led, in 2012, to a procedure of adaptation of legal measures in Europe \cite{directive2012eu}: the previous european law on the protection of privacy and data sharing dated back from 1995 \cite{directive199595}, long before the era of what is now called ``Big Data". \\

The procedure often used when a company shares private data with a third party such as a research group is the  following: the company keeps on secured machines the exact private information such as names, addresses or phone numbers on their customers, as well as the CDRs, which contain the phone number of the caller, the callee, the time stamp of the call, the tower at which the caller was connected, idem for the callee, and additional information such as special service usage and so on. The anonymization procedure consists then in replacing each phone number by a randomly generated number, such that each user has a unique random ID, from which it is impossible to retrieve the original phone number by reverse engineering procedures. The CDRs are then modified such that phone numbers are replaced by the corresponding ID. After this procedure, the CDRs are anonymized, and can be transferred to a third party. The standard procedure then implies that the third party signs a non-disclosure agreement, stipulating that they cannot make the CDR data available, and the agreement usually also restricts the range of potential research questions to be explored with the data. The safety of users privacy is then guaranteed both by the removal of information allowing to identify users and by the assumption that the third party doesn't make use of the data for any malicious intent.\\

\subsection*{De-anonymization attacks}
Some research has been produced on mobile phone datasets to challenge this apparent feeling of security, however, recent results are opening new ways of considering the privacy problem. Using CDR data containing mobility traces, Zang and Bolot \cite{zang2011anonymization} show how it is possible to uniquely identify a large fraction of users with a small number of preferred locations. Their methodology goes as follows: for each user, it is possible to list the top N locations at which calls have been recorded. The authors show then that depending on the granularity of the locations, a non-negligible fraction of users may be uniquely identified by only 2 locations. For example, if locations are taken at cell level, up to 35\% of the users of a 25 million communication network can be uniquely identified with 2 locations, which will be likely to correspond to home and work. Thus, while the anonymization procedure is intended to impeach any linkage between the dataset and individuals, using this procedure allows to potentially retrieve the mobility and calling pattern of targeted users given the access to as little information as home and work addresses. If additional data, such as year of birth or gender of users would be available -- which is common in most datasets -- it would be possible to identify very large fractions of the network. However, in this attack scheme, one has to know quite well the profile of the user for them to be found in the database. Using a different approach, de Montjoye \textit{et al.}\ \cite{demontjoye2013unique} show that knowing only four points in space and time where a user was allows to uniquely re-identify the user with $95\%$ probability. Using only very little information that could be available easily to an attacker, the authors thus show how unique each user's trajectory is. They further show that blurring the resolution of space or time does not reduce much the information needed to re-identify a user in the database, thus keeping the database very vulnerable if faced with this type of de-anonymization attack. \\

Other possible attacks have also been considered on anonymized online social networks. Although those attacks are not likely to be applied in the case of mobile phone data, we quickly mention some of them, as it is likely that breaches found in different applications might be similar to potential breaches in mobile phone datasets.\\
For example, Backstrom \textit{et al.}\ \cite{backstrom2007wherefore} describe a family of local attacks, which enable to retrieve the position of some targets in the network, and hence to uncover the connections between those patterns. The authors showed that on a network of 4.4 million nodes, by controlling the links of 7 dummy nodes they manage to uncover the presence or absence of 2,400 links between 70 target nodes, without being detected by the database manager. On a wider scale, Narayannan and Shmatikov \cite{narayanan2009anonymizing} show that it is possible to retrieve the identity of a large part of a social network by combining it with an auxiliary network. Such a situation happens when users are present in two separate datasets. The authors show then that even if this overlap is available for only a fraction of the users, it is still possible to retrieve the information for a large part of the network.\\

Against these possible threats of privacy breach, one may wonder if solutions are proposed to counter such attacks. If research on mobile datasets only considers average behaviors, rather than exact patterns, a simple countermeasure is to perform small modifications of the dataset, that would not alter the general aspect of it but that would have dramatic consequences on the algorithms used by attackers, who search for exact matchings between statistics on the network and a priori known properties of the targets.\\
Another protection against such attacks, and particularly when mobility data is involved, is to produce new random identifiers for each user at regular time intervals. By regenerating random identifiers, it makes it impossible to use longitudinal information in order to assess the preferred locations of a user. As shown by Zang and Bolot \cite{zang2011anonymization}, by changing every day the ID of each user, only 3\% of the nodes can still be identified using their top 2 locations. While this method seems efficient to protect the privacy of users, it reduces substantially the possible information to retrieve from such a dataset for research purposes. Using a similar approach also proved useful against the attack scheme considered by de Montjoye \textit{et al.}, as Song \textit{et al.} show in \cite{song2014not} that changing the ID of each user every six hours reduces substantially the fraction of unique trajectories in the dataset. A compromise between preserving the anonymity and keeping enough information in the dataset is difficult to achieve. In collaboration with the Universit\'e catholique de Louvain, the provider Orange tried to achieve this for their first D4D challenge before releasing a dataset to a wide community of researchers (more than 150 research teams participated). Through releasing four different datasets anonymized differently \cite{blondel2012data} and containing information of different spatio-temporal resolutions, they could guarantee the preservation of the anonymity of users. Yet, the loss of information was not too dramatic, as many studies showed very good results using the provided aggregated information. The challenge was such a success that a second one is currently in process, using a wider dataset from Senegal \cite{demontjoye2014d4d}. \\
Another question that is closely linked to this research is how to quantify the anonymity of a database. Latanya Sweeney proposed a measure that is \textit{k-anonymity} \cite{sweeney2002kanonymity}, defining that a database achieves $k$-anonymity if for any tuples of previously defined entries of the database, there are at least $k$ users corresponding to it, making it impossible to re-identify a single user with only information on these entries of the database. Of course, the larger $k$ is, the most difficult it becomes to achieve this, especially in a CDR database containing spatio-temporal information about each call. Moreover, when the attacker is looking for a particular person in the database, enabling him to reduce the number of potential corresponding users to a small number is sometimes already a lot of information, and too big a risk to release the database publicly. 
Another potential solution to preserving privacy was suggested by Isaacman \textit{et al.}\ \cite{isaacman2012human} who suggest using synthetic data to model the mobility of people. They used mobile phone data from two american cities to validate their model, showing that their model, based on only aggregated data and probability distributions, could reproduce many of the features of mobility of users, without any of them corresponding to a real person. Mir \textit{et al.}\ further proposed an evolved version called DP-WHERE \cite{mir2013dpwhere} of the previous model, adding controlled noise to the set of empirical probability distributions. This noise then guarantees that the model achieves \textit{differential privacy}, that is, that the analyses will not be significantly different whether or not a single individual is in the database from which the model is derived, even if this individual has an unusual behavior. However, on may wonder if these synthetic data could be used to carry out analyses that were not previously tested on the real database, as no guarantee exists on the outcome of analyses that were not foreseen by the researchers that tested the model for compatibility with empirical data. \\ 

\subsection*{Personal data: ownership, usage, privacy}
Phone companies collect data about their users, about their habits, their mobility, their acquaintances. Still, the legislation up to 2013 was fuzzy \cite{madanreality}, chilling companies to share such data for research and making customers feel that George Orwell's predictions are coming true, especially after the scandal in 2013 revealing how much personal information the NSA was collecting from many sources \cite{landau2013making}. \\
Such data represents an enormous added value, both to companies, for marketing purposes and client screening, and to authorities for traffic management or epidemic outbreak prevention. It is often forgotten, but the use of mobile phone datasets also has a huge positive potential in the developing world, as many of the proposed project to the Data for Development challenge showed \cite{D4Dbook}, may it be for supervising the health status of populations, generating census data or optimizing public transport.  \\
Such opportunities, both for corporates and authorities need to develop standardized procedures for the acquisition, conservation and usage of personal data, which is not yet the case. The communication about these procedures to customers hasn't been clear, as are the possibilities for a user to ``opt-out'' if they don't want to have their personal data released.\\
With this intent, several voices have recently been raised in order to urge authorities to develop a ``New Deal'' \cite{pentland2009reality} on data ownership, in which users would own their personal data as well as the decisions to provide it --in exchange of payment-- to companies interested in their usage. A transparent system armed with the necessary protocols and regulation for a transparent use of personal data would also facilitate the access to data for researchers \cite{eagle2009engineering}, and could so benefit to the entire society.

\section{Conclusion and research questions}
The first analyses of mobile phone datasets appeared in the late 90's, and the result of this decade of research contains a large number of surprises and several promising directions for the future. In this paper, we have reviewed the most prominent results obtained so far, in particular in the analysis of the structure of our social networks, and human mobility. We decided not to cover some closely related questions, such as churn prediction (see \cite{hung2006applying,dasgupta2008social,richter2010predicting,dierkes2011estimating}) or dynamic pricing \cite{fitkov2001dynamic,kim2010empirical}, which are rather business-related topics, and for which a vast literature is available.\\

The recent availability of mobile phone datasets have led to many discoveries on human behavior. We are not all similar in our ways of communicating, and differences between users can range to several orders of magnitudes. Our networks are clustered in well-structured groups, which are spatially well-located. With the raise of communication technology, some have predicted that the barrier of distance would fall, shrinking the world into a small village. However, mobile phone data suggests instead that distance still plays a role, but that its impact is nuanced by the varying population density. Regarding our mobility behavior, individuals appear to have highly predictable movements \cite{barabasi2010you}, while as populations we act and react in a remarkable synchronized way. In this context, the availability of mobile phone data has for the first time allowed to observe populations from a God-eye point of view, monitoring the pace of daily life or the response to catastrophes. \\
The ubiquity of mobile phones -- there are nowadays more mobile phones than personal computers in use -- which allows us to obtain such precise results raises also the thread of viral outbreaks, from which mobile phones have been safe until now. Mobile viruses could be a potential risk for users' privacy, as it is also the case that the anonymized datasets provided by operators to third parties for research could potentially be de-anonymized too.\\

The availability of such enormous datasets creates a huge potential that could benefit to society, up to the point of saving lives. The research that has been conducted so far only represents the tip of the iceberg of what could potentially be done, when adequately exploited. However, it is the necessity of authorities to ensure that such datasets could not be misused.\\

\subsection*{Further research}
The number of possible research questions on mobile phone datasets is gigantic. In this last part, we will present one research direction that we believe to be highly important and still not addressed in its most general form.\\

A large number of research has been conducted on the analysis of social networks, based on CDRs. As it appears from the different publications on this topic, there exist some common features but also many differences in the structure of the constructed network. Recall as simplest example the degree distributions, which show different functional forms for most datasets.\\
These differences may, of course, be linked to cultural differences between the different countries of interest, but there are probably other, quantifiable, reasons. The datasets differ greatly in the market shares of the operators, in the time span of the data collection period, in the size of the network and in the geographical span of the considered country. The method of network construction is also always different and has a tangible impact on the network structure. The use of directed or undirected links, weights and thresholds for removing low-intensity or non-mutual links all greatly impact the structure and hence the statistical features of the obtained network.\\

Hence, we believe that a serious analysis, both on theoretical and on empirical side of the influence of these factors on the general structure of mobile phone networks may lead to a general framework, allowing to interpret differences between results obtained on several datasets with the knowledge of potential side-effects.\\
This question is closely related to the even more general question of the significance of information provided by CDR data. Recalling what was said in Section \ref{sec:sn}, CDR datasets are noisy data, some links appear there by chance, while other have not been captured in the dataset. It would thus be interesting to question the stability of the obtained results, provided that the real network is different from what has been observed in the data. This links with the work of Gourab \cite{ghoshal2011ranking}, who analyzed the stability of PageRank under random noise on the network structure. Again, in this framework, no real theoretical result has yet been achieved, allowing to characterize which results are significant, and which are not.


\begin{backmatter}

\section*{Acknowledgements}
We would like to thank Franscesco Calabrese, Yves-Alexandre de Montjoye, Vanessa Frias-Martinez, Marta Gonz\'a{}lez, Jukka-Pekka Onnela, Jari Saram\"a{}ki and Zbigniew Smoreda for their valuable comments and advice in finalizing this survey.  AD is a research fellow with the Fonds de la Recherche Scientifique - FNRS.


\bibliographystyle{bmc-mathphys} 
\bibliography{mobphones}      


\begin{thebibliography}{184}
\ifx \bisbn   \undefined \def \bisbn  #1{ISBN #1}\fi
\ifx \binits  \undefined \def \binits#1{#1}\fi
\ifx \bauthor  \undefined \def \bauthor#1{#1}\fi
\ifx \batitle  \undefined \def \batitle#1{#1}\fi
\ifx \bjtitle  \undefined \def \bjtitle#1{#1}\fi
\ifx \bvolume  \undefined \def \bvolume#1{\textbf{#1}}\fi
\ifx \byear  \undefined \def \byear#1{#1}\fi
\ifx \bissue  \undefined \def \bissue#1{#1}\fi
\ifx \bfpage  \undefined \def \bfpage#1{#1}\fi
\ifx \blpage  \undefined \def \blpage #1{#1}\fi
\ifx \burl  \undefined \def \burl#1{\textsf{#1}}\fi
\ifx \doiurl  \undefined \def \doiurl#1{\textsf{#1}}\fi
\ifx \betal  \undefined \def \betal{\textit{et al.}}\fi
\ifx \binstitute  \undefined \def \binstitute#1{#1}\fi
\ifx \binstitutionaled  \undefined \def \binstitutionaled#1{#1}\fi
\ifx \bctitle  \undefined \def \bctitle#1{#1}\fi
\ifx \beditor  \undefined \def \beditor#1{#1}\fi
\ifx \bpublisher  \undefined \def \bpublisher#1{#1}\fi
\ifx \bbtitle  \undefined \def \bbtitle#1{#1}\fi
\ifx \bedition  \undefined \def \bedition#1{#1}\fi
\ifx \bseriesno  \undefined \def \bseriesno#1{#1}\fi
\ifx \blocation  \undefined \def \blocation#1{#1}\fi
\ifx \bsertitle  \undefined \def \bsertitle#1{#1}\fi
\ifx \bsnm \undefined \def \bsnm#1{#1}\fi
\ifx \bsuffix \undefined \def \bsuffix#1{#1}\fi
\ifx \bparticle \undefined \def \bparticle#1{#1}\fi
\ifx \barticle \undefined \def \barticle#1{#1}\fi
\ifx \bconfdate \undefined \def \bconfdate #1{#1}\fi
\ifx \botherref \undefined \def \botherref #1{#1}\fi
\ifx \url \undefined \def \url#1{\textsf{#1}}\fi
\ifx \bchapter \undefined \def \bchapter#1{#1}\fi
\ifx \bbook \undefined \def \bbook#1{#1}\fi
\ifx \bcomment \undefined \def \bcomment#1{#1}\fi
\ifx \oauthor \undefined \def \oauthor#1{#1}\fi
\ifx \citeauthoryear \undefined \def \citeauthoryear#1{#1}\fi
\ifx \endbibitem  \undefined \def \endbibitem {}\fi
\ifx \bconflocation  \undefined \def \bconflocation#1{#1}\fi
\ifx \arxivurl  \undefined \def \arxivurl#1{\textsf{#1}}\fi
\csname PreBibitemsHook\endcsname

\bibitem{ITU}
\begin{botherref}
The world in 2014 : {ICT} Facts and Figures.
International Telecommunication Union.
\url{http://www.itu.int/}
(2014)
\end{botherref}
\endbibitem

\bibitem{kwok2009personal}
\begin{barticle}
\bauthor{\bsnm{Kwok}, \binits{R.}}:
\batitle{Personal technology: Phoning in data}.
\bjtitle{Nature}
\bvolume{458}(\bissue{7241}),
\bfpage{959}
(\byear{2009})
\end{barticle}
\endbibitem

\bibitem{zipf1949human}
\begin{bbook}
\bauthor{\bsnm{Zipf}, \binits{G.K.}}:
\bbtitle{{Human Behavior and the Principle of Least Effort: An Introduction to
  Human Ecology}}.
\bpublisher{Addison-Wesley Press}, \blocation{???}
(\byear{1949})
\end{bbook}
\endbibitem

\bibitem{cortes2001communities}
\begin{botherref}
\oauthor{\bsnm{Cortes}, \binits{C.}},
\oauthor{\bsnm{Pregibon}, \binits{D.}},
\oauthor{\bsnm{Volinsky}, \binits{C.}}:
Communities of interest.
Advances in Intelligent Data Analysis,
105--114
(2001)
\end{botherref}
\endbibitem

\bibitem{Krings2012thesis}
\begin{botherref}
\oauthor{\bsnm{Krings}, \binits{G.}}:
Extraction of information from large networks.
PhD thesis,
Universit\'e catholique de Louvain
(2012)
\end{botherref}
\endbibitem

\bibitem{abello1999maximum}
\begin{barticle}
\bauthor{\bsnm{Abello}, \binits{J.}},
\bauthor{\bsnm{Pardalos}, \binits{P.M.}},
\bauthor{\bsnm{Resende}, \binits{M.G.C.}}:
\batitle{On maximum clique problems in very large graphs}.
\bjtitle{External memory algorithms}
\bvolume{50},
\bfpage{119}--\blpage{130}
(\byear{1999})
\end{barticle}
\endbibitem

\bibitem{aiello2000random}
\begin{bchapter}
\bauthor{\bsnm{Aiello}, \binits{W.}},
\bauthor{\bsnm{Chung}, \binits{F.}},
\bauthor{\bsnm{Lu}, \binits{L.}}:
\bctitle{A random graph model for massive graphs}.
In: \bbtitle{Proceedings of the Thirty-second Annual ACM Symposium on Theory of
  Computing},
pp. \bfpage{171}--\blpage{180}
(\byear{2000}).
\bcomment{{ACM}}
\end{bchapter}
\endbibitem

\bibitem{onnela2007analysis}
\begin{barticle}
\bauthor{\bsnm{Onnela}, \binits{J.P.}},
\bauthor{\bsnm{Saramaki}, \binits{J.}},
\bauthor{\bsnm{Hyvonen}, \binits{J.}},
\bauthor{\bsnm{Szabo}, \binits{G.}},
\bauthor{\bparticle{de} \bsnm{Menezes}, \binits{M.A.}},
\bauthor{\bsnm{Kaski}, \binits{K.}},
\bauthor{\bsnm{Barabasi}, \binits{A.L.}},
\bauthor{\bsnm{Kertesz}, \binits{J.}}:
\batitle{{Analysis of a large-scale weighted network of one-to-one human
  communication}}.
\bjtitle{New Journal of Physics}
\bvolume{9}(\bissue{6}),
\bfpage{179}
(\byear{2007})
\end{barticle}
\endbibitem

\bibitem{lambiotte2008gdm}
\begin{barticle}
\bauthor{\bsnm{Lambiotte}, \binits{R.}},
\bauthor{\bsnm{Blondel}, \binits{V.D.}},
\bauthor{\bparticle{de} \bsnm{Kerchove}, \binits{C.}},
\bauthor{\bsnm{Huens}, \binits{E.}},
\bauthor{\bsnm{Prieur}, \binits{C.}},
\bauthor{\bsnm{Smoreda}, \binits{Z.}},
\bauthor{\bsnm{Van~Dooren}, \binits{P.}}:
\batitle{{Geographical dispersal of mobile communication networks}}.
\bjtitle{Physica A: Statistical Mechanics and its Applications}
\bvolume{387}(\bissue{21}),
\bfpage{5317}--\blpage{5325}
(\byear{2008})
\end{barticle}
\endbibitem

\bibitem{onnela2007structure}
\begin{barticle}
\bauthor{\bsnm{Onnela}, \binits{J.P.}},
\bauthor{\bsnm{Saramaki}, \binits{J.}},
\bauthor{\bsnm{Hyvonen}, \binits{J.}},
\bauthor{\bsnm{Szabo}, \binits{G.}},
\bauthor{\bsnm{Lazer}, \binits{D.}},
\bauthor{\bsnm{Kaski}, \binits{K.}},
\bauthor{\bsnm{Kertesz}, \binits{J.}},
\bauthor{\bsnm{Barabasi}, \binits{A.L.}}:
\batitle{{Structure and tie strengths in mobile communication networks}}.
\bjtitle{Proceedings of the National Academy of Sciences}
\bvolume{104}(\bissue{18}),
\bfpage{7332}
(\byear{2007})
\end{barticle}
\endbibitem

\bibitem{li2014statistically}
\begin{botherref}
\oauthor{\bsnm{Li}, \binits{M.-X.}},
\oauthor{\bsnm{Palchykov}, \binits{V.}},
\oauthor{\bsnm{Jiang}, \binits{Z.-Q.}},
\oauthor{\bsnm{Kaski}, \binits{K.}},
\oauthor{\bsnm{Kert{\'e}sz}, \binits{J.}},
\oauthor{\bsnm{Miccich{\`e}}, \binits{S.}},
\oauthor{\bsnm{Tumminello}, \binits{M.}},
\oauthor{\bsnm{Zhou}, \binits{W.-X.}},
\oauthor{\bsnm{Mantegna}, \binits{R.N.}}:
Statistically validated mobile communication networks: Evolution of motifs in
  european and chinese data.
arXiv preprint arXiv:1403.3785
(2014)
\end{botherref}
\endbibitem

\bibitem{kovanen2011reciprocity}
\begin{barticle}
\bauthor{\bsnm{Kovanen}, \binits{L.}},
\bauthor{\bsnm{Saram}, \binits{J.}},
\bauthor{\bsnm{Kaski}, \binits{K.}}:
\batitle{Reciprocity of mobile phone calls}.
\bjtitle{JDySES}
\bvolume{2}(\bissue{2}),
\bfpage{138}--\blpage{151}
(\byear{2011})
\end{barticle}
\endbibitem

\bibitem{ling2012sociodemographics}
\begin{barticle}
\bauthor{\bsnm{Ling}, \binits{R.}},
\bauthor{\bsnm{Bertel}, \binits{T.F.}},
\bauthor{\bsnm{Sunds{\o}y}, \binits{P.R.}}:
\batitle{The socio-demographics of texting: An analysis of traffic data}.
\bjtitle{New Media \& Society}
\bvolume{14}(\bissue{2}),
\bfpage{281}--\blpage{298}
(\byear{2012}).
doi:\doiurl{10.1177/1461444811412711}.
\arxivurl{http://nms.sagepub.com/content/14/2/281.full.pdf+html}
\end{barticle}
\endbibitem

\bibitem{nanavati2006structural}
\begin{bchapter}
\bauthor{\bsnm{Nanavati}, \binits{A.A.}},
\bauthor{\bsnm{Gurumurthy}, \binits{S.}},
\bauthor{\bsnm{Das}, \binits{G.}},
\bauthor{\bsnm{Chakraborty}, \binits{D.}},
\bauthor{\bsnm{Dasgupta}, \binits{K.}},
\bauthor{\bsnm{Mukherjea}, \binits{S.}},
\bauthor{\bsnm{Joshi}, \binits{A.}}:
\bctitle{On the structural properties of massive telecom call graphs: findings
  and implications}.
In: \bbtitle{Proceedings of the 15th ACM International Conference on
  Information and Knowledge Management},
pp. \bfpage{435}--\blpage{444}
(\byear{2006}).
\bcomment{ACM}
\end{bchapter}
\endbibitem

\bibitem{barabasi2009scale}
\begin{barticle}
\bauthor{\bsnm{Barab{\'a}si}, \binits{A.L.}}:
\batitle{Scale-free networks: a decade and beyond}.
\bjtitle{Science}
\bvolume{325}(\bissue{5939}),
\bfpage{412}
(\byear{2009})
\end{barticle}
\endbibitem

\bibitem{watts1998collective}
\begin{barticle}
\bauthor{\bsnm{Watts}, \binits{D.J.}},
\bauthor{\bsnm{Strogatz}, \binits{S.H.}}:
\batitle{{Collective dynamics of small-world networks}}.
\bjtitle{Nature}
\bvolume{393}(\bissue{6684}),
\bfpage{440}--\blpage{442}
(\byear{1998})
\end{barticle}
\endbibitem

\bibitem{seshadri2008mobile}
\begin{bchapter}
\bauthor{\bsnm{Seshadri}, \binits{M.}},
\bauthor{\bsnm{Machiraju}, \binits{S.}},
\bauthor{\bsnm{Sridharan}, \binits{A.}},
\bauthor{\bsnm{Bolot}, \binits{J.}},
\bauthor{\bsnm{Faloutsos}, \binits{C.}},
\bauthor{\bsnm{Leskovec}, \binits{J.}}:
\bctitle{Mobile call graphs: beyond power-law and lognormal distributions}.
In: \bbtitle{Proceeding of the 14th ACM SIGKDD International Conference on
  Knowledge Discovery and Data Mining},
pp. \bfpage{596}--\blpage{604}
(\byear{2008}).
\bcomment{ACM}
\end{bchapter}
\endbibitem

\bibitem{krings2012effects}
\begin{barticle}
\bauthor{\bsnm{Krings}, \binits{G.}},
\bauthor{\bsnm{Karsai}, \binits{M.}},
\bauthor{\bsnm{Bernhardsson}, \binits{S.}},
\bauthor{\bsnm{Blondel}, \binits{V.D.}},
\bauthor{\bsnm{Saram\"aki}, \binits{J.}}:
\batitle{Effects of time window size and placement on the structure of an
  aggregated communication network}.
\bjtitle{EPJ Data Science}
\bvolume{1}(\bissue{4}),
\bfpage{1}--\blpage{16}
(\byear{2012})
\end{barticle}
\endbibitem

\bibitem{granovetter1973strength}
\begin{barticle}
\bauthor{\bsnm{Granovetter}, \binits{M.S.}}:
\batitle{{The Strength of Weak Ties}}.
\bjtitle{American Journal of Sociology}
\bvolume{78},
\bfpage{1360}--\blpage{1380}
(\byear{1973})
\end{barticle}
\endbibitem

\bibitem{onnela2005intensity}
\begin{barticle}
\bauthor{\bsnm{Onnela}, \binits{J.-P.}},
\bauthor{\bsnm{Saram{\"a}ki}, \binits{J.}},
\bauthor{\bsnm{Kert{\'e}sz}, \binits{J.}},
\bauthor{\bsnm{Kaski}, \binits{K.}}:
\batitle{Intensity and coherence of motifs in weighted complex networks}.
\bjtitle{Physical Review E}
\bvolume{71}(\bissue{6}),
\bfpage{065103}
(\byear{2005})
\end{barticle}
\endbibitem

\bibitem{du2009large}
\begin{bchapter}
\bauthor{\bsnm{Du}, \binits{N.}},
\bauthor{\bsnm{Faloutsos}, \binits{C.}},
\bauthor{\bsnm{Wang}, \binits{B.}},
\bauthor{\bsnm{Akoglu}, \binits{L.}}:
\bctitle{Large human communication networks: patterns and a utility-driven
  generator}.
In: \bbtitle{Proceedings of the 15th ACM SIGKDD International Conference on
  Knowledge Discovery and Data Mining},
pp. \bfpage{269}--\blpage{278}
(\byear{2009}).
\bcomment{ACM}
\end{bchapter}
\endbibitem

\bibitem{kianmehr2009calling}
\begin{barticle}
\bauthor{\bsnm{Kianmehr}, \binits{K.}},
\bauthor{\bsnm{Alhajj}, \binits{R.}}:
\batitle{Calling communities analysis and identification using machine learning
  techniques}.
\bjtitle{Expert Systems with Applications}
\bvolume{36}(\bissue{3}),
\bfpage{6218}--\blpage{6226}
(\byear{2009})
\end{barticle}
\endbibitem

\bibitem{zhang2008discovery}
\begin{bchapter}
\bauthor{\bsnm{Zhang}, \binits{H.}},
\bauthor{\bsnm{Dantu}, \binits{R.}}:
\bctitle{Discovery of social groups using call detail records}.
In: \bbtitle{On the Move to Meaningful Internet Systems: OTM 2008 Workshops},
pp. \bfpage{489}--\blpage{498}
(\byear{2008}).
\bcomment{Springer}
\end{bchapter}
\endbibitem

\bibitem{tibely2011communities}
\begin{barticle}
\bauthor{\bsnm{Tib{\'e}ly}, \binits{G.}},
\bauthor{\bsnm{Kovanen}, \binits{L.}},
\bauthor{\bsnm{Karsai}, \binits{M.}},
\bauthor{\bsnm{Kaski}, \binits{K.}},
\bauthor{\bsnm{Kert{\'e}sz}, \binits{J.}},
\bauthor{\bsnm{Saram{\"a}ki}, \binits{J.}}:
\batitle{Communities and beyond: mesoscopic analysis of a large social network
  with complementary methods}.
\bjtitle{Physical Review E}
\bvolume{83}(\bissue{5}),
\bfpage{056125}
(\byear{2011})
\end{barticle}
\endbibitem

\bibitem{blondel2008fuc}
\begin{botherref}
\oauthor{\bsnm{Blondel}, \binits{V.D.}},
\oauthor{\bsnm{Guillaume}, \binits{J.L.}},
\oauthor{\bsnm{Lambiotte}, \binits{R.}},
\oauthor{\bsnm{Mech}, \binits{E.L.J.S.}}:
{Fast unfolding of communities in large networks}.
J. Stat. Mech,
10008
(2008)
\end{botherref}
\endbibitem

\bibitem{rosvall2008maps}
\begin{barticle}
\bauthor{\bsnm{Rosvall}, \binits{M.}},
\bauthor{\bsnm{Bergstrom}, \binits{C.T.}}:
\batitle{{Maps of random walks on complex networks reveal community
  structure}}.
\bjtitle{Proceedings of the National Academy of Sciences}
\bvolume{105}(\bissue{4}),
\bfpage{1118}
(\byear{2008})
\end{barticle}
\endbibitem

\bibitem{palla2007quantifying}
\begin{barticle}
\bauthor{\bsnm{Palla}, \binits{G.}},
\bauthor{\bsnm{Barab{\'a}si}, \binits{A.}},
\bauthor{\bsnm{Vicsek}, \binits{T.}}:
\batitle{Quantifying social group evolution.}
\bjtitle{Nature}
\bvolume{446}(\bissue{7136}),
\bfpage{664}
(\byear{2007})
\end{barticle}
\endbibitem

\bibitem{ahn2010link}
\begin{barticle}
\bauthor{\bsnm{Ahn}, \binits{Y.Y.}},
\bauthor{\bsnm{Bagrow}, \binits{J.P.}},
\bauthor{\bsnm{Lehmann}, \binits{S.}}:
\batitle{{Link communities reveal multiscale complexity in networks}}.
\bjtitle{Nature}
\bvolume{466}(\bissue{7307}),
\bfpage{761}--\blpage{764}
(\byear{2010})
\end{barticle}
\endbibitem

\bibitem{Lazer2009computational}
\begin{barticle}
\bauthor{\bsnm{Lazer}, \binits{D.}},
\bauthor{\bsnm{Pentland}, \binits{A.}},
\bauthor{\bsnm{Adamic}, \binits{L.}},
\bauthor{\bsnm{Aral}, \binits{S.}},
\bauthor{\bsnm{Barab{\'a}si}, \binits{A.-L.}},
\bauthor{\bsnm{Brewer}, \binits{D.}},
\bauthor{\bsnm{Christakis}, \binits{N.}},
\bauthor{\bsnm{Contractor}, \binits{N.}},
\bauthor{\bsnm{Fowler}, \binits{J.}},
\bauthor{\bsnm{Gutmann}, \binits{M.}},
\bauthor{\bsnm{Jebara}, \binits{T.}},
\bauthor{\bsnm{King}, \binits{G.}},
\bauthor{\bsnm{Macy}, \binits{M.}},
\bauthor{\bsnm{Roy}, \binits{D.}},
\bauthor{\bsnm{Van~Alstyne}, \binits{M.}}:
\batitle{Computational social science}.
\bjtitle{Science}
\bvolume{323}(\bissue{5915}),
\bfpage{721}--\blpage{723}
(\byear{2009}).
doi:\doiurl{10.1126/science.1167742}.
\arxivurl{http://www.sciencemag.org/content/323/5915/721.full.pdf}
\end{barticle}
\endbibitem

\bibitem{eagle2009inferring}
\begin{barticle}
\bauthor{\bsnm{Eagle}, \binits{N.}},
\bauthor{\bsnm{Pentland}, \binits{A.S.}},
\bauthor{\bsnm{Lazer}, \binits{D.}}:
\batitle{Inferring friendship network structure by using mobile phone data}.
\bjtitle{Proceedings of the National Academy of Sciences}
\bvolume{106}(\bissue{36}),
\bfpage{15274}
(\byear{2009})
\end{barticle}
\endbibitem

\bibitem{wiese2015you}
\begin{bchapter}
\bauthor{\bsnm{Wiese}, \binits{J.}},
\bauthor{\bsnm{Min}, \binits{J.-K.}},
\bauthor{\bsnm{Hong}, \binits{J.I.}},
\bauthor{\bsnm{Zimmerman}, \binits{J.}}:
\bctitle{``you never call, you never write'': Call and sms logs do not always
  indicate tie strength}.
In: \bbtitle{Proceedings of the 2015 Conference on Computer Supported
  Cooperative work-CSCW'15}
(\byear{2015})
\end{bchapter}
\endbibitem

\bibitem{blumenstock2010who}
\begin{barticle}
\bauthor{\bsnm{Blumenstock}, \binits{J.E.}},
\bauthor{\bsnm{Gillick}, \binits{D.}},
\bauthor{\bsnm{Eagle}, \binits{N.}}:
\batitle{Who's calling? demographics of mobile phone use in rwanda}.
\bjtitle{Transportation}
\bvolume{32},
\bfpage{2}--\blpage{5}
(\byear{2010})
\end{barticle}
\endbibitem

\bibitem{smoreda2000gender}
\begin{barticle}
\bauthor{\bsnm{Smoreda}, \binits{Z.}},
\bauthor{\bsnm{Licoppe}, \binits{C.}}:
\batitle{Gender-specific use of the domestic telephone}.
\bjtitle{Social Psychology Quarterly}
\bvolume{63}(\bissue{3}),
\bfpage{238}--\blpage{252}
(\byear{2000})
\end{barticle}
\endbibitem

\bibitem{kovanen2013temporal}
\begin{barticle}
\bauthor{\bsnm{Kovanen}, \binits{L.}},
\bauthor{\bsnm{Kaski}, \binits{K.}},
\bauthor{\bsnm{Kert\'{e}sz}, \binits{J.}},
\bauthor{\bsnm{Saram\"aki}, \binits{J.}}:
\batitle{Temporal motifs reveal homophily, gender-specific patterns, and group
  talk in call sequences}.
\bjtitle{Proceedings of the National Academy of Sciences}
\bvolume{110}(\bissue{45}),
\bfpage{18070}--\blpage{18075}
(\byear{2013})
\end{barticle}
\endbibitem

\bibitem{frias2010gender}
\begin{bchapter}
\bauthor{\bsnm{Frias-Martinez}, \binits{V.}},
\bauthor{\bsnm{Frias-Martinez}, \binits{E.}},
\bauthor{\bsnm{Oliver}, \binits{N.}}:
\bctitle{A gender-centric analysis of calling behavior in a developing economy
  using call detail records.}
In: \bbtitle{AAAI Spring Symposium: Artificial Intelligence for Development}
(\byear{2010})
\end{bchapter}
\endbibitem

\bibitem{blumenstock2012divided}
\begin{barticle}
\bauthor{\bsnm{Blumenstock}, \binits{J.E.}},
\bauthor{\bsnm{Eagle}, \binits{N.}}:
\batitle{Divided we call: Disparities in access and use of mobile phones in
  rwanda}.
\bjtitle{Information Technologies \& International Development}
\bvolume{8}(\bissue{2}),
\bfpage{1}
(\byear{2012})
\end{barticle}
\endbibitem

\bibitem{chawla2011reciprocity}
\begin{botherref}
\oauthor{\bsnm{Chawla}, \binits{N.V.}},
\oauthor{\bsnm{Hachen}, \binits{D.}},
\oauthor{\bsnm{Lizardo}, \binits{O.}},
\oauthor{\bsnm{Toroczkai}, \binits{Z.}},
\oauthor{\bsnm{Strathman}, \binits{A.}},
\oauthor{\bsnm{Wang}, \binits{C.}}:
Weighted reciprocity in human communication networks.
Technical Report arXiv:1108.2822
(2011)
\end{botherref}
\endbibitem

\bibitem{motahari2012affinity}
\begin{bchapter}
\bauthor{\bsnm{Motahari}, \binits{S.}},
\bauthor{\bsnm{Mengshoel}, \binits{O.J.}},
\bauthor{\bsnm{Reuther}, \binits{P.}},
\bauthor{\bsnm{Appala}, \binits{S.}},
\bauthor{\bsnm{Zoia}, \binits{L.}},
\bauthor{\bsnm{Shah}, \binits{J.}}:
\bctitle{The impact of social affinity on phone calling patterns: Categorizing
  social ties from call data records}.
In: \bbtitle{The 6th SNA-KDD Workshop '12}
(\byear{2012})
\end{bchapter}
\endbibitem

\bibitem{barthelemy2011spatial}
\begin{barticle}
\bauthor{\bsnm{Barth{\'e}lemy}, \binits{M.}}:
\batitle{Spatial networks}.
\bjtitle{Physics Reports}
\bvolume{499}(\bissue{1}),
\bfpage{1}--\blpage{101}
(\byear{2011})
\end{barticle}
\endbibitem

\bibitem{deville2014dynamic}
\begin{barticle}
\bauthor{\bsnm{Deville}, \binits{P.}},
\bauthor{\bsnm{Linard}, \binits{C.}},
\bauthor{\bsnm{Martin}, \binits{S.}},
\bauthor{\bsnm{Gilbert}, \binits{M.}},
\bauthor{\bsnm{Stevens}, \binits{F.R.}},
\bauthor{\bsnm{Gaughan}, \binits{A.E.}},
\bauthor{\bsnm{Blondel}, \binits{V.D.}},
\bauthor{\bsnm{Tatem}, \binits{A.J.}}:
\batitle{Dynamic population mapping using mobile phone data}.
\bjtitle{Proceedings of the National Academy of Sciences}
\bvolume{111}(\bissue{45}),
\bfpage{15888}--\blpage{15893}
(\byear{2014})
\end{barticle}
\endbibitem

\bibitem{sterlyD4Ddensity}
\begin{bchapter}
\bauthor{\bsnm{Sterly}, \binits{H.}},
\bauthor{\bsnm{Hennig}, \binits{B.}},
\bauthor{\bsnm{Dongo}, \binits{K.}}:
\bctitle{``calling abidjan" - improving population estimations with mobile
  communication data}.
In: \bbtitle{Mobile Phone Data for Development - Analysis of Mobile Phone
  Datasets for the Development of Ivory Coast},
pp. \bfpage{108}--\blpage{114}.
\bpublisher{Orange D4D Challenge}, \blocation{???}
(\byear{2013})
\end{bchapter}
\endbibitem

\bibitem{afripop}
\begin{botherref}
Afripop Project.
\url{http://www.worldpop.org.uk}
\end{botherref}
\endbibitem

\bibitem{krings2009urban}
\begin{barticle}
\bauthor{\bsnm{Krings}, \binits{G.}},
\bauthor{\bsnm{Calabrese}, \binits{F.}},
\bauthor{\bsnm{Ratti}, \binits{C.}},
\bauthor{\bsnm{Blondel}, \binits{V.D.}}:
\batitle{Urban gravity: a model for inter-city telecommunication flows}.
\bjtitle{Journal of Statistical Mechanics: Theory and Experiment}
\bvolume{2009},
\bfpage{07003}
(\byear{2009})
\end{barticle}
\endbibitem

\bibitem{krings2009scaling}
\begin{bchapter}
\bauthor{\bsnm{Krings}, \binits{G.}},
\bauthor{\bsnm{Calabrese}, \binits{F.}},
\bauthor{\bsnm{Ratti}, \binits{C.}},
\bauthor{\bsnm{Blondel}, \binits{V.D.}}:
\bctitle{Scaling behaviors in the communication network between cities}.
In: \bbtitle{2009 International Conference on Computational Science and
  Engineering},
pp. \bfpage{936}--\blpage{939}
(\byear{2009}).
\bcomment{IEEE}
\end{bchapter}
\endbibitem

\bibitem{onnela2011geographic}
\begin{barticle}
\bauthor{\bsnm{Onnela}, \binits{J.P.}},
\bauthor{\bsnm{Arbesman}, \binits{S.}},
\bauthor{\bsnm{Gonz{\'a}lez}, \binits{M.C.}},
\bauthor{\bsnm{Barab{\'a}si}, \binits{A.L.}},
\bauthor{\bsnm{Christakis}, \binits{N.A.}}:
\batitle{Geographic constraints on social network groups}.
\bjtitle{PloS one}
\bvolume{6}(\bissue{4}),
\bfpage{16939}
(\byear{2011})
\end{barticle}
\endbibitem

\bibitem{bucicovschiD4Dsocdiv}
\begin{bchapter}
\bauthor{\bsnm{Bucicovschi}, \binits{O.}},
\bauthor{\bsnm{Douglass}, \binits{R.W.}},
\bauthor{\bsnm{Meyer}, \binits{D.A.}},
\bauthor{\bsnm{Ram}, \binits{M.}},
\bauthor{\bsnm{Rideout}, \binits{D.}},
\bauthor{\bsnm{Song}, \binits{D.}}:
\bctitle{Analyzing social divisions using cell phone data}.
In: \bbtitle{D4D Book: Mobile Phone Data for Development. Analysis of Mobile
  Phone Datasets for the Development of Ivory Coast}
(\byear{2013})
\end{bchapter}
\endbibitem

\bibitem{noulas2012tale}
\begin{barticle}
\bauthor{\bsnm{Noulas}, \binits{A.}},
\bauthor{\bsnm{Scellato}, \binits{S.}},
\bauthor{\bsnm{Lambiotte}, \binits{R.}},
\bauthor{\bsnm{Pontil}, \binits{M.}},
\bauthor{\bsnm{Mascolo}, \binits{C.}}:
\batitle{A tale of many cities: universal patterns in human urban mobility}.
\bjtitle{PloS one}
\bvolume{7}(\bissue{5}),
\bfpage{37027}
(\byear{2012})
\end{barticle}
\endbibitem

\bibitem{liben2005geographic}
\begin{barticle}
\bauthor{\bsnm{Liben-Nowell}, \binits{D.}},
\bauthor{\bsnm{Novak}, \binits{J.}},
\bauthor{\bsnm{Kumar}, \binits{R.}},
\bauthor{\bsnm{Raghavan}, \binits{P.}},
\bauthor{\bsnm{Tomkins}, \binits{A.}}:
\batitle{Geographic routing in social networks}.
\bjtitle{Proceedings of the National Academy of Sciences of the United States
  of America}
\bvolume{102}(\bissue{33}),
\bfpage{11623}
(\byear{2005})
\end{barticle}
\endbibitem

\bibitem{carolan2012issc}
\begin{bchapter}
\bauthor{\bsnm{Carolan}, \binits{E.}},
\bauthor{\bsnm{McLoone}, \binits{S.C.}},
\bauthor{\bsnm{McLoone}, \binits{S.F.}},
\bauthor{\bsnm{Farrell}, \binits{R.}}:
\bctitle{Analysing ireland's interurban communication network using call data
  records}.
In: \bbtitle{Signals and Systems Conference (ISSC 2012), IET Irish}
(\byear{2012})
\end{bchapter}
\endbibitem

\bibitem{schlapfer2014scaling}
\begin{barticle}
\bauthor{\bsnm{Schl{\"a}pfer}, \binits{M.}},
\bauthor{\bsnm{Bettencourt}, \binits{L.}},
\bauthor{\bsnm{Grauwin}, \binits{S.}},
\bauthor{\bsnm{Raschke}, \binits{M.}},
\bauthor{\bsnm{Claxton}, \binits{R.}},
\bauthor{\bsnm{Smoreda}, \binits{Z.}},
\bauthor{\bsnm{West}, \binits{G.B.}},
\bauthor{\bsnm{Ratti}, \binits{C.}}:
\batitle{The scaling of human interactions with city size}.
\bjtitle{Journal of the Royal Society Interface}
\bvolume{11},
\bfpage{20130789}
(\byear{2014})
\end{barticle}
\endbibitem

\bibitem{jo2014spatial}
\begin{botherref}
\oauthor{\bsnm{Jo}, \binits{H.-H.}},
\oauthor{\bsnm{Saram{\"a}ki}, \binits{J.}},
\oauthor{\bsnm{Dunbar}, \binits{R.I.}},
\oauthor{\bsnm{Kaski}, \binits{K.}}:
Spatial patterns of close relationships across the lifespan.
Scientific reports
\textbf{4}
(2014)
\end{botherref}
\endbibitem

\bibitem{herrera2014elliptic}
\begin{barticle}
\bauthor{\bsnm{Herrera-Yag\"ue}, \binits{C.}},
\bauthor{\bsnm{Schneider}, \binits{C.M.}},
\bauthor{\bsnm{Smoreda}, \binits{Z.}},
\bauthor{\bsnm{Couronn\'e}, \binits{T.}},
\bauthor{\bsnm{Zufiria}, \binits{P.J.}},
\bauthor{\bsnm{Gonz\'alez}, \binits{M.C.}}:
\batitle{The elliptic model for communication fluxes}.
\bjtitle{Journal of Statistical Mechanics: Theory and Experiment}
\bvolume{2014}(\bissue{4}),
\bfpage{04022}
(\byear{2014})
\end{barticle}
\endbibitem

\bibitem{grady2012modularity}
\begin{bchapter}
\bauthor{\bsnm{Grady}, \binits{D.}},
\bauthor{\bsnm{Brune}, \binits{R.}},
\bauthor{\bsnm{Thiemann}, \binits{C.}},
\bauthor{\bsnm{Theis}, \binits{F.}},
\bauthor{\bsnm{Brockmann}, \binits{D.}}:
\bctitle{Modularity maximization and tree clustering: Novel ways to determine
  effective geographic borders}.
In: \bbtitle{Handbook of Optimization in Complex Networks},
pp. \bfpage{169}--\blpage{208}.
\bpublisher{Springer}, \blocation{???}
(\byear{2012})
\end{bchapter}
\endbibitem

\bibitem{blondel2011voice}
\begin{botherref}
\oauthor{\bsnm{Blondel}, \binits{V.D.}},
\oauthor{\bsnm{Deville}, \binits{P.}},
\oauthor{\bsnm{Morlot}, \binits{F.}},
\oauthor{\bsnm{Smoreda}, \binits{Z.}},
\oauthor{\bsnm{Van~Dooren}, \binits{P.}},
\oauthor{\bsnm{Ziemlicki}, \binits{C.}}:
Voice on the border: do cellphones redraw the maps?
Paris Tech Review
(2011)
\end{botherref}
\endbibitem

\bibitem{blondel2010regions}
\begin{botherref}
\oauthor{\bsnm{Blondel}, \binits{V.}},
\oauthor{\bsnm{Krings}, \binits{G.}},
\oauthor{\bsnm{Thomas}, \binits{I.}}:
Regions and borders of mobile telephony in belgium and in the brussels
  metropolitan zone.
Brussels Studies
\textbf{42}(4)
(2010)
\end{botherref}
\endbibitem

\bibitem{expert2011uncovering}
\begin{barticle}
\bauthor{\bsnm{Expert}, \binits{P.}},
\bauthor{\bsnm{Evans}, \binits{T.S.}},
\bauthor{\bsnm{Blondel}, \binits{V.D.}},
\bauthor{\bsnm{Lambiotte}, \binits{R.}}:
\batitle{Uncovering space-independent communities in spatial networks}.
\bjtitle{Proceedings of the National Academy of Sciences}
\bvolume{108}(\bissue{19}),
\bfpage{7663}
(\byear{2011})
\end{barticle}
\endbibitem

\bibitem{ratti2010redrawing}
\begin{barticle}
\bauthor{\bsnm{Ratti}, \binits{C.}},
\bauthor{\bsnm{Sobolevsky}, \binits{S.}},
\bauthor{\bsnm{Calabrese}, \binits{F.}},
\bauthor{\bsnm{Andris}, \binits{C.}},
\bauthor{\bsnm{Reades}, \binits{J.}},
\bauthor{\bsnm{Martino}, \binits{M.}},
\bauthor{\bsnm{Claxton}, \binits{R.}},
\bauthor{\bsnm{Strogatz}, \binits{S.H.}}:
\batitle{Redrawing the map of great britain from a network of human
  interactions}.
\bjtitle{PLoS One}
\bvolume{5}(\bissue{12}),
\bfpage{14248}
(\byear{2010})
\end{barticle}
\endbibitem

\bibitem{blumenstock2013social}
\begin{bchapter}
\bauthor{\bsnm{Blumenstock}, \binits{J.E.}},
\bauthor{\bsnm{Fratamico}, \binits{L.}}:
\bctitle{Social and spatial ethnic segregation: A framework for analyzing
  segregation with large-scale spatial network data}.
In: \bbtitle{Proceedings of the 4th Annual Symposium on Computing for
  Development}.
\bsertitle{ACM DEV-4 '13},
pp. \bfpage{11}--\blpage{11110}.
\bpublisher{ACM},
\blocation{New York, NY, USA}
(\byear{2013})
\end{bchapter}
\endbibitem

\bibitem{eagle2010network}
\begin{barticle}
\bauthor{\bsnm{Eagle}, \binits{N.}},
\bauthor{\bsnm{Macy}, \binits{M.}},
\bauthor{\bsnm{Claxton}, \binits{R.}}:
\batitle{Network diversity and economic development}.
\bjtitle{Science}
\bvolume{328}(\bissue{5981}),
\bfpage{1029}
(\byear{2010})
\end{barticle}
\endbibitem

\bibitem{maoD4Dregeco}
\begin{bchapter}
\bauthor{\bsnm{Mao}, \binits{H.}},
\bauthor{\bsnm{Shuai}, \binits{X.}},
\bauthor{\bsnm{Ahn}, \binits{Y.Y.}},
\bauthor{\bsnm{Bollen}, \binits{J.}}:
\bctitle{Mobile communications reveal the regional economy in c{\^o}te
  d'ivoire}.
In: \bbtitle{Mobile Phone Data for Development - Analysis of Mobile Phone
  Datasets for the Development of Ivory Coast}.
\bpublisher{Orange D4D Challenge}, \blocation{???}
(\byear{2013})
\end{bchapter}
\endbibitem

\bibitem{smithclarke2014poverty}
\begin{bchapter}
\bauthor{\bsnm{Smith-Clarke}, \binits{C.}},
\bauthor{\bsnm{Mashhadi}, \binits{A.}},
\bauthor{\bsnm{Capra}, \binits{L.}}:
\bctitle{Poverty on the cheap: Estimating poverty maps using aggregated mobile
  communication networks}.
In: \bbtitle{Proceedings of the 32nd Annual ACM Conference on Human Factors in
  Computing Systems},
pp. \bfpage{511}--\blpage{520}
(\byear{2014}).
\bcomment{ACM}
\end{bchapter}
\endbibitem

\bibitem{friasmartinez2010socio}
\begin{bchapter}
\bauthor{\bsnm{Frias-Martinez}, \binits{V.}},
\bauthor{\bsnm{Virseda}, \binits{J.}},
\bauthor{\bsnm{Frias-Martinez}, \binits{E.}}:
\bctitle{Socio-economic levels and human mobility}.
In: \bbtitle{Qual Meets Quant Workshop-QMQ}
(\byear{2010})
\end{bchapter}
\endbibitem

\bibitem{frias2013forecasting}
\begin{bchapter}
\bauthor{\bsnm{Frias-Martinez}, \binits{V.}},
\bauthor{\bsnm{Soguero-Ruiz}, \binits{C.}},
\bauthor{\bsnm{Frias-Martinez}, \binits{E.}},
\bauthor{\bsnm{Josephidou}, \binits{M.}}:
\bctitle{Forecasting socioeconomic trends with cell phone records}.
In: \bbtitle{Proceedings of the 3rd ACM Symposium on Computing for
  Development},
p. \bfpage{15}
(\byear{2013}).
\bcomment{ACM}
\end{bchapter}
\endbibitem

\bibitem{gutierrez2013evaluating}
\begin{botherref}
\oauthor{\bsnm{Gutierrez}, \binits{T.}},
\oauthor{\bsnm{Krings}, \binits{G.}},
\oauthor{\bsnm{Blondel}, \binits{V.D.}}:
Evaluating socio-economic state of a country analyzing airtime credit and
  mobile phone datasets.
arXiv preprint arXiv:1309.4496
(2013)
\end{botherref}
\endbibitem

\bibitem{holme2012temporal}
\begin{barticle}
\bauthor{\bsnm{Holme}, \binits{P.}},
\bauthor{\bsnm{Saram\"aki}, \binits{J.}}:
\batitle{Temporal networks}.
\bjtitle{Physics reports}
\bvolume{519}(\bissue{3}),
\bfpage{97}--\blpage{125}
(\byear{2012})
\end{barticle}
\endbibitem

\bibitem{hidalgo2008dynamics}
\begin{barticle}
\bauthor{\bsnm{Hidalgo}, \binits{C.A.}},
\bauthor{\bsnm{Rodriguez-Sickert}, \binits{C.}}:
\batitle{The dynamics of a mobile phone network}.
\bjtitle{Physica A: Statistical Mechanics and its Applications}
\bvolume{387}(\bissue{12}),
\bfpage{3017}--\blpage{3024}
(\byear{2008})
\end{barticle}
\endbibitem

\bibitem{raeder2011predictors}
\begin{barticle}
\bauthor{\bsnm{Raeder}, \binits{T.}},
\bauthor{\bsnm{Lizardo}, \binits{O.}},
\bauthor{\bsnm{Hachen}, \binits{D.}},
\bauthor{\bsnm{Chawla}, \binits{N.V.}}:
\batitle{Predictors of short-term decay of cell phone contacts in a large scale
  communication network}.
\bjtitle{Social Networks}
\bvolume{33}(\bissue{4}),
\bfpage{245}--\blpage{257}
(\byear{2011})
\end{barticle}
\endbibitem

\bibitem{karsai2014time}
\begin{botherref}
\oauthor{\bsnm{Karsai}, \binits{M.}},
\oauthor{\bsnm{Perra}, \binits{N.}},
\oauthor{\bsnm{Vespignani}, \binits{A.}}:
Time varying networks and the weakness of strong ties.
Scientific reports
\textbf{4}
(2014)
\end{botherref}
\endbibitem

\bibitem{miritello2013limited}
\begin{botherref}
\oauthor{\bsnm{Miritello}, \binits{G.}},
\oauthor{\bsnm{Rub{\'e}n}, \binits{L.}},
\oauthor{\bsnm{Cebrian}, \binits{M.}},
\oauthor{\bsnm{Moro}, \binits{E.}}:
Limited communication capacity unveils strategies for human interaction.
Scientific Reports
\textbf{3}
(2013)
\end{botherref}
\endbibitem

\bibitem{miritello2013time}
\begin{barticle}
\bauthor{\bsnm{Miritello}, \binits{G.}},
\bauthor{\bsnm{Moro}, \binits{E.}},
\bauthor{\bsnm{Lara}, \binits{R.}},
\bauthor{\bsnm{Mart\'inez-L\'opez}, \binits{R.}},
\bauthor{\bsnm{Belchamber}, \binits{J.}},
\bauthor{\bsnm{Roberts}, \binits{S.G.B.}},
\bauthor{\bsnm{Dunbar}, \binits{R.I.M.}}:
\batitle{Time as a limited resource: Communication strategy in mobile phone
  networks}.
\bjtitle{Social Networks}
\bvolume{35}(\bissue{1}),
\bfpage{89}--\blpage{95}
(\byear{2013})
\end{barticle}
\endbibitem

\bibitem{saramaki2014persistence}
\begin{barticle}
\bauthor{\bsnm{Saram\"aki}, \binits{J.}},
\bauthor{\bsnm{Leicht}, \binits{E.A.}},
\bauthor{\bsnm{L{\'o}pez}, \binits{E.}},
\bauthor{\bsnm{Roberts}, \binits{S.G.B.}},
\bauthor{\bsnm{Reed-Tsochas}, \binits{F.}},
\bauthor{\bsnm{Dunbar}, \binits{R.I.M.}}:
\batitle{The persistence of social signatures in human communication}.
\bjtitle{Proceedings of the National Academy of Sciences}
\bvolume{111}(\bissue{3}),
\bfpage{942}--\blpage{947}
(\byear{2014})
\end{barticle}
\endbibitem

\bibitem{kovanen2011temporal}
\begin{barticle}
\bauthor{\bsnm{Kovanen}, \binits{L.}},
\bauthor{\bsnm{Karsai}, \binits{M.}},
\bauthor{\bsnm{Kaski}, \binits{K.}},
\bauthor{\bsnm{Kert{\'e}sz}, \binits{J.}},
\bauthor{\bsnm{Saram{\"a}ki}, \binits{J.}}:
\batitle{Temporal motifs in time-dependent networks}.
\bjtitle{Journal of Statistical Mechanics: Theory and Experiment}
\bvolume{2011}(\bissue{11}),
\bfpage{11005}
(\byear{2011})
\end{barticle}
\endbibitem

\bibitem{cebrian2010disentangling}
\begin{botherref}
\oauthor{\bsnm{Cebrian}, \binits{M.}},
\oauthor{\bsnm{Pentland}, \binits{A.}},
\oauthor{\bsnm{Kirkpatrick}, \binits{S.}}:
Disentangling social networks inferred from call logs.
Arxiv preprint arXiv:1008.1357
(2010)
\end{botherref}
\endbibitem

\bibitem{barabasi2005origin}
\begin{barticle}
\bauthor{\bsnm{Barab{\'a}si}, \binits{A.}}:
\batitle{The origin of bursts and heavy tails in human activity}.
\bjtitle{Nature}
\bvolume{435},
\bfpage{207}
(\byear{2005})
\end{barticle}
\endbibitem

\bibitem{karsai2011small}
\begin{barticle}
\bauthor{\bsnm{Karsai}, \binits{M.}},
\bauthor{\bsnm{Kivel{\"a}}, \binits{M.}},
\bauthor{\bsnm{Pan}, \binits{R.}},
\bauthor{\bsnm{Kaski}, \binits{K.}},
\bauthor{\bsnm{Kert{\'e}sz}, \binits{J.}},
\bauthor{\bsnm{Barab{\'a}si}, \binits{A.L.}},
\bauthor{\bsnm{Saram{\"a}ki}, \binits{J.}}:
\batitle{Small but slow world: How network topology and burstiness slow down
  spreading}.
\bjtitle{Physical Review E}
\bvolume{83}(\bissue{2}),
\bfpage{025102}
(\byear{2011})
\end{barticle}
\endbibitem

\bibitem{karsai2012universal}
\begin{botherref}
\oauthor{\bsnm{Karsai}, \binits{M.}},
\oauthor{\bsnm{Kaski}, \binits{K.}},
\oauthor{\bsnm{Barab\'a{}si}, \binits{A.L.}},
\oauthor{\bsnm{Kert\'{e}sz}, \binits{J.}}:
Universal features of correlated bursty behaviour.
Scientific Reports
\textbf{2}
(2012)
\end{botherref}
\endbibitem

\bibitem{wu2010evidence}
\begin{barticle}
\bauthor{\bsnm{Wu}, \binits{Y.}},
\bauthor{\bsnm{Zhou}, \binits{C.}},
\bauthor{\bsnm{Xiao}, \binits{J.}},
\bauthor{\bsnm{Kurths}, \binits{J.}},
\bauthor{\bsnm{Schellnhuber}, \binits{H.J.}}:
\batitle{Evidence for a bimodal distribution in human communication}.
\bjtitle{Proceedings of the National Academy of Sciences}
\bvolume{107}(\bissue{44}),
\bfpage{18803}--\blpage{18808}
(\byear{2010})
\end{barticle}
\endbibitem

\bibitem{candia2008uncovering}
\begin{barticle}
\bauthor{\bsnm{Candia}, \binits{J.}},
\bauthor{\bsnm{Gonz{\'a}lez}, \binits{M.C.}},
\bauthor{\bsnm{Wang}, \binits{P.}},
\bauthor{\bsnm{Schoenharl}, \binits{T.}},
\bauthor{\bsnm{Madey}, \binits{G.}},
\bauthor{\bsnm{Barab{\'a}si}, \binits{A.L.}}:
\batitle{Uncovering individual and collective human dynamics from mobile phone
  records}.
\bjtitle{Journal of Physics A: Mathematical and Theoretical}
\bvolume{41},
\bfpage{224015}
(\byear{2008})
\end{barticle}
\endbibitem

\bibitem{jo2012burstiness}
\begin{barticle}
\bauthor{\bsnm{Jo}, \binits{H.-H.}},
\bauthor{\bsnm{Karsai}, \binits{M.}},
\bauthor{\bsnm{Kert\'{e}sz}, \binits{J.}},
\bauthor{\bsnm{Kaski}, \binits{K.}}:
\batitle{Circadian pattern and burstiness in mobile phone communication}.
\bjtitle{New Journal of Physics}
\bvolume{14}(\bissue{1}),
\bfpage{013055}
(\byear{2012})
\end{barticle}
\endbibitem

\bibitem{gonzalez2008understanding}
\begin{barticle}
\bauthor{\bsnm{Gonz{\'a}lez}, \binits{M.C.}},
\bauthor{\bsnm{Hidalgo}, \binits{C.A.}},
\bauthor{\bsnm{Barab{\'a}si}, \binits{A.L.}}:
\batitle{{Understanding individual human mobility patterns}}.
\bjtitle{Nature}
\bvolume{453}(\bissue{7196}),
\bfpage{779}--\blpage{782}
(\byear{2008})
\end{barticle}
\endbibitem

\bibitem{song2010modelling}
\begin{botherref}
\oauthor{\bsnm{Song}, \binits{C.}},
\oauthor{\bsnm{Koren}, \binits{T.}},
\oauthor{\bsnm{Wang}, \binits{P.}},
\oauthor{\bsnm{Barab{\'a}si}, \binits{A.L.}}:
Modelling the scaling properties of human mobility.
Nature Physics
(2010)
\end{botherref}
\endbibitem

\bibitem{csaji2013exploring}
\begin{barticle}
\bauthor{\bsnm{Cs{\'a}ji}, \binits{B.}},
\bauthor{\bsnm{Browet}, \binits{A.}},
\bauthor{\bsnm{Traag}, \binits{V.A.}},
\bauthor{\bsnm{Delvenne}, \binits{J.-C.}},
\bauthor{\bsnm{Huens}, \binits{E.}},
\bauthor{\bsnm{Van~Dooren}, \binits{P.}},
\bauthor{\bsnm{Smoreda}, \binits{Z.}},
\bauthor{\bsnm{Blondel}, \binits{V.D.}}:
\batitle{Exploring the mobility of mobile phone users}.
\bjtitle{Physica A: Statistical Mechanics and its Applications}
\bvolume{392}(\bissue{6}),
\bfpage{1459}--\blpage{1473}
(\byear{2013})
\end{barticle}
\endbibitem

\bibitem{bagrow2012mesoscopic}
\begin{barticle}
\bauthor{\bsnm{Bagrow}, \binits{J.P.}},
\bauthor{\bsnm{Lin}, \binits{Y.-R.}}:
\batitle{Mesoscopic structure and social aspects of human mobility}.
\bjtitle{PloS one}
\bvolume{7}(\bissue{5}),
\bfpage{37676}
(\byear{2012})
\end{barticle}
\endbibitem

\bibitem{amini2013differing}
\begin{bchapter}
\bauthor{\bsnm{Amini}, \binits{A.}},
\bauthor{\bsnm{Kung}, \binits{K.}},
\bauthor{\bsnm{Kang}, \binits{C.}},
\bauthor{\bsnm{Sobolevsky}, \binits{S.}},
\bauthor{\bsnm{Ratti}, \binits{C.}}:
\bctitle{The differing tribal and infrastructural influences on mobility in
  developing and industrialized regions}.
In: \bbtitle{Mobile Phone Data for Development - Analysis of Mobile Phone
  Datasets for the Development of Ivory Coast},
pp. \bfpage{330}--\blpage{339}.
\bpublisher{Orange D4D Challenge}, \blocation{???}
(\byear{2013})
\end{bchapter}
\endbibitem

\bibitem{song2010limits}
\begin{barticle}
\bauthor{\bsnm{Song}, \binits{C.}},
\bauthor{\bsnm{Qu}, \binits{Z.}},
\bauthor{\bsnm{Blumm}, \binits{N.}},
\bauthor{\bsnm{Barab{\'a}si}, \binits{A.L.}}:
\batitle{Limits of predictability in human mobility}.
\bjtitle{Science}
\bvolume{327}(\bissue{5968}),
\bfpage{1018}
(\byear{2010})
\end{barticle}
\endbibitem

\bibitem{calabrese2010human}
\begin{bchapter}
\bauthor{\bsnm{Calabrese}, \binits{F.}},
\bauthor{\bsnm{Di~Lorenzo}, \binits{G.}},
\bauthor{\bsnm{Ratti}, \binits{C.}}:
\bctitle{Human mobility prediction based on individual and collective
  geographical preferences}.
In: \bbtitle{Intelligent Transportation Systems (ITSC), 2010 13th International
  IEEE Conference On},
pp. \bfpage{312}--\blpage{317}
(\byear{2010}).
\bcomment{IEEE}
\end{bchapter}
\endbibitem

\bibitem{park2010eigenmode}
\begin{barticle}
\bauthor{\bsnm{Park}, \binits{J.}},
\bauthor{\bsnm{Lee}, \binits{D.S.}},
\bauthor{\bsnm{Gonz{\'a}lez}, \binits{M.C.}}:
\batitle{The eigenmode analysis of human motion}.
\bjtitle{Journal of Statistical Mechanics: Theory and Experiment}
\bvolume{2010},
\bfpage{11021}
(\byear{2010})
\end{barticle}
\endbibitem

\bibitem{simini2012universal}
\begin{barticle}
\bauthor{\bsnm{Simini}, \binits{F.}},
\bauthor{\bsnm{Gonzalez}, \binits{M.C.}},
\bauthor{\bsnm{Maritan}, \binits{A.}},
\bauthor{\bsnm{Barabasi}, \binits{A.-L.}}:
\batitle{A universal model for mobility and migration patterns}.
\bjtitle{Nature}
\bvolume{484}(\bissue{7392}),
\bfpage{96}--\blpage{100}
(\byear{2012})
\end{barticle}
\endbibitem

\bibitem{palchykov2014inferring}
\begin{botherref}
\oauthor{\bsnm{Palchykov}, \binits{V.}},
\oauthor{\bsnm{Mitrovic}, \binits{M.}},
\oauthor{\bsnm{Jo}, \binits{H.-H.}},
\oauthor{\bsnm{Saramaki}, \binits{J.}},
\oauthor{\bsnm{Pan}, \binits{R.K.}}:
Inferring human mobility using communication patterns.
Scientific reports
\textbf{4}
(2014)
\end{botherref}
\endbibitem

\bibitem{martino2010ocean}
\begin{bchapter}
\bauthor{\bsnm{Martino}, \binits{M.}},
\bauthor{\bsnm{Calabrese}, \binits{F.}},
\bauthor{\bsnm{Di~Lorenzo}, \binits{G.}},
\bauthor{\bsnm{Andris}, \binits{C.}},
\bauthor{\bsnm{Liang}, \binits{L.}},
\bauthor{\bsnm{Ratti}, \binits{C.}}:
\bctitle{Ocean of information: fusing aggregate \& individual dynamics for
  metropolitan analysis}.
In: \bbtitle{Proceedings of the 15th International Conference on Intelligent
  User Interfaces},
pp. \bfpage{357}--\blpage{360}
(\byear{2010}).
\bcomment{ACM}
\end{bchapter}
\endbibitem

\bibitem{eagle2009eigenbehaviors}
\begin{barticle}
\bauthor{\bsnm{Eagle}, \binits{N.}},
\bauthor{\bsnm{Pentland}, \binits{A.S.}}:
\batitle{Eigenbehaviors: Identifying structure in routine}.
\bjtitle{Behavioral Ecology and Sociobiology}
\bvolume{63}(\bissue{7}),
\bfpage{1057}--\blpage{1066}
(\byear{2009})
\end{barticle}
\endbibitem

\bibitem{ratti2006mobile}
\begin{barticle}
\bauthor{\bsnm{Ratti}, \binits{C.}},
\bauthor{\bsnm{Williams}, \binits{S.}},
\bauthor{\bsnm{Frenchman}, \binits{D.}},
\bauthor{\bsnm{Pulselli}, \binits{R.}}:
\batitle{Mobile landscapes: using location data from cell phones for urban
  analysis}.
\bjtitle{Environment and Planning B: Planning and Design}
\bvolume{33}(\bissue{5}),
\bfpage{727}
(\byear{2006})
\end{barticle}
\endbibitem

\bibitem{calabrese2011real}
\begin{barticle}
\bauthor{\bsnm{Calabrese}, \binits{F.}},
\bauthor{\bsnm{Colonna}, \binits{M.}},
\bauthor{\bsnm{Lovisolo}, \binits{P.}},
\bauthor{\bsnm{Parata}, \binits{D.}},
\bauthor{\bsnm{Ratti}, \binits{C.}}:
\batitle{Real-time urban monitoring using cell phones: A case study in rome}.
\bjtitle{Intelligent Transportation Systems, IEEE Transactions on}
\bvolume{12}(\bissue{1}),
\bfpage{141}--\blpage{151}
(\byear{2011})
\end{barticle}
\endbibitem

\bibitem{reades2007cellular}
\begin{botherref}
\oauthor{\bsnm{Reades}, \binits{J.}},
\oauthor{\bsnm{Calabrese}, \binits{F.}},
\oauthor{\bsnm{Sevtsuk}, \binits{A.}},
\oauthor{\bsnm{Ratti}, \binits{C.}}:
Cellular census: Explorations in urban data collection.
IEEE Pervasive Computing,
30--38
(2007)
\end{botherref}
\endbibitem

\bibitem{reades2009eigenplaces}
\begin{barticle}
\bauthor{\bsnm{Reades}, \binits{J.}},
\bauthor{\bsnm{Calabrese}, \binits{F.}},
\bauthor{\bsnm{Ratti}, \binits{C.}}:
\batitle{Eigenplaces: analysing cities using the space- time structure of the
  mobile phone network}.
\bjtitle{Environment and Planning B: Planning and Design}
\bvolume{36}(\bissue{5}),
\bfpage{824}--\blpage{836}
(\byear{2009})
\end{barticle}
\endbibitem

\bibitem{isaacman2010tale}
\begin{bchapter}
\bauthor{\bsnm{Isaacman}, \binits{S.}},
\bauthor{\bsnm{Becker}, \binits{R.}},
\bauthor{\bsnm{C{\'a}ceres}, \binits{R.}},
\bauthor{\bsnm{Kobourov}, \binits{S.}},
\bauthor{\bsnm{Rowland}, \binits{J.}},
\bauthor{\bsnm{Varshavsky}, \binits{A.}}:
\bctitle{A tale of two cities}.
In: \bbtitle{Proceedings of the Eleventh Workshop on Mobile Computing Systems
  \& Applications},
pp. \bfpage{19}--\blpage{24}
(\byear{2010}).
\bcomment{ACM}
\end{bchapter}
\endbibitem

\bibitem{louail2014mobile}
\begin{botherref}
\oauthor{\bsnm{Louail}, \binits{T.}},
\oauthor{\bsnm{Lenormand}, \binits{M.}},
\oauthor{\bsnm{Cant{\'u}}, \binits{O.G.}},
\oauthor{\bsnm{Picornell}, \binits{M.}},
\oauthor{\bsnm{Herranz}, \binits{R.}},
\oauthor{\bsnm{Frias-Martinez}, \binits{E.}},
\oauthor{\bsnm{Ramasco}, \binits{J.J.}},
\oauthor{\bsnm{Barthelemy}, \binits{M.}}:
From mobile phone data to the spatial structure of cities.
arXiv preprint arXiv:1401.4540
(2014)
\end{botherref}
\endbibitem

\bibitem{trasarti2013discovering}
\begin{botherref}
\oauthor{\bsnm{Trasarti}, \binits{R.}},
\oauthor{\bsnm{Olteanu-Raimond}, \binits{A.-M.}},
\oauthor{\bsnm{Nanni}, \binits{M.}},
\oauthor{\bsnm{Couronn{\'e}}, \binits{T.}},
\oauthor{\bsnm{Furletti}, \binits{B.}},
\oauthor{\bsnm{Giannotti}, \binits{F.}},
\oauthor{\bsnm{Smoreda}, \binits{Z.}},
\oauthor{\bsnm{Ziemlicki}, \binits{C.}}:
Discovering urban and country dynamics from mobile phone data with spatial
  correlation patterns.
Telecommunications Policy
(2014)
\end{botherref}
\endbibitem

\bibitem{jiang2013review}
\begin{bchapter}
\bauthor{\bsnm{Jiang}, \binits{S.}},
\bauthor{\bsnm{Fiore}, \binits{G.A.}},
\bauthor{\bsnm{Yang}, \binits{Y.}},
\bauthor{\bsnm{Ferreira}, \binits{J.}},
\bauthor{\bsnm{Frazzoli}, \binits{E.}},
\bauthor{\bsnm{Gonz{\'a}lez}, \binits{M.C.}}:
\bctitle{A review of urban computing for mobile phone traces: current methods,
  challenges and opportunities}.
In: \bbtitle{Proceedings of the 2nd ACM SIGKDD International Workshop on Urban
  Computing},
p. \bfpage{2}
(\byear{2013})
\end{bchapter}
\endbibitem

\bibitem{berlingerio2013allaboard}
\begin{bchapter}
\bauthor{\bsnm{Berlingerio}, \binits{M.}},
\bauthor{\bsnm{Calabrese}, \binits{F.}},
\bauthor{\bsnm{Di~Lorenzo}, \binits{G.}},
\bauthor{\bsnm{Nair}, \binits{R.}},
\bauthor{\bsnm{Pinelli}, \binits{F.}},
\bauthor{\bsnm{Sbodio}, \binits{M.}}:
\bctitle{Allaboard: A system for exploring urban mobility and optimizing public
  transport using cellphone data}.
In: \beditor{\bsnm{Blockeel}, \binits{H.}},
\beditor{\bsnm{Kersting}, \binits{K.}},
\beditor{\bsnm{Nijssen}, \binits{S.}},
\beditor{\bsnm{{\v Z}elezn{\'y}}, \binits{F.}} (eds.)
\bbtitle{Machine Learning and Knowledge Discovery in Databases}.
\bsertitle{Lecture Notes in Computer Science},
vol. \bseriesno{8190},
pp. \bfpage{663}--\blpage{666}.
\bpublisher{Springer}, \blocation{???}
(\byear{2013})
\end{bchapter}
\endbibitem

\bibitem{nanni2013mp4a}
\begin{bchapter}
\bauthor{\bsnm{Nanni}, \binits{M.}},
\bauthor{\bsnm{Trasarti}, \binits{R.}},
\bauthor{\bsnm{Furletti}, \binits{B.}},
\bauthor{\bsnm{Gabrielli}, \binits{L.}},
\bauthor{\bsnm{Van Der~Mede}, \binits{P.}},
\bauthor{\bsnm{De~Bruijn}, \binits{J.}},
\bauthor{\bsnm{De~Romph}, \binits{E.}},
\bauthor{\bsnm{Bruil}, \binits{G.}}:
\bctitle{Mp4-a project: Mobility planning for africa}.
In: \bbtitle{Mobile Phone Data for Development - Analysis of Mobile Phone
  Datasets for the Development of Ivory Coast},
pp. \bfpage{423}--\blpage{446}.
\bpublisher{Orange D4D Challenge}, \blocation{???}
(\byear{2013})
\end{bchapter}
\endbibitem

\bibitem{angelakis2013mobility}
\begin{bchapter}
\bauthor{\bsnm{Angelakis}, \binits{V.}},
\bauthor{\bsnm{Gundleg{\aa}rd}, \binits{D.}},
\bauthor{\bsnm{Rajna}, \binits{B.}},
\bauthor{\bsnm{Rydergren}, \binits{C.}},
\bauthor{\bsnm{Vrotsou}, \binits{K.}},
\bauthor{\bsnm{Carlsson}, \binits{R.}},
\bauthor{\bsnm{Forgeat}, \binits{J.}},
\bauthor{\bsnm{Hu}, \binits{T.H.}},
\bauthor{\bsnm{Liu}, \binits{E.L.}},
\bauthor{\bsnm{Moritz}, \binits{S.}},
\bauthor{\bsnm{Zhao}, \binits{S.}},
\bauthor{\bsnm{Zheng}, \binits{Y.}}:
\bctitle{Mobility modeling for transport efficiency - analysis of travel
  characteristics based on mobile phone data}.
In: \bbtitle{Mobile Phone Data for Development - Analysis of Mobile Phone
  Datasets for the Development of Ivory Coast},
pp. \bfpage{412}--\blpage{422}.
\bpublisher{Orange D4D Challenge}, \blocation{???}
(\byear{2013})
\end{bchapter}
\endbibitem

\bibitem{bagrow2011collective}
\begin{barticle}
\bauthor{\bsnm{Bagrow}, \binits{J.P.}},
\bauthor{\bsnm{Wang}, \binits{D.}},
\bauthor{\bsnm{Barab{\'a}si}, \binits{A.L.}}:
\batitle{Collective response of human populations to large-scale emergencies}.
\bjtitle{PloS one}
\bvolume{6}(\bissue{3}),
\bfpage{17680}
(\byear{2011})
\end{barticle}
\endbibitem

\bibitem{gao2014quantifying}
\begin{botherref}
\oauthor{\bsnm{Gao}, \binits{L.}},
\oauthor{\bsnm{Song}, \binits{C.}},
\oauthor{\bsnm{Gao}, \binits{Z.}},
\oauthor{\bsnm{Barab\'a{}si}, \binits{A.L.}},
\oauthor{\bsnm{Bagrow}, \binits{J.P.}},
\oauthor{\bsnm{Wang}, \binits{D.}}:
Quantifying information flow during emergencies.
Scientific Reports
\textbf{4}
(2014)
\end{botherref}
\endbibitem

\bibitem{xavier2013understanding}
\begin{bchapter}
\bauthor{\bsnm{Xavier}, \binits{F.H.Z.}},
\bauthor{\bsnm{Silveira}, \binits{L.M.}},
\bauthor{\bsnm{Almeida}, \binits{J.M.}},
\bauthor{\bsnm{Malab}, \binits{C.H.S.}},
\bauthor{\bsnm{Ziviani}, \binits{A.}},
\bauthor{\bsnm{Marques-Neto}, \binits{H.T.}}:
\bctitle{Understanding human mobility due to large-scale events}.
In: \bbtitle{NetMob 2013 - Third International Conference on the Analysis of
  Mobile Phone Datasets}
(\byear{2013})
\end{bchapter}
\endbibitem

\bibitem{altshuler2013social}
\begin{barticle}
\bauthor{\bsnm{Altshuler}, \binits{Y.}},
\bauthor{\bsnm{Fire}, \binits{M.}},
\bauthor{\bsnm{Shmueli}, \binits{E.}},
\bauthor{\bsnm{Elovici}, \binits{Y.}},
\bauthor{\bsnm{Bruckstein}, \binits{A.}},
\bauthor{\bsnm{Pentland}, \binits{A.S.}},
\bauthor{\bsnm{Lazer}, \binits{D.}}:
\batitle{The social amplifier - reaction of human communities to emergencies}.
\bjtitle{Journal of Statistical Physics}
\bvolume{152}(\bissue{3}),
\bfpage{399}--\blpage{418}
(\byear{2013})
\end{barticle}
\endbibitem

\bibitem{lu2012predictability}
\begin{barticle}
\bauthor{\bsnm{Lu}, \binits{X.}},
\bauthor{\bsnm{Bengtsson}, \binits{L.}},
\bauthor{\bsnm{Holme}, \binits{P.}}:
\batitle{Predictability of population displacement after the 2010 haiti
  earthquake}.
\bjtitle{Proceedings of the National Academy of Sciences}
\bvolume{109}(\bissue{29}),
\bfpage{11576}--\blpage{11581}
(\byear{2012})
\end{barticle}
\endbibitem

\bibitem{calabrese2011interplay}
\begin{barticle}
\bauthor{\bsnm{Calabrese}, \binits{F.}},
\bauthor{\bsnm{Smoreda}, \binits{Z.}},
\bauthor{\bsnm{Blondel}, \binits{V.D.}},
\bauthor{\bsnm{Ratti}, \binits{C.}}:
\batitle{Interplay between telecommunications and face-to-face interactions: A
  study using mobile phone data}.
\bjtitle{PloS one}
\bvolume{6}(\bissue{7}),
\bfpage{20814}
(\byear{2011})
\end{barticle}
\endbibitem

\bibitem{wang2011human}
\begin{bchapter}
\bauthor{\bsnm{Wang}, \binits{D.}},
\bauthor{\bsnm{Pedreschi}, \binits{D.}},
\bauthor{\bsnm{Song}, \binits{C.}},
\bauthor{\bsnm{Giannotti}, \binits{F.}},
\bauthor{\bsnm{Barab{\'a}si}, \binits{A.L.}}:
\bctitle{Human mobility, social ties, and link prediction}.
In: \bbtitle{17th ACM SIGKDD Conference on Knowledge Discovery and Data Mining
  (KDD'11)}
(\byear{2011})
\end{bchapter}
\endbibitem

\bibitem{eagle2009community}
\begin{bchapter}
\bauthor{\bsnm{Eagle}, \binits{N.}},
\bauthor{\bparticle{de} \bsnm{Montjoye}, \binits{Y.A.}},
\bauthor{\bsnm{Bettencourt}, \binits{L.M.A.}}:
\bctitle{Community computing: Comparisons between rural and urban societies
  using mobile phone data}.
In: \bbtitle{2009 International Conference on Computational Science and
  Engineering},
pp. \bfpage{144}--\blpage{150}
(\byear{2009}).
\bcomment{IEEE}
\end{bchapter}
\endbibitem

\bibitem{peruani2011directedness}
\begin{barticle}
\bauthor{\bsnm{Peruani}, \binits{F.}},
\bauthor{\bsnm{Tabourier}, \binits{L.}}:
\batitle{Directedness of information flow in mobile phone communication
  networks}.
\bjtitle{PLoS One}
\bvolume{6}(\bissue{12}),
\bfpage{28860}
(\byear{2011})
\end{barticle}
\endbibitem

\bibitem{tabourier2012how}
\begin{bchapter}
\bauthor{\bsnm{Tabourier}, \binits{L.}},
\bauthor{\bsnm{Stoica}, \binits{A.}},
\bauthor{\bsnm{Peruani}, \binits{F.}}:
\bctitle{How to detect causality effects on large dynamical communication
  networks: a case study}.
In: \bbtitle{Communication Systems and Networks (COMSNETS), 2012 Fourth
  International Conference On},
pp. \bfpage{1}--\blpage{7}
(\byear{2012}).
\bcomment{IEEE}
\end{bchapter}
\endbibitem

\bibitem{miritello2011dynamical}
\begin{barticle}
\bauthor{\bsnm{Miritello}, \binits{G.}},
\bauthor{\bsnm{Moro}, \binits{E.}},
\bauthor{\bsnm{Lara}, \binits{R.}}:
\batitle{Dynamical strength of social ties in information spreading}.
\bjtitle{Physical Review E}
\bvolume{83}(\bissue{4}),
\bfpage{045102}
(\byear{2011})
\end{barticle}
\endbibitem

\bibitem{newman2002spread}
\begin{barticle}
\bauthor{\bsnm{Newman}, \binits{M.E.J.}}:
\batitle{Spread of epidemic disease on networks}.
\bjtitle{Physical Review E}
\bvolume{66},
\bfpage{016128}
(\byear{2002}).
doi:\doiurl{10.1103/PhysRevE.66.016128}
\end{barticle}
\endbibitem

\bibitem{newman2006structure}
\begin{bbook}
\bauthor{\bsnm{Newman}, \binits{M.E.J.}},
\bauthor{\bsnm{Barabasi}, \binits{A.L.}},
\bauthor{\bsnm{Watts}, \binits{D.J.}}:
\bbtitle{The Structure and Dynamics of Networks}.
\bpublisher{Princeton University Press},
\blocation{Princeton}
(\byear{2006})
\end{bbook}
\endbibitem

\bibitem{kivela2012multiscale}
\begin{barticle}
\bauthor{\bsnm{Kivel\"{a}}, \binits{M.}},
\bauthor{\bsnm{Pan}, \binits{R.}},
\bauthor{\bsnm{Kaski}, \binits{K.}},
\bauthor{\bsnm{Kert\'{e}sz}, \binits{J.}},
\bauthor{\bsnm{Saram\"aki}, \binits{J.}},
\bauthor{\bsnm{Karsai}, \binits{M.}}:
\batitle{Multiscale analysis of spreading in a large communication network}.
\bjtitle{Journal of Statistical Mechanics: Theory and Experiment}
\bvolume{2012}(\bissue{03}),
\bfpage{03005}
(\byear{2012})
\end{barticle}
\endbibitem

\bibitem{liu2011controllability}
\begin{barticle}
\bauthor{\bsnm{Liu}, \binits{Y.Y.}},
\bauthor{\bsnm{Slotine}, \binits{J.J.}},
\bauthor{\bsnm{Barab{\'a}si}, \binits{A.L.}}:
\batitle{Controllability of complex networks}.
\bjtitle{Nature}
\bvolume{473}(\bissue{7346}),
\bfpage{167}--\blpage{173}
(\byear{2011})
\end{barticle}
\endbibitem

\bibitem{cebrian2010measuring}
\begin{barticle}
\bauthor{\bsnm{Cebrian}, \binits{M.}},
\bauthor{\bsnm{Lahiri}, \binits{M.}},
\bauthor{\bsnm{Oliver}, \binits{N.}},
\bauthor{\bsnm{Pentland}, \binits{A.}}:
\batitle{Measuring the collective potential of populations from dynamic social
  interaction data}.
\bjtitle{Selected Topics in Signal Processing, IEEE Journal of}
\bvolume{4}(\bissue{4}),
\bfpage{677}--\blpage{686}
(\byear{2010})
\end{barticle}
\endbibitem

\bibitem{wang2009understanding}
\begin{barticle}
\bauthor{\bsnm{Wang}, \binits{P.}},
\bauthor{\bsnm{Gonz{\'a}lez}, \binits{M.C.}},
\bauthor{\bsnm{Hidalgo}, \binits{C.A.}},
\bauthor{\bsnm{Barab{\'a}si}, \binits{A.L.}}:
\batitle{Understanding the spreading patterns of mobile phone viruses}.
\bjtitle{Science}
\bvolume{324}(\bissue{5930}),
\bfpage{1071}
(\byear{2009})
\end{barticle}
\endbibitem

\bibitem{wang2010new}
\begin{botherref}
\oauthor{\bsnm{Wang}, \binits{P.}},
\oauthor{\bsnm{Gonz{\'a}lez}, \binits{M.C.}},
\oauthor{\bsnm{Menezes}, \binits{R.}},
\oauthor{\bsnm{Barab{\'a}si}, \binits{A.L.}}:
New generation of mobile phone viruses and corresponding countermeasures.
Arxiv preprint arXiv:1012.3156
(2010)
\end{botherref}
\endbibitem

\bibitem{wang2013understanding}
\begin{barticle}
\bauthor{\bsnm{Wang}, \binits{P.}},
\bauthor{\bsnm{Gonz{\'a}lez}, \binits{M.C.}},
\bauthor{\bsnm{Menezes}, \binits{R.}},
\bauthor{\bsnm{Barab\'a{}si}, \binits{A.L.}}:
\batitle{Understanding the spread of malicious mobile-phone programs and their
  damage potential}.
\bjtitle{International journal of information security}
\bvolume{12}(\bissue{5}),
\bfpage{383}--\blpage{392}
(\byear{2013})
\end{barticle}
\endbibitem

\bibitem{baccelli2011modeling}
\begin{botherref}
\oauthor{\bsnm{Baccelli}, \binits{F.}},
\oauthor{\bsnm{Bolot}, \binits{J.}}:
Modeling the economic value of location and preference data of mobile users.
Proc. IEEE Infocom 2011
(2011)
\end{botherref}
\endbibitem

\bibitem{calabrese2014urban}
\begin{barticle}
\bauthor{\bsnm{Calabrese}, \binits{F.}},
\bauthor{\bsnm{Ferrari}, \binits{L.}},
\bauthor{\bsnm{Blondel}, \binits{V.D.}}:
\batitle{Urban sensing using mobile phone network data: A survey of research}.
\bjtitle{ACM Computing Surveys (CSUR)}
\bvolume{47}(\bissue{2}),
\bfpage{25}
(\byear{2014})
\end{barticle}
\endbibitem

\bibitem{isaacman2011identifying}
\begin{botherref}
\oauthor{\bsnm{Isaacman}, \binits{S.}},
\oauthor{\bsnm{Becker}, \binits{R.}},
\oauthor{\bsnm{C{\'a}ceres}, \binits{R.}},
\oauthor{\bsnm{Kobourov}, \binits{S.}},
\oauthor{\bsnm{Martonosi}, \binits{M.}},
\oauthor{\bsnm{Rowland}, \binits{J.}},
\oauthor{\bsnm{Varshavsky}, \binits{A.}}:
Identifying important places in people's lives from cellular network data.
Pervasive Computing,
133--151
(2011)
\end{botherref}
\endbibitem

\bibitem{steenbruggen2011mobile}
\begin{botherref}
\oauthor{\bsnm{Steenbruggen}, \binits{J.}},
\oauthor{\bsnm{Borzacchiello}, \binits{M.T.}},
\oauthor{\bsnm{Nijkamp}, \binits{P.}},
\oauthor{\bsnm{Scholten}, \binits{H.}}:
Mobile phone data from gsm networks for traffic parameter and urban spatial
  pattern assessment: a review of applications and opportunities.
GeoJournal,
1--21
(2011)
\end{botherref}
\endbibitem

\bibitem{toole2014path}
\begin{botherref}
\oauthor{\bsnm{Toole}, \binits{J.L.}},
\oauthor{\bsnm{Colak}, \binits{S.}},
\oauthor{\bsnm{Alhasoun}, \binits{F.}},
\oauthor{\bsnm{Evsukoff}, \binits{A.}},
\oauthor{\bsnm{Gonzalez}, \binits{M.C.}}:
The path most travelled: Mining road usage patterns from massive call data.
arXiv preprint arXiv:1403.0636
(2014)
\end{botherref}
\endbibitem

\bibitem{wang2012understanding}
\begin{botherref}
\oauthor{\bsnm{Wang}, \binits{P.}},
\oauthor{\bsnm{Hunter}, \binits{T.}},
\oauthor{\bsnm{Bayen}, \binits{A.M.}},
\oauthor{\bsnm{Schechtner}, \binits{K.}},
\oauthor{\bsnm{Gonz{\'a}lez}, \binits{M.C.}}:
Understanding road usage patterns in urban areas.
Scientific reports
\textbf{2}
(2012)
\end{botherref}
\endbibitem

\bibitem{mcinerney2013crowdsourcing}
\begin{bchapter}
\bauthor{\bsnm{McInerney}, \binits{J.}},
\bauthor{\bsnm{Roger}, \binits{A.}},
\bauthor{\bsnm{Jennings}, \binits{N.R.}}:
\bctitle{Crowdsourcing physical package delivery using the existing routine
  mobility of a local population}.
In: \bbtitle{Mobile Phone Data for Development - Analysis of Mobile Phone
  Datasets for the Development of Ivory Coast},
pp. \bfpage{447}--\blpage{456}.
\bpublisher{Orange D4D Challenge}, \blocation{???}
(\byear{2013})
\end{bchapter}
\endbibitem

\bibitem{gambs2013towards}
\begin{bchapter}
\bauthor{\bsnm{Gambs}, \binits{S.}},
\bauthor{\bsnm{Killijian}, \binits{M.-O.}}:
\bctitle{Towards a recomender system for bush taxis}.
In: \bbtitle{Mobile Phone Data for Development - Analysis of Mobile Phone
  Datasets for the Development of Ivory Coast},
pp. \bfpage{457}--\blpage{466}.
\bpublisher{Orange D4D Challenge}, \blocation{???}
(\byear{2013})
\end{bchapter}
\endbibitem

\bibitem{calabrese2010geography}
\begin{botherref}
\oauthor{\bsnm{Calabrese}, \binits{F.}},
\oauthor{\bsnm{Pereira}, \binits{F.}},
\oauthor{\bsnm{Di~Lorenzo}, \binits{G.}},
\oauthor{\bsnm{Liu}, \binits{L.}},
\oauthor{\bsnm{Ratti}, \binits{C.}}:
The geography of taste: analyzing cell-phone mobility and social events.
Pervasive Computing,
22--37
(2010)
\end{botherref}
\endbibitem

\bibitem{quercia2010recommending}
\begin{bchapter}
\bauthor{\bsnm{Quercia}, \binits{D.}},
\bauthor{\bsnm{Lathia}, \binits{N.}},
\bauthor{\bsnm{Calabrese}, \binits{F.}},
\bauthor{\bsnm{Di~Lorenzo}, \binits{G.}},
\bauthor{\bsnm{Crowcroft}, \binits{J.}}:
\bctitle{Recommending social events from mobile phone location data}.
In: \bbtitle{2010 IEEE International Conference on Data Mining},
pp. \bfpage{971}--\blpage{976}
(\byear{2010}).
\bcomment{IEEE}
\end{bchapter}
\endbibitem

\bibitem{cloquet2014forecasting}
\begin{botherref}
\oauthor{\bsnm{Cloquet}, \binits{C.}},
\oauthor{\bsnm{Blondel}, \binits{V.D.}}:
Forecasting event attendance with anonymized mobile phone data.
submitted to Big Data Research, Elsevier
(2014)
\end{botherref}
\endbibitem

\bibitem{xavier2012analyzing}
\begin{bchapter}
\bauthor{\bsnm{Xavier}, \binits{F.H.Z.}},
\bauthor{\bsnm{Silveira}, \binits{L.M.}},
\bauthor{\bsnm{Almeida}, \binits{J.M.}},
\bauthor{\bsnm{Ziviani}, \binits{A.}},
\bauthor{\bsnm{Malab}, \binits{C.H.S.}},
\bauthor{\bsnm{Marques-Neto}, \binits{H.T.}}:
\bctitle{Analyzing the workload dynamics of a mobile phone network in large
  scale events}.
In: \bbtitle{Proceedings of the First Workshop on Urban Networking},
pp. \bfpage{37}--\blpage{42}
(\byear{2012}).
\bcomment{ACM}
\end{bchapter}
\endbibitem

\bibitem{manfredini2011monitoring}
\begin{botherref}
\oauthor{\bsnm{Manfredini}, \binits{F.}},
\oauthor{\bsnm{Tagliolato}, \binits{P.}},
\oauthor{\bsnm{Di~Rosa}, \binits{C.}}:
Monitoring temporary populations through cellular core network data.
Computational Science and Its Applications-ICCSA 2011,
151--161
(2011)
\end{botherref}
\endbibitem

\bibitem{kuusik2009analysing}
\begin{bchapter}
\bauthor{\bsnm{Kuusik}, \binits{A.}},
\bauthor{\bsnm{Ahas}, \binits{R.}},
\bauthor{\bsnm{Tiru}, \binits{M.}}:
\bctitle{Analysing repeat visitation on country level with passive mobile
  positioning method: an estonian case study}.
In: \bbtitle{XVII Scientific Conference on Economic Policy},
pp. \bfpage{1}--\blpage{3}
(\byear{2009})
\end{bchapter}
\endbibitem

\bibitem{wesolowski2012quantifying}
\begin{barticle}
\bauthor{\bsnm{Wesolowski}, \binits{A.}},
\bauthor{\bsnm{Eagle}, \binits{N.}},
\bauthor{\bsnm{Tatem}, \binits{A.J.}},
\bauthor{\bsnm{Smith}, \binits{D.L.}},
\bauthor{\bsnm{Noor}, \binits{A.M.}},
\bauthor{\bsnm{Snow}, \binits{R.W.}},
\bauthor{\bsnm{Buckee}, \binits{C.O.}}:
\batitle{Quantifying the impact of human mobility on malaria}.
\bjtitle{Science}
\bvolume{338}(\bissue{6104}),
\bfpage{267}--\blpage{270}
(\byear{2012})
\end{barticle}
\endbibitem

\bibitem{tizzoni2014use}
\begin{barticle}
\bauthor{\bsnm{Tizzoni}, \binits{M.}},
\bauthor{\bsnm{Bajardi}, \binits{P.}},
\bauthor{\bsnm{Decuyper}, \binits{A.}},
\bauthor{\bsnm{Kon Kam~King}, \binits{G.}},
\bauthor{\bsnm{Schneider}, \binits{C.M.}},
\bauthor{\bsnm{Blondel}, \binits{V.D.}},
\bauthor{\bsnm{Smoreda}, \binits{Z.}},
\bauthor{\bsnm{Gonz{\'a}lez}, \binits{M.C.}},
\bauthor{\bsnm{Colizza}, \binits{V.}}:
\batitle{On the use of human mobility proxies for modeling epidemics}.
\bjtitle{PLoS computational biology}
\bvolume{10}(\bissue{7}),
\bfpage{1003716}
(\byear{2014})
\end{barticle}
\endbibitem

\bibitem{frias2011agent}
\begin{bchapter}
\bauthor{\bsnm{Frias-Martinez}, \binits{E.}},
\bauthor{\bsnm{Williamson}, \binits{G.}},
\bauthor{\bsnm{Frias-Martinez}, \binits{V.}}:
\bctitle{An agent-based model of epidemic spread using human mobility and
  social network information}.
In: \bbtitle{Privacy, Security, Risk and Trust (PASSAT) and 2011 IEEE Third
  Inernational Conference on Social Computing (SocialCom), 2011 IEEE Third
  International Conference On},
pp. \bfpage{57}--\blpage{64}
(\byear{2011}).
doi:\doiurl{10.1109/PASSAT/SocialCom.2011.142}
\end{bchapter}
\endbibitem

\bibitem{blondel2012data}
\begin{botherref}
\oauthor{\bsnm{Blondel}, \binits{V.D.}},
\oauthor{\bsnm{Esch}, \binits{M.}},
\oauthor{\bsnm{Chan}, \binits{C.}},
\oauthor{\bsnm{Cl{\'e}rot}, \binits{F.}},
\oauthor{\bsnm{Deville}, \binits{P.}},
\oauthor{\bsnm{Huens}, \binits{E.}},
\oauthor{\bsnm{Morlot}, \binits{F.}},
\oauthor{\bsnm{Smoreda}, \binits{Z.}},
\oauthor{\bsnm{Ziemlicki}, \binits{C.}}:
Data for development: the {D4D} challenge on mobile phone data.
arXiv preprint arXiv:1210.0137
(2012)
\end{botherref}
\endbibitem

\bibitem{kafsi2013mitigating}
\begin{botherref}
\oauthor{\bsnm{Kafsi}, \binits{M.}},
\oauthor{\bsnm{Kazemi}, \binits{E.}},
\oauthor{\bsnm{Maystre}, \binits{L.}},
\oauthor{\bsnm{Yartseva}, \binits{L.}},
\oauthor{\bsnm{Grossglauser}, \binits{M.}},
\oauthor{\bsnm{Thiran}, \binits{P.}}:
Mitigating epidemics through mobile micro-measures.
arXiv preprint arXiv:1307.2084
(2013)
\end{botherref}
\endbibitem

\bibitem{lima2013exploiting}
\begin{botherref}
\oauthor{\bsnm{Lima}, \binits{A.}},
\oauthor{\bsnm{De~Domenico}, \binits{M.}},
\oauthor{\bsnm{Pejovic}, \binits{V.}},
\oauthor{\bsnm{Musolesi}, \binits{M.}}:
Exploiting cellular data for disease containment and information campaigns
  strategies in country-wide epidemics.
arXiv preprint arXiv:1306.4534
(2013)
\end{botherref}
\endbibitem

\bibitem{wesolowski2014containing}
\begin{botherref}
\oauthor{\bsnm{Wesolowski}, \binits{A.}},
\oauthor{\bsnm{Buckee}, \binits{C.O.}},
\oauthor{\bsnm{Bengtsson}, \binits{L.}},
\oauthor{\bsnm{Wetter}, \binits{E.}},
\oauthor{\bsnm{Lu}, \binits{X.}},
\oauthor{\bsnm{Tatem}, \binits{A.J.}}:
Commentary: Containing the ebola outbreak--the potential and challenge of
  mobile network data.
PLOS Currents Outbreaks
(2014)
\end{botherref}
\endbibitem

\bibitem{demontjoye2014enabling}
\begin{botherref}
\oauthor{\bparticle{de} \bsnm{Montjoye}, \binits{Y.A.}},
\oauthor{\bsnm{Kendall}, \binits{J.}},
\oauthor{\bsnm{Kerry}, \binits{C.F.}}:
Enabling humanitarian use of mobile phone data.
Issues in Technology Innovation
(26)
(2014)
\end{botherref}
\endbibitem

\bibitem{katz1970personal}
\begin{bbook}
\bauthor{\bsnm{Katz}, \binits{E.}},
\bauthor{\bsnm{Lazarsfeld}, \binits{P.F.}}:
\bbtitle{Personal Influence, The Part Played by People in the Flow of Mass
  Communications}.
\bpublisher{Transaction Publishers}, \blocation{???}
(\byear{1970})
\end{bbook}
\endbibitem

\bibitem{watts2007influentials}
\begin{barticle}
\bauthor{\bsnm{Watts}, \binits{D.J.}},
\bauthor{\bsnm{Dodds}, \binits{P.S.}}:
\batitle{Influentials, networks, and public opinion formation}.
\bjtitle{Journal of consumer research}
\bvolume{34}(\bissue{4}),
\bfpage{441}--\blpage{458}
(\byear{2007})
\end{barticle}
\endbibitem

\bibitem{szabo2006network}
\begin{botherref}
\oauthor{\bsnm{Szab{\'o}}, \binits{G.}},
\oauthor{\bsnm{Barab\'a{}si}, \binits{A.L.}}:
Network effects in service usage.
Arxiv preprint physics/0611177
(2006)
\end{botherref}
\endbibitem

\bibitem{hill2006network}
\begin{barticle}
\bauthor{\bsnm{Hill}, \binits{S.}},
\bauthor{\bsnm{Provost}, \binits{F.}},
\bauthor{\bsnm{Volinsky}, \binits{C.}}:
\batitle{Network-based marketing: Identifying likely adopters via consumer
  networks}.
\bjtitle{Statistical Science}
\bvolume{21}(\bissue{2}),
\bfpage{256}--\blpage{276}
(\byear{2006})
\end{barticle}
\endbibitem

\bibitem{aharony2010tracing}
\begin{bchapter}
\bauthor{\bsnm{Aharony}, \binits{N.}},
\bauthor{\bsnm{Pan}, \binits{W.}},
\bauthor{\bsnm{Ip}, \binits{C.}},
\bauthor{\bsnm{Pentland}, \binits{A.}}:
\bctitle{Tracing mobile phone app installations in the" friends and family"
  study}.
In: \bbtitle{Proceedings of the 2010 Workshop on Information in Networks
  (WIN'10)}
(\byear{2010})
\end{bchapter}
\endbibitem

\bibitem{risselada2014dynamic}
\begin{barticle}
\bauthor{\bsnm{Risselada}, \binits{H.}},
\bauthor{\bsnm{Verhoef}, \binits{P.C.}},
\bauthor{\bsnm{Bijmolt}, \binits{T.H.A.}}:
\batitle{Dynamic effects of social influence and direct marketing on the
  adoption of high-technology products}.
\bjtitle{Journal of Marketing}
\bvolume{78}(\bissue{2}),
\bfpage{52}--\blpage{68}
(\byear{2014})
\end{barticle}
\endbibitem

\bibitem{backlund2014effects}
\begin{barticle}
\bauthor{\bsnm{Backlund}, \binits{V.-P.}},
\bauthor{\bsnm{Saram\"aki}, \binits{J.}},
\bauthor{\bsnm{Pan}, \binits{R.K.}}:
\batitle{Effects of temporal correlations on cascades: Threshold models on
  temporal networks}.
\bjtitle{Phys. Rev. E}
\bvolume{89},
\bfpage{062815}
(\byear{2014}).
doi:\doiurl{10.1103/PhysRevE.89.062815}
\end{barticle}
\endbibitem

\bibitem{blondel2006social}
\begin{barticle}
\bauthor{\bsnm{Blondel}, \binits{V.}},
\bauthor{\bparticle{de} \bsnm{Kerchove}, \binits{C.}},
\bauthor{\bsnm{Huens}, \binits{E.}},
\bauthor{\bsnm{Van~Dooren}, \binits{P.}}:
\batitle{Social leaders in graphs}.
\bjtitle{Lecture notes in control and information sciences}
\bvolume{341},
\bfpage{231}
(\byear{2006})
\end{barticle}
\endbibitem

\bibitem{deranking}
\begin{botherref}
\oauthor{\bparticle{de} \bsnm{Kerchove~d'Exaerde}, \binits{C.}}:
Ranking large networks: Leadership, optimization and distrust (phd thesis)
(2009)
\end{botherref}
\endbibitem

\bibitem{D4Dbook}
\begin{bbook}
\beditor{\bsnm{Blondel}, \binits{V.D.}},
\beditor{\bparticle{de} \bsnm{Cordes}, \binits{N.}},
\beditor{\bsnm{Decuyper}, \binits{A.}},
\beditor{\bsnm{Deville}, \binits{P.}},
\beditor{\bsnm{Raguenez}, \binits{J.}},
\beditor{\bsnm{Smoreda}, \binits{Z.}} (eds.):
\bbtitle{Mobile Phone Data for Development - Analysis of Mobile Phone Datasets
  for the Development of Ivory Coast}.
\bpublisher{Orange D4D Challenge}, \blocation{???}
(\byear{2013})
\end{bbook}
\endbibitem

\bibitem{salnikov2014geography}
\begin{barticle}
\bauthor{\bsnm{Salnikov}, \binits{V.}},
\bauthor{\bsnm{Schien}, \binits{D.}},
\bauthor{\bsnm{Youn}, \binits{H.}},
\bauthor{\bsnm{Lambiotte}, \binits{R.}},
\bauthor{\bsnm{Gastner}, \binits{M.T.}}:
\batitle{The geography and carbon footprint of mobile phone use in c\^ote
  d'ivoire}.
\bjtitle{EPJ Data Science}
\bvolume{3}(\bissue{1}),
\bfpage{1}--\blpage{15}
(\byear{2014})
\end{barticle}
\endbibitem

\bibitem{frias2012relationship}
\begin{bchapter}
\bauthor{\bsnm{Frias-Martinez}, \binits{V.}},
\bauthor{\bsnm{Virseda}, \binits{J.}}:
\bctitle{On the relationship between socio-economic factors and cell phone
  usage}.
In: \bbtitle{Proceedings of the Fifth International Conference on Information
  and Communication Technologies and Development},
pp. \bfpage{76}--\blpage{84}
(\byear{2012}).
\bcomment{ACM}
\end{bchapter}
\endbibitem

\bibitem{buckee2013mobile}
\begin{barticle}
\bauthor{\bsnm{Buckee}, \binits{C.O.}},
\bauthor{\bsnm{Wesolowski}, \binits{A.}},
\bauthor{\bsnm{Eagle}, \binits{N.N.}},
\bauthor{\bsnm{Hansen}, \binits{E.}},
\bauthor{\bsnm{Snow}, \binits{R.W.}}:
\batitle{Mobile phones and malaria: modeling human and parasite travel}.
\bjtitle{Travel medicine and infectious disease}
\bvolume{11}(\bissue{1}),
\bfpage{15}--\blpage{22}
(\byear{2013})
\end{barticle}
\endbibitem

\bibitem{onnela2012spreading}
\begin{barticle}
\bauthor{\bsnm{Onnela}, \binits{J.P.}},
\bauthor{\bsnm{Christakis}, \binits{N.A.}}:
\batitle{Spreading paths in partially observed social networks}.
\bjtitle{Physical Review E}
\bvolume{85}(\bissue{3}),
\bfpage{036106}
(\byear{2012})
\end{barticle}
\endbibitem

\bibitem{ranjan2012are}
\begin{barticle}
\bauthor{\bsnm{Ranjan}, \binits{G.}},
\bauthor{\bsnm{Zang}, \binits{H.}},
\bauthor{\bsnm{Zhang}, \binits{Z.L.}},
\bauthor{\bsnm{Bolot}, \binits{J.}}:
\batitle{Are call detail records biased for sampling human mobility?}
\bjtitle{ACM SIGMOBILE Mobile Computing and Communications Review}
\bvolume{16}(\bissue{3}),
\bfpage{33}--\blpage{44}
(\byear{2012})
\end{barticle}
\endbibitem

\bibitem{stopczynski2014measuring}
\begin{barticle}
\bauthor{\bsnm{Stopczynski}, \binits{A.}},
\bauthor{\bsnm{Sekara}, \binits{V.}},
\bauthor{\bsnm{Sapiezynski}, \binits{P.}},
\bauthor{\bsnm{Cuttone}, \binits{A.}},
\bauthor{\bsnm{Madsen}, \binits{M.M.}},
\bauthor{\bsnm{Larsen}, \binits{J.E.}},
\bauthor{\bsnm{Lehmann}, \binits{S.}}:
\batitle{Measuring large-scale social networks with high resolution}.
\bjtitle{PloS one}
\bvolume{9}(\bissue{4}),
\bfpage{95978}
(\byear{2014})
\end{barticle}
\endbibitem

\bibitem{singel2009netflix}
\begin{botherref}
\oauthor{\bsnm{Singel}, \binits{R.}}:
Netflix spilled your brokeback mountain secret, lawsuit claims.
Threat Level (blog), Wired
(2009)
\end{botherref}
\endbibitem

\bibitem{barth2012re}
\begin{botherref}
\oauthor{\bsnm{Barth-Jones}, \binits{D.C.}}:
The're-identification'of governor william weld's medical information: A
  critical re-examination of health data identification risks and privacy
  protections, then and now.
Then and Now
(2012)
\end{botherref}
\endbibitem

\bibitem{directive2012eu}
\begin{botherref}
\oauthor{\bsnm{Commission}, \binits{E.}}:
Commission proposes a comprehensive reform of data protection rules to increase
  users' control of their data and to cut costs for businesses.
Reference: IP/12/46
(2012)
\end{botherref}
\endbibitem

\bibitem{directive199595}
\begin{botherref}
\oauthor{\bsnm{Directive}, \binits{E.}}:
95/46/ec of the european parliament and of the council of 24 october 1995 on
  the protection of individuals with regard to the processing of personal data
  and on the free movement of such data.
Official Journal of the EC
\textbf{23}(6)
(1995)
\end{botherref}
\endbibitem

\bibitem{zang2011anonymization}
\begin{botherref}
\oauthor{\bsnm{Zang}, \binits{H.}},
\oauthor{\bsnm{Bolot}, \binits{J.}}:
Anonymization of location data does not work: a large-scale measurement study.
submitted to ACM Mobicom
\textbf{11}
(2011)
\end{botherref}
\endbibitem

\bibitem{demontjoye2013unique}
\begin{botherref}
\oauthor{\bparticle{de} \bsnm{Montjoye}, \binits{Y.A.}},
\oauthor{\bsnm{Hidalgo}, \binits{C.A.}},
\oauthor{\bsnm{Verleysen}, \binits{M.}},
\oauthor{\bsnm{Blondel}, \binits{V.D.}}:
Unique in the crowd: The privacy bounds of human mobility.
Scientific Reports
\textbf{3}
(2013)
\end{botherref}
\endbibitem

\bibitem{backstrom2007wherefore}
\begin{bchapter}
\bauthor{\bsnm{Backstrom}, \binits{L.}},
\bauthor{\bsnm{Dwork}, \binits{C.}},
\bauthor{\bsnm{Kleinberg}, \binits{J.}}:
\bctitle{Wherefore art thou r3579x?: anonymized social networks, hidden
  patterns, and structural steganography}.
In: \bbtitle{Proceedings of the 16th International Conference on World Wide
  Web},
pp. \bfpage{181}--\blpage{190}
(\byear{2007}).
\bcomment{ACM}
\end{bchapter}
\endbibitem

\bibitem{narayanan2009anonymizing}
\begin{bchapter}
\bauthor{\bsnm{Narayanan}, \binits{A.}},
\bauthor{\bsnm{Shmatikov}, \binits{V.}}:
\bctitle{De-anonymizing social networks}.
In: \bbtitle{2009 30th IEEE Symposium on Security and Privacy},
pp. \bfpage{173}--\blpage{187}
(\byear{2009}).
\bcomment{IEEE}
\end{bchapter}
\endbibitem

\bibitem{song2014not}
\begin{bchapter}
\bauthor{\bsnm{Song}, \binits{Y.}},
\bauthor{\bsnm{Dahlmeier}, \binits{D.}},
\bauthor{\bsnm{Bressan}, \binits{S.}}:
\bctitle{Not so unique in the crowd: a simple and effective algorithm for
  anonymizing location data}.
In: \bbtitle{Proceeding of the 1st International Workshop on Privacy-Preserving
  IR: When Information Retrieval Meets Privacy and Security (PIR 2014)},
p. \bfpage{19}
(\byear{2014})
\end{bchapter}
\endbibitem

\bibitem{demontjoye2014d4d}
\begin{botherref}
\oauthor{\bparticle{de} \bsnm{Montjoye}, \binits{Y.A.}},
\oauthor{\bsnm{Smoreda}, \binits{Z.}},
\oauthor{\bsnm{Trinquart}, \binits{R.}},
\oauthor{\bsnm{Ziemlicki}, \binits{C.}},
\oauthor{\bsnm{Blondel}, \binits{V.D.}}:
{D4D-Senegal}: The second mobile phone data for development challenge.
arXiv preprint arXiv:1407.4885
(2014)
\end{botherref}
\endbibitem

\bibitem{sweeney2002kanonymity}
\begin{barticle}
\bauthor{\bsnm{Sweeney}, \binits{L.}}:
\batitle{\textit{k}-anonymity: a model for protecting privacy}.
\bjtitle{International Journal of Uncertainty, Fuzziness and Knowledge-Based
  Systems}
\bvolume{10}(\bissue{05}),
\bfpage{557}--\blpage{570}
(\byear{2002})
\end{barticle}
\endbibitem

\bibitem{isaacman2012human}
\begin{bchapter}
\bauthor{\bsnm{Isaacman}, \binits{S.}},
\bauthor{\bsnm{Becker}, \binits{R.}},
\bauthor{\bsnm{C{\'a}ceres}, \binits{R.}},
\bauthor{\bsnm{Martonosi}, \binits{M.}},
\bauthor{\bsnm{Rowland}, \binits{J.}},
\bauthor{\bsnm{Varshavsky}, \binits{A.}},
\bauthor{\bsnm{Willinger}, \binits{W.}}:
\bctitle{Human mobility modeling at metropolitan scales}.
In: \bbtitle{Proceedings of the 10th International Conference on Mobile
  Systems, Applications, and Services},
pp. \bfpage{239}--\blpage{252}
(\byear{2012}).
\bcomment{ACM}
\end{bchapter}
\endbibitem

\bibitem{mir2013dpwhere}
\begin{bchapter}
\bauthor{\bsnm{Mir}, \binits{D.J.}},
\bauthor{\bsnm{Isaacman}, \binits{S.}},
\bauthor{\bsnm{C{\'a}ceres}, \binits{R.}},
\bauthor{\bsnm{Martonosi}, \binits{M.}},
\bauthor{\bsnm{Wright}, \binits{R.N.}}:
\bctitle{Dp-where: Differentially private modeling of human mobility}.
In: \bbtitle{Big Data, 2013 IEEE International Conference On},
pp. \bfpage{580}--\blpage{588}
(\byear{2013}).
\bcomment{IEEE}
\end{bchapter}
\endbibitem

\bibitem{madanreality}
\begin{botherref}
\oauthor{\bsnm{Madan}, \binits{A.}},
\oauthor{\bsnm{Waber}, \binits{B.N.}},
\oauthor{\bsnm{Ding}, \binits{M.}},
\oauthor{\bsnm{Kominers}, \binits{P.}},
\oauthor{\bsnm{Pentland}, \binits{A.S.}}:
Reality mining and personal privacy
(2009)
\end{botherref}
\endbibitem

\bibitem{landau2013making}
\begin{botherref}
\oauthor{\bsnm{Landau}, \binits{S.}}:
Making sense from snowden.
IEEE Security \& Privacy Magazine
(3),
5463
(2013)
\end{botherref}
\endbibitem

\bibitem{pentland2009reality}
\begin{botherref}
\oauthor{\bsnm{Pentland}, \binits{A.}}:
Reality mining of mobile communications: Toward a new deal on data.
The Global Information Technology Report 2008--2009,
1981
(2009)
\end{botherref}
\endbibitem

\bibitem{eagle2009engineering}
\begin{bchapter}
\bauthor{\bsnm{Eagle}, \binits{N.}}:
\bctitle{Engineering a common good: Fair use of aggregated, anonymized
  behavioral data}.
In: \bbtitle{First International Forum on the Application and Management of
  Personal Electronic Information}
(\byear{2009})
\end{bchapter}
\endbibitem

\bibitem{hung2006applying}
\begin{barticle}
\bauthor{\bsnm{Hung}, \binits{S.Y.}},
\bauthor{\bsnm{Yen}, \binits{D.C.}},
\bauthor{\bsnm{Wang}, \binits{H.Y.}}:
\batitle{Applying data mining to telecom churn management}.
\bjtitle{Expert Systems with Applications}
\bvolume{31}(\bissue{3}),
\bfpage{515}--\blpage{524}
(\byear{2006})
\end{barticle}
\endbibitem

\bibitem{dasgupta2008social}
\begin{bchapter}
\bauthor{\bsnm{Dasgupta}, \binits{K.}},
\bauthor{\bsnm{Singh}, \binits{R.}},
\bauthor{\bsnm{Viswanathan}, \binits{B.}},
\bauthor{\bsnm{Chakraborty}, \binits{D.}},
\bauthor{\bsnm{Mukherjea}, \binits{S.}},
\bauthor{\bsnm{Nanavati}, \binits{A.A.}},
\bauthor{\bsnm{Joshi}, \binits{A.}}:
\bctitle{Social ties and their relevance to churn in mobile telecom networks}.
In: \bbtitle{Proceedings of the 11th International Conference on Extending
  Database Technology: Advances in Database Technology},
pp. \bfpage{668}--\blpage{677}
(\byear{2008}).
\bcomment{ACM}
\end{bchapter}
\endbibitem

\bibitem{richter2010predicting}
\begin{bchapter}
\bauthor{\bsnm{Richter}, \binits{Y.}},
\bauthor{\bsnm{Yom-Tov}, \binits{E.}},
\bauthor{\bsnm{Slonim}, \binits{N.}}:
\bctitle{Predicting customer churn in mobile networks through analysis of
  social groups}.
In: \bbtitle{Proceedings of the 2010 SIAM International Conference on Data
  Mining (SDM 2010)}
(\byear{2010})
\end{bchapter}
\endbibitem

\bibitem{dierkes2011estimating}
\begin{botherref}
\oauthor{\bsnm{Dierkes}, \binits{T.}},
\oauthor{\bsnm{Bichler}, \binits{M.}},
\oauthor{\bsnm{Krishnan}, \binits{R.}}:
Estimating the effect of word of mouth on churn and cross-buying in the mobile
  phone market with markov logic networks.
Decision Support Systems
(2011)
\end{botherref}
\endbibitem

\bibitem{fitkov2001dynamic}
\begin{bchapter}
\bauthor{\bsnm{Fitkov-Norris}, \binits{E.}},
\bauthor{\bsnm{Khanifar}, \binits{A.}}:
\bctitle{Dynamic pricing in cellular networks, a mobility model with a
  provider-oriented approach}.
In: \bbtitle{3G Mobile Communication Technologies, 2001. Second International
  Conference on (Conf. Publ. No. 477)},
pp. \bfpage{63}--\blpage{67}
(\byear{2001}).
\bcomment{IET}
\end{bchapter}
\endbibitem

\bibitem{kim2010empirical}
\begin{barticle}
\bauthor{\bsnm{Kim}, \binits{Y.}},
\bauthor{\bsnm{Telang}, \binits{R.}},
\bauthor{\bsnm{Vogt}, \binits{W.B.}},
\bauthor{\bsnm{Krishnan}, \binits{R.}}:
\batitle{An empirical analysis of mobile voice service and sms: A structural
  model}.
\bjtitle{Management Science}
\bvolume{56}(\bissue{2}),
\bfpage{234}--\blpage{252}
(\byear{2010})
\end{barticle}
\endbibitem

\bibitem{barabasi2010you}
\begin{botherref}
\oauthor{\bsnm{Barab{\'a}si}, \binits{A.L.}}:
You're so predictable.
Physics World,
22--26
(2010)
\end{botherref}
\endbibitem

\bibitem{ghoshal2011ranking}
\begin{barticle}
\bauthor{\bsnm{Ghoshal}, \binits{G.}},
\bauthor{\bsnm{Barab{\'a}si}, \binits{A.L.}}:
\batitle{Ranking stability and super-stable nodes in complex networks}.
\bjtitle{Nature Communications}
\bvolume{2},
\bfpage{394}
(\byear{2011})
\end{barticle}
\endbibitem

\end{thebibliography}

\newcommand{\BMCxmlcomment}[1]{}

\BMCxmlcomment{

<refgrp>

<bibl id="B1">
  <title><p>The world in 2014 : {ICT} Facts and Figures</p></title>
  <source>International Telecommunication Union</source>
  <pubdate>2014</pubdate>
  <note>\url{http://www.itu.int/}</note>
</bibl>

<bibl id="B2">
  <title><p>Personal technology: Phoning in data</p></title>
  <aug>
    <au><snm>Kwok</snm><fnm>R.</fnm></au>
  </aug>
  <source>Nature</source>
  <pubdate>2009</pubdate>
  <volume>458</volume>
  <issue>7241</issue>
  <fpage>959</fpage>
</bibl>

<bibl id="B3">
  <title><p>{Human Behavior and the Principle of Least Effort: An Introduction
  to Human Ecology}</p></title>
  <aug>
    <au><snm>Zipf</snm><fnm>G.K.</fnm></au>
  </aug>
  <publisher>Addison-Wesley Press</publisher>
  <pubdate>1949</pubdate>
</bibl>

<bibl id="B4">
  <title><p>Communities of interest</p></title>
  <aug>
    <au><snm>Cortes</snm><fnm>C.</fnm></au>
    <au><snm>Pregibon</snm><fnm>D.</fnm></au>
    <au><snm>Volinsky</snm><fnm>C.</fnm></au>
  </aug>
  <source>Advances in Intelligent Data Analysis</source>
  <publisher>Springer</publisher>
  <pubdate>2001</pubdate>
  <fpage>105</fpage>
  <lpage>-114</lpage>
</bibl>

<bibl id="B5">
  <title><p>Extraction of information from large networks.</p></title>
  <aug>
    <au><snm>Krings</snm><fnm>G.</fnm></au>
  </aug>
  <source>PhD thesis</source>
  <publisher>Universit\'e catholique de Louvain</publisher>
  <pubdate>2012</pubdate>
  <issue>364</issue>
</bibl>

<bibl id="B6">
  <title><p>On maximum clique problems in very large graphs</p></title>
  <aug>
    <au><snm>Abello</snm><fnm>J.</fnm></au>
    <au><snm>Pardalos</snm><fnm>P.M.</fnm></au>
    <au><snm>Resende</snm><fnm>M.G.C.</fnm></au>
  </aug>
  <source>External memory algorithms</source>
  <publisher>Citeseer</publisher>
  <pubdate>1999</pubdate>
  <volume>50</volume>
  <fpage>119</fpage>
  <lpage>-130</lpage>
</bibl>

<bibl id="B7">
  <title><p>A random graph model for massive graphs</p></title>
  <aug>
    <au><snm>Aiello</snm><fnm>W.</fnm></au>
    <au><snm>Chung</snm><fnm>F.</fnm></au>
    <au><snm>Lu</snm><fnm>L.</fnm></au>
  </aug>
  <source>Proceedings of the thirty-second annual ACM symposium on Theory of
  computing</source>
  <pubdate>2000</pubdate>
  <fpage>171</fpage>
  <lpage>-180</lpage>
</bibl>

<bibl id="B8">
  <title><p>{Analysis of a large-scale weighted network of one-to-one human
  communication}</p></title>
  <aug>
    <au><snm>Onnela</snm><fnm>J.P.</fnm></au>
    <au><snm>Saramaki</snm><fnm>J.</fnm></au>
    <au><snm>Hyvonen</snm><fnm>J.</fnm></au>
    <au><snm>Szabo</snm><fnm>G.</fnm></au>
    <au><snm>Menezes</snm><fnm>M.A.</fnm></au>
    <au><snm>Kaski</snm><fnm>K.</fnm></au>
    <au><snm>Barabasi</snm><fnm>A.L.</fnm></au>
    <au><snm>Kertesz</snm><fnm>J.</fnm></au>
  </aug>
  <source>New Journal of Physics</source>
  <publisher>Institute of Physics Publishing</publisher>
  <pubdate>2007</pubdate>
  <volume>9</volume>
  <issue>6</issue>
  <fpage>179</fpage>
</bibl>

<bibl id="B9">
  <title><p>{Geographical dispersal of mobile communication
  networks}</p></title>
  <aug>
    <au><snm>Lambiotte</snm><fnm>R.</fnm></au>
    <au><snm>Blondel</snm><fnm>V.D.</fnm></au>
    <au><snm>Kerchove</snm><fnm>C.</fnm></au>
    <au><snm>Huens</snm><fnm>E.</fnm></au>
    <au><snm>Prieur</snm><fnm>C.</fnm></au>
    <au><snm>Smoreda</snm><fnm>Z.</fnm></au>
    <au><snm>Van Dooren</snm><fnm>P.</fnm></au>
  </aug>
  <source>Physica A: Statistical Mechanics and its Applications</source>
  <publisher>Elsevier</publisher>
  <pubdate>2008</pubdate>
  <volume>387</volume>
  <issue>21</issue>
  <fpage>5317</fpage>
  <lpage>-5325</lpage>
</bibl>

<bibl id="B10">
  <title><p>{Structure and tie strengths in mobile communication
  networks}</p></title>
  <aug>
    <au><snm>Onnela</snm><fnm>J.P.</fnm></au>
    <au><snm>Saramaki</snm><fnm>J.</fnm></au>
    <au><snm>Hyvonen</snm><fnm>J.</fnm></au>
    <au><snm>Szabo</snm><fnm>G.</fnm></au>
    <au><snm>Lazer</snm><fnm>D.</fnm></au>
    <au><snm>Kaski</snm><fnm>K.</fnm></au>
    <au><snm>Kertesz</snm><fnm>J.</fnm></au>
    <au><snm>Barabasi</snm><fnm>A.L.</fnm></au>
  </aug>
  <source>Proceedings of the National Academy of Sciences</source>
  <publisher>National Acad Sciences</publisher>
  <pubdate>2007</pubdate>
  <volume>104</volume>
  <issue>18</issue>
  <fpage>7332</fpage>
</bibl>

<bibl id="B11">
  <title><p>Statistically validated mobile communication networks: Evolution of
  motifs in European and Chinese data</p></title>
  <aug>
    <au><snm>Li</snm><fnm>MX</fnm></au>
    <au><snm>Palchykov</snm><fnm>V</fnm></au>
    <au><snm>Jiang</snm><fnm>ZQ</fnm></au>
    <au><snm>Kaski</snm><fnm>K</fnm></au>
    <au><snm>Kert{\'e}sz</snm><fnm>J</fnm></au>
    <au><snm>Miccich{\`e}</snm><fnm>S</fnm></au>
    <au><snm>Tumminello</snm><fnm>M</fnm></au>
    <au><snm>Zhou</snm><fnm>WX</fnm></au>
    <au><snm>Mantegna</snm><fnm>RN</fnm></au>
  </aug>
  <source>arXiv preprint arXiv:1403.3785</source>
  <pubdate>2014</pubdate>
</bibl>

<bibl id="B12">
  <title><p>Reciprocity of mobile phone calls</p></title>
  <aug>
    <au><snm>Kovanen</snm><fnm>L.</fnm></au>
    <au><snm>Saram</snm><fnm>J.</fnm></au>
    <au><snm>Kaski</snm><fnm>K.</fnm></au>
  </aug>
  <source>JDySES</source>
  <pubdate>2011</pubdate>
  <volume>2</volume>
  <issue>2</issue>
  <fpage>138</fpage>
  <lpage>-151</lpage>
</bibl>

<bibl id="B13">
  <title><p>The socio-demographics of texting: An analysis of traffic
  data</p></title>
  <aug>
    <au><snm>Ling</snm><fnm>R</fnm></au>
    <au><snm>Bertel</snm><fnm>TF</fnm></au>
    <au><snm>Sunds{\o}y</snm><fnm>PR</fnm></au>
  </aug>
  <source>New Media \& Society</source>
  <pubdate>2012</pubdate>
  <volume>14</volume>
  <issue>2</issue>
  <fpage>281</fpage>
  <lpage>298</lpage>
  <url>http://nms.sagepub.com/content/14/2/281.abstract</url>
</bibl>

<bibl id="B14">
  <title><p>On the structural properties of massive telecom call graphs:
  findings and implications</p></title>
  <aug>
    <au><snm>Nanavati</snm><fnm>A.A.</fnm></au>
    <au><snm>Gurumurthy</snm><fnm>S.</fnm></au>
    <au><snm>Das</snm><fnm>G.</fnm></au>
    <au><snm>Chakraborty</snm><fnm>D.</fnm></au>
    <au><snm>Dasgupta</snm><fnm>K.</fnm></au>
    <au><snm>Mukherjea</snm><fnm>S.</fnm></au>
    <au><snm>Joshi</snm><fnm>A.</fnm></au>
  </aug>
  <source>Proceedings of the 15th ACM international conference on Information
  and knowledge management</source>
  <pubdate>2006</pubdate>
  <fpage>435</fpage>
  <lpage>-444</lpage>
</bibl>

<bibl id="B15">
  <title><p>Scale-free networks: a decade and beyond</p></title>
  <aug>
    <au><snm>Barab{\'a}si</snm><fnm>A.L.</fnm></au>
  </aug>
  <source>Science</source>
  <publisher>American Association for the Advancement of Science</publisher>
  <pubdate>2009</pubdate>
  <volume>325</volume>
  <issue>5939</issue>
  <fpage>412</fpage>
</bibl>

<bibl id="B16">
  <title><p>{Collective dynamics of small-world networks}</p></title>
  <aug>
    <au><snm>Watts</snm><fnm>D.J.</fnm></au>
    <au><snm>Strogatz</snm><fnm>S.H.</fnm></au>
  </aug>
  <source>Nature</source>
  <pubdate>1998</pubdate>
  <volume>393</volume>
  <issue>6684</issue>
  <fpage>440</fpage>
  <lpage>-442</lpage>
</bibl>

<bibl id="B17">
  <title><p>Mobile call graphs: beyond power-law and lognormal
  distributions</p></title>
  <aug>
    <au><snm>Seshadri</snm><fnm>M.</fnm></au>
    <au><snm>Machiraju</snm><fnm>S.</fnm></au>
    <au><snm>Sridharan</snm><fnm>A.</fnm></au>
    <au><snm>Bolot</snm><fnm>J.</fnm></au>
    <au><snm>Faloutsos</snm><fnm>C.</fnm></au>
    <au><snm>Leskovec</snm><fnm>J.</fnm></au>
  </aug>
  <source>Proceeding of the 14th ACM SIGKDD international conference on
  Knowledge discovery and data mining</source>
  <pubdate>2008</pubdate>
  <fpage>596</fpage>
  <lpage>-604</lpage>
</bibl>

<bibl id="B18">
  <title><p>Effects of time window size and placement on the structure of an
  aggregated communication network</p></title>
  <aug>
    <au><snm>Krings</snm><fnm>G.</fnm></au>
    <au><snm>Karsai</snm><fnm>M.</fnm></au>
    <au><snm>Bernhardsson</snm><fnm>S.</fnm></au>
    <au><snm>Blondel</snm><fnm>V.D.</fnm></au>
    <au><snm>Saram\"aki</snm><fnm>J.</fnm></au>
  </aug>
  <source>EPJ Data Science</source>
  <pubdate>2012</pubdate>
  <volume>1</volume>
  <issue>4</issue>
  <fpage>1</fpage>
  <lpage>-16</lpage>
</bibl>

<bibl id="B19">
  <title><p>{The Strength of Weak Ties}</p></title>
  <aug>
    <au><snm>Granovetter</snm><fnm>MS</fnm></au>
  </aug>
  <source>American Journal of Sociology</source>
  <pubdate>1973</pubdate>
  <volume>78</volume>
  <fpage>1360</fpage>
  <lpage>-1380</lpage>
</bibl>

<bibl id="B20">
  <title><p>Intensity and coherence of motifs in weighted complex
  networks</p></title>
  <aug>
    <au><snm>Onnela</snm><fnm>JP</fnm></au>
    <au><snm>Saram{\"a}ki</snm><fnm>J</fnm></au>
    <au><snm>Kert{\'e}sz</snm><fnm>J</fnm></au>
    <au><snm>Kaski</snm><fnm>K</fnm></au>
  </aug>
  <source>Physical Review E</source>
  <publisher>APS</publisher>
  <pubdate>2005</pubdate>
  <volume>71</volume>
  <issue>6</issue>
  <fpage>065103</fpage>
</bibl>

<bibl id="B21">
  <title><p>Large human communication networks: patterns and a utility-driven
  generator</p></title>
  <aug>
    <au><snm>Du</snm><fnm>N.</fnm></au>
    <au><snm>Faloutsos</snm><fnm>C.</fnm></au>
    <au><snm>Wang</snm><fnm>B.</fnm></au>
    <au><snm>Akoglu</snm><fnm>L.</fnm></au>
  </aug>
  <source>Proceedings of the 15th ACM SIGKDD international conference on
  Knowledge discovery and data mining</source>
  <pubdate>2009</pubdate>
  <fpage>269</fpage>
  <lpage>-278</lpage>
</bibl>

<bibl id="B22">
  <title><p>Calling communities analysis and identification using machine
  learning techniques</p></title>
  <aug>
    <au><snm>Kianmehr</snm><fnm>K.</fnm></au>
    <au><snm>Alhajj</snm><fnm>R.</fnm></au>
  </aug>
  <source>Expert Systems with Applications</source>
  <publisher>Elsevier</publisher>
  <pubdate>2009</pubdate>
  <volume>36</volume>
  <issue>3</issue>
  <fpage>6218</fpage>
  <lpage>-6226</lpage>
</bibl>

<bibl id="B23">
  <title><p>Discovery of Social Groups Using Call Detail Records</p></title>
  <aug>
    <au><snm>Zhang</snm><fnm>H.</fnm></au>
    <au><snm>Dantu</snm><fnm>R.</fnm></au>
  </aug>
  <source>On the Move to Meaningful Internet Systems: OTM 2008
  Workshops</source>
  <pubdate>2008</pubdate>
  <fpage>489</fpage>
  <lpage>-498</lpage>
</bibl>

<bibl id="B24">
  <title><p>Communities and beyond: mesoscopic analysis of a large social
  network with complementary methods</p></title>
  <aug>
    <au><snm>Tib{\'e}ly</snm><fnm>G.</fnm></au>
    <au><snm>Kovanen</snm><fnm>L.</fnm></au>
    <au><snm>Karsai</snm><fnm>M.</fnm></au>
    <au><snm>Kaski</snm><fnm>K.</fnm></au>
    <au><snm>Kert{\'e}sz</snm><fnm>J.</fnm></au>
    <au><snm>Saram{\"a}ki</snm><fnm>J.</fnm></au>
  </aug>
  <source>Physical Review E</source>
  <publisher>APS</publisher>
  <pubdate>2011</pubdate>
  <volume>83</volume>
  <issue>5</issue>
  <fpage>056125</fpage>
</bibl>

<bibl id="B25">
  <title><p>{Fast unfolding of communities in large networks}</p></title>
  <aug>
    <au><snm>Blondel</snm><fnm>V.D.</fnm></au>
    <au><snm>Guillaume</snm><fnm>J.L.</fnm></au>
    <au><snm>Lambiotte</snm><fnm>R.</fnm></au>
    <au><snm>Mech</snm><fnm>E.L.J.S.</fnm></au>
  </aug>
  <source>J. Stat. Mech</source>
  <pubdate>2008</pubdate>
  <fpage>P10008</fpage>
</bibl>

<bibl id="B26">
  <title><p>{Maps of random walks on complex networks reveal community
  structure}</p></title>
  <aug>
    <au><snm>Rosvall</snm><fnm>M.</fnm></au>
    <au><snm>Bergstrom</snm><fnm>C.T.</fnm></au>
  </aug>
  <source>Proceedings of the National Academy of Sciences</source>
  <publisher>National Acad Sciences</publisher>
  <pubdate>2008</pubdate>
  <volume>105</volume>
  <issue>4</issue>
  <fpage>1118</fpage>
</bibl>

<bibl id="B27">
  <title><p>Quantifying social group evolution.</p></title>
  <aug>
    <au><snm>Palla</snm><fnm>G.</fnm></au>
    <au><snm>Barab{\'a}si</snm><fnm>AL</fnm></au>
    <au><snm>Vicsek</snm><fnm>T.</fnm></au>
  </aug>
  <source>Nature</source>
  <pubdate>2007</pubdate>
  <volume>446</volume>
  <issue>7136</issue>
  <fpage>664</fpage>
</bibl>

<bibl id="B28">
  <title><p>{Link communities reveal multiscale complexity in
  networks}</p></title>
  <aug>
    <au><snm>Ahn</snm><fnm>Y.Y.</fnm></au>
    <au><snm>Bagrow</snm><fnm>J.P.</fnm></au>
    <au><snm>Lehmann</snm><fnm>S.</fnm></au>
  </aug>
  <source>Nature</source>
  <publisher>Nature Publishing Group</publisher>
  <pubdate>2010</pubdate>
  <volume>466</volume>
  <issue>7307</issue>
  <fpage>761</fpage>
  <lpage>-764</lpage>
</bibl>

<bibl id="B29">
  <title><p>Computational Social Science</p></title>
  <aug>
    <au><snm>Lazer</snm><fnm>D</fnm></au>
    <au><snm>Pentland</snm><fnm>A</fnm></au>
    <au><snm>Adamic</snm><fnm>L</fnm></au>
    <au><snm>Aral</snm><fnm>S</fnm></au>
    <au><snm>Barab{\'a}si</snm><fnm>AL</fnm></au>
    <au><snm>Brewer</snm><fnm>D</fnm></au>
    <au><snm>Christakis</snm><fnm>N</fnm></au>
    <au><snm>Contractor</snm><fnm>N</fnm></au>
    <au><snm>Fowler</snm><fnm>J</fnm></au>
    <au><snm>Gutmann</snm><fnm>M</fnm></au>
    <au><snm>Jebara</snm><fnm>T</fnm></au>
    <au><snm>King</snm><fnm>G</fnm></au>
    <au><snm>Macy</snm><fnm>M</fnm></au>
    <au><snm>Roy</snm><fnm>D</fnm></au>
    <au><snm>Van Alstyne</snm><fnm>M</fnm></au>
  </aug>
  <source>Science</source>
  <pubdate>2009</pubdate>
  <volume>323</volume>
  <issue>5915</issue>
  <fpage>721</fpage>
  <lpage>723</lpage>
  <url>http://www.sciencemag.org/content/323/5915/721.short</url>
</bibl>

<bibl id="B30">
  <title><p>Inferring friendship network structure by using mobile phone
  data</p></title>
  <aug>
    <au><snm>Eagle</snm><fnm>N.</fnm></au>
    <au><snm>Pentland</snm><fnm>A.S.</fnm></au>
    <au><snm>Lazer</snm><fnm>D.</fnm></au>
  </aug>
  <source>Proceedings of the National Academy of Sciences</source>
  <publisher>National Acad Sciences</publisher>
  <pubdate>2009</pubdate>
  <volume>106</volume>
  <issue>36</issue>
  <fpage>15274</fpage>
</bibl>

<bibl id="B31">
  <title><p>``You Never Call, You Never Write'': Call and SMS Logs Do Not
  Always Indicate Tie Strength</p></title>
  <aug>
    <au><snm>Wiese</snm><fnm>J</fnm></au>
    <au><snm>Min</snm><fnm>JK</fnm></au>
    <au><snm>Hong</snm><fnm>JI</fnm></au>
    <au><snm>Zimmerman</snm><fnm>J</fnm></au>
  </aug>
  <source>Proceedings of the 2015 conference on Computer supported cooperative
  work-CSCW'15</source>
  <pubdate>2015</pubdate>
</bibl>

<bibl id="B32">
  <title><p>Who's calling? Demographics of mobile phone use in
  Rwanda</p></title>
  <aug>
    <au><snm>Blumenstock</snm><fnm>J.E.</fnm></au>
    <au><snm>Gillick</snm><fnm>D.</fnm></au>
    <au><snm>Eagle</snm><fnm>N.</fnm></au>
  </aug>
  <source>Transportation</source>
  <pubdate>2010</pubdate>
  <volume>32</volume>
  <fpage>2</fpage>
  <lpage>-5</lpage>
</bibl>

<bibl id="B33">
  <title><p>Gender-Specific Use of the Domestic Telephone</p></title>
  <aug>
    <au><snm>Smoreda</snm><fnm>Z.</fnm></au>
    <au><snm>Licoppe</snm><fnm>C.</fnm></au>
  </aug>
  <source>Social Psychology Quarterly</source>
  <pubdate>2000</pubdate>
  <volume>63</volume>
  <issue>3</issue>
  <fpage>238</fpage>
  <lpage>252</lpage>
</bibl>

<bibl id="B34">
  <title><p>Temporal motifs reveal homophily, gender-specific patterns, and
  group talk in call sequences</p></title>
  <aug>
    <au><snm>Kovanen</snm><fnm>L.</fnm></au>
    <au><snm>Kaski</snm><fnm>K.</fnm></au>
    <au><snm>Kert\'{e}sz</snm><fnm>J.</fnm></au>
    <au><snm>Saram\"aki</snm><fnm>J.</fnm></au>
  </aug>
  <source>Proceedings of the National Academy of Sciences</source>
  <pubdate>2013</pubdate>
  <volume>110</volume>
  <issue>45</issue>
  <fpage>18070</fpage>
  <lpage>18075</lpage>
</bibl>

<bibl id="B35">
  <title><p>A Gender-Centric Analysis of Calling Behavior in a Developing
  Economy Using Call Detail Records.</p></title>
  <aug>
    <au><snm>Frias Martinez</snm><fnm>V</fnm></au>
    <au><snm>Frias Martinez</snm><fnm>E</fnm></au>
    <au><snm>Oliver</snm><fnm>N</fnm></au>
  </aug>
  <source>AAAI Spring Symposium: Artificial Intelligence for
  Development</source>
  <pubdate>2010</pubdate>
</bibl>

<bibl id="B36">
  <title><p>Divided we call: Disparities in access and use of mobile phones in
  Rwanda</p></title>
  <aug>
    <au><snm>Blumenstock</snm><fnm>JE</fnm></au>
    <au><snm>Eagle</snm><fnm>N</fnm></au>
  </aug>
  <source>Information Technologies \& International Development</source>
  <pubdate>2012</pubdate>
  <volume>8</volume>
  <issue>2</issue>
  <fpage>pp</fpage>
  <lpage>-1</lpage>
</bibl>

<bibl id="B37">
  <title><p>Weighted reciprocity in human communication networks</p></title>
  <aug>
    <au><snm>Chawla</snm><fnm>N.V.</fnm></au>
    <au><snm>Hachen</snm><fnm>D.</fnm></au>
    <au><snm>Lizardo</snm><fnm>O.</fnm></au>
    <au><snm>Toroczkai</snm><fnm>Z.</fnm></au>
    <au><snm>Strathman</snm><fnm>A.</fnm></au>
    <au><snm>Wang</snm><fnm>C.</fnm></au>
  </aug>
  <pubdate>2011</pubdate>
  <issue>arXiv:1108.2822</issue>
</bibl>

<bibl id="B38">
  <title><p>The Impact of Social Affinity on Phone Calling Patterns:
  Categorizing Social Ties from Call Data Records</p></title>
  <aug>
    <au><snm>Motahari</snm><fnm>S.</fnm></au>
    <au><snm>Mengshoel</snm><fnm>O. J.</fnm></au>
    <au><snm>Reuther</snm><fnm>P.</fnm></au>
    <au><snm>Appala</snm><fnm>S.</fnm></au>
    <au><snm>Zoia</snm><fnm>L.</fnm></au>
    <au><snm>Shah</snm><fnm>J.</fnm></au>
  </aug>
  <source>The 6th SNA-KDD Workshop '12</source>
  <pubdate>2012</pubdate>
</bibl>

<bibl id="B39">
  <title><p>Spatial Networks</p></title>
  <aug>
    <au><snm>Barth{\'e}lemy</snm><fnm>M.</fnm></au>
  </aug>
  <source>Physics Reports</source>
  <pubdate>2011</pubdate>
  <volume>499</volume>
  <issue>1</issue>
  <fpage>1</fpage>
  <lpage>-101</lpage>
</bibl>

<bibl id="B40">
  <title><p>Dynamic population mapping using mobile phone data</p></title>
  <aug>
    <au><snm>Deville</snm><fnm>P.</fnm></au>
    <au><snm>Linard</snm><fnm>C.</fnm></au>
    <au><snm>Martin</snm><fnm>S.</fnm></au>
    <au><snm>Gilbert</snm><fnm>M.</fnm></au>
    <au><snm>Stevens</snm><fnm>F.R.</fnm></au>
    <au><snm>Gaughan</snm><fnm>A.E.</fnm></au>
    <au><snm>Blondel</snm><fnm>V. D.</fnm></au>
    <au><snm>Tatem</snm><fnm>A. J.</fnm></au>
  </aug>
  <source>Proceedings of the National Academy of Sciences</source>
  <pubdate>2014</pubdate>
  <volume>111</volume>
  <issue>45</issue>
  <fpage>15888</fpage>
  <lpage>15893</lpage>
</bibl>

<bibl id="B41">
  <title><p>``Calling Abidjan" - Improving Population Estimations with Mobile
  Communication Data</p></title>
  <aug>
    <au><snm>Sterly</snm><fnm>H.</fnm></au>
    <au><snm>Hennig</snm><fnm>B.</fnm></au>
    <au><snm>Dongo</snm><fnm>K.</fnm></au>
  </aug>
  <source>Mobile Phone Data for Development - Analysis of mobile phone datasets
  for the development of Ivory Coast</source>
  <publisher>Orange D4D Challenge</publisher>
  <pubdate>2013</pubdate>
  <fpage>108</fpage>
  <lpage>-114</lpage>
</bibl>

<bibl id="B42">
  <title><p>Afripop Project</p></title>
  <note>\url{http://www.worldpop.org.uk}</note>
</bibl>

<bibl id="B43">
  <title><p>Urban gravity: a model for inter-city telecommunication
  flows</p></title>
  <aug>
    <au><snm>Krings</snm><fnm>G.</fnm></au>
    <au><snm>Calabrese</snm><fnm>F.</fnm></au>
    <au><snm>Ratti</snm><fnm>C.</fnm></au>
    <au><snm>Blondel</snm><fnm>V.D.</fnm></au>
  </aug>
  <source>Journal of Statistical Mechanics: Theory and Experiment</source>
  <publisher>IOP Publishing</publisher>
  <pubdate>2009</pubdate>
  <volume>2009</volume>
  <fpage>L07003</fpage>
</bibl>

<bibl id="B44">
  <title><p>Scaling behaviors in the communication network between
  cities</p></title>
  <aug>
    <au><snm>Krings</snm><fnm>G.</fnm></au>
    <au><snm>Calabrese</snm><fnm>F.</fnm></au>
    <au><snm>Ratti</snm><fnm>C.</fnm></au>
    <au><snm>Blondel</snm><fnm>V.D.</fnm></au>
  </aug>
  <source>2009 International Conference on Computational Science and
  Engineering</source>
  <pubdate>2009</pubdate>
  <fpage>936</fpage>
  <lpage>-939</lpage>
</bibl>

<bibl id="B45">
  <title><p>Geographic constraints on social network groups</p></title>
  <aug>
    <au><snm>Onnela</snm><fnm>J.P.</fnm></au>
    <au><snm>Arbesman</snm><fnm>S.</fnm></au>
    <au><snm>Gonz{\'a}lez</snm><fnm>M.C.</fnm></au>
    <au><snm>Barab{\'a}si</snm><fnm>A.L.</fnm></au>
    <au><snm>Christakis</snm><fnm>N.A.</fnm></au>
  </aug>
  <source>PloS one</source>
  <publisher>Public Library of Science</publisher>
  <pubdate>2011</pubdate>
  <volume>6</volume>
  <issue>4</issue>
  <fpage>e16939</fpage>
</bibl>

<bibl id="B46">
  <title><p>Analyzing Social Divisions Using Cell Phone Data</p></title>
  <aug>
    <au><snm>Bucicovschi</snm><fnm>O.</fnm></au>
    <au><snm>Douglass</snm><fnm>R. W.</fnm></au>
    <au><snm>Meyer</snm><fnm>D. A.</fnm></au>
    <au><snm>Ram</snm><fnm>M.</fnm></au>
    <au><snm>Rideout</snm><fnm>D.</fnm></au>
    <au><snm>Song</snm><fnm>D.</fnm></au>
  </aug>
  <source>D4D Book: Mobile Phone Data for Development. Analysis of mobile phone
  datasets for the development of Ivory Coast</source>
  <pubdate>2013</pubdate>
</bibl>

<bibl id="B47">
  <title><p>A tale of many cities: universal patterns in human urban
  mobility</p></title>
  <aug>
    <au><snm>Noulas</snm><fnm>A</fnm></au>
    <au><snm>Scellato</snm><fnm>S</fnm></au>
    <au><snm>Lambiotte</snm><fnm>R</fnm></au>
    <au><snm>Pontil</snm><fnm>M</fnm></au>
    <au><snm>Mascolo</snm><fnm>C</fnm></au>
  </aug>
  <source>PloS one</source>
  <publisher>Public Library of Science</publisher>
  <pubdate>2012</pubdate>
  <volume>7</volume>
  <issue>5</issue>
  <fpage>e37027</fpage>
</bibl>

<bibl id="B48">
  <title><p>Geographic routing in social networks</p></title>
  <aug>
    <au><snm>Liben Nowell</snm><fnm>D.</fnm></au>
    <au><snm>Novak</snm><fnm>J.</fnm></au>
    <au><snm>Kumar</snm><fnm>R.</fnm></au>
    <au><snm>Raghavan</snm><fnm>P.</fnm></au>
    <au><snm>Tomkins</snm><fnm>A.</fnm></au>
  </aug>
  <source>Proceedings of the National Academy of Sciences of the United States
  of America</source>
  <publisher>National Acad Sciences</publisher>
  <pubdate>2005</pubdate>
  <volume>102</volume>
  <issue>33</issue>
  <fpage>11623</fpage>
</bibl>

<bibl id="B49">
  <title><p>Analysing Ireland's Interurban Communication Network using Call
  Data Records</p></title>
  <aug>
    <au><snm>Carolan</snm><fnm>E.</fnm></au>
    <au><snm>McLoone</snm><fnm>S. C.</fnm></au>
    <au><snm>McLoone</snm><fnm>S. F.</fnm></au>
    <au><snm>Farrell</snm><fnm>R.</fnm></au>
  </aug>
  <source>Signals and Systems Conference (ISSC 2012), IET Irish</source>
  <pubdate>2012</pubdate>
</bibl>

<bibl id="B50">
  <title><p>The scaling of human interactions with city size</p></title>
  <aug>
    <au><snm>Schl{\"a}pfer</snm><fnm>M.</fnm></au>
    <au><snm>Bettencourt</snm><fnm>L.</fnm></au>
    <au><snm>Grauwin</snm><fnm>S.</fnm></au>
    <au><snm>Raschke</snm><fnm>M.</fnm></au>
    <au><snm>Claxton</snm><fnm>R.</fnm></au>
    <au><snm>Smoreda</snm><fnm>Z.</fnm></au>
    <au><snm>West</snm><fnm>G.B.</fnm></au>
    <au><snm>Ratti</snm><fnm>C.</fnm></au>
  </aug>
  <source>Journal of the Royal Society Interface</source>
  <pubdate>2014</pubdate>
  <volume>11</volume>
  <fpage>20130789</fpage>
</bibl>

<bibl id="B51">
  <title><p>Spatial patterns of close relationships across the
  lifespan</p></title>
  <aug>
    <au><snm>Jo</snm><fnm>HH</fnm></au>
    <au><snm>Saram{\"a}ki</snm><fnm>J</fnm></au>
    <au><snm>Dunbar</snm><fnm>RI</fnm></au>
    <au><snm>Kaski</snm><fnm>K</fnm></au>
  </aug>
  <source>Scientific reports</source>
  <publisher>Nature Publishing Group</publisher>
  <pubdate>2014</pubdate>
  <volume>4</volume>
</bibl>

<bibl id="B52">
  <title><p>The elliptic model for communication fluxes</p></title>
  <aug>
    <au><snm>Herrera Yag\"ue</snm><fnm>C.</fnm></au>
    <au><snm>Schneider</snm><fnm>C. M.</fnm></au>
    <au><snm>Smoreda</snm><fnm>Z.</fnm></au>
    <au><snm>Couronn\'e</snm><fnm>T.</fnm></au>
    <au><snm>Zufiria</snm><fnm>P.J.</fnm></au>
    <au><snm>Gonz\'alez</snm><fnm>M.C.</fnm></au>
  </aug>
  <source>Journal of Statistical Mechanics: Theory and Experiment</source>
  <pubdate>2014</pubdate>
  <volume>2014</volume>
  <issue>4</issue>
  <fpage>P04022</fpage>
</bibl>

<bibl id="B53">
  <title><p>Modularity maximization and tree clustering: Novel ways to
  determine effective geographic borders</p></title>
  <aug>
    <au><snm>Grady</snm><fnm>D</fnm></au>
    <au><snm>Brune</snm><fnm>R</fnm></au>
    <au><snm>Thiemann</snm><fnm>C</fnm></au>
    <au><snm>Theis</snm><fnm>F</fnm></au>
    <au><snm>Brockmann</snm><fnm>D</fnm></au>
  </aug>
  <source>Handbook of Optimization in Complex Networks</source>
  <publisher>Springer</publisher>
  <pubdate>2012</pubdate>
  <fpage>169</fpage>
  <lpage>-208</lpage>
</bibl>

<bibl id="B54">
  <title><p>Voice on the border: do cellphones redraw the maps?</p></title>
  <aug>
    <au><snm>Blondel</snm><fnm>V.D.</fnm></au>
    <au><snm>Deville</snm><fnm>P.</fnm></au>
    <au><snm>Morlot</snm><fnm>F.</fnm></au>
    <au><snm>Smoreda</snm><fnm>Z.</fnm></au>
    <au><snm>Van Dooren</snm><fnm>P.</fnm></au>
    <au><snm>Ziemlicki</snm><fnm>C.</fnm></au>
  </aug>
  <source>Paris Tech Review</source>
  <pubdate>2011</pubdate>
</bibl>

<bibl id="B55">
  <title><p>Regions and borders of mobile telephony in Belgium and in the
  Brussels metropolitan zone</p></title>
  <aug>
    <au><snm>Blondel</snm><fnm>V.</fnm></au>
    <au><snm>Krings</snm><fnm>G.</fnm></au>
    <au><snm>Thomas</snm><fnm>I.</fnm></au>
  </aug>
  <source>Brussels Studies</source>
  <pubdate>2010</pubdate>
  <volume>42</volume>
  <issue>4</issue>
</bibl>

<bibl id="B56">
  <title><p>Uncovering space-independent communities in spatial
  networks</p></title>
  <aug>
    <au><snm>Expert</snm><fnm>P.</fnm></au>
    <au><snm>Evans</snm><fnm>T.S.</fnm></au>
    <au><snm>Blondel</snm><fnm>V.D.</fnm></au>
    <au><snm>Lambiotte</snm><fnm>R.</fnm></au>
  </aug>
  <source>Proceedings of the National Academy of Sciences</source>
  <publisher>National Acad Sciences</publisher>
  <pubdate>2011</pubdate>
  <volume>108</volume>
  <issue>19</issue>
  <fpage>7663</fpage>
</bibl>

<bibl id="B57">
  <title><p>Redrawing the map of Great Britain from a network of human
  interactions</p></title>
  <aug>
    <au><snm>Ratti</snm><fnm>C.</fnm></au>
    <au><snm>Sobolevsky</snm><fnm>S.</fnm></au>
    <au><snm>Calabrese</snm><fnm>F.</fnm></au>
    <au><snm>Andris</snm><fnm>C.</fnm></au>
    <au><snm>Reades</snm><fnm>J.</fnm></au>
    <au><snm>Martino</snm><fnm>M.</fnm></au>
    <au><snm>Claxton</snm><fnm>R.</fnm></au>
    <au><snm>Strogatz</snm><fnm>S.H.</fnm></au>
  </aug>
  <source>PLoS One</source>
  <pubdate>2010</pubdate>
  <volume>5</volume>
  <issue>12</issue>
  <fpage>e14248</fpage>
</bibl>

<bibl id="B58">
  <title><p>Social and Spatial Ethnic Segregation: A Framework for Analyzing
  Segregation With Large-Scale Spatial Network Data</p></title>
  <aug>
    <au><snm>Blumenstock</snm><fnm>J.E.</fnm></au>
    <au><snm>Fratamico</snm><fnm>L.</fnm></au>
  </aug>
  <source>Proceedings of the 4th Annual Symposium on Computing for
  Development</source>
  <publisher>New York, NY, USA: ACM</publisher>
  <series><title><p>ACM DEV-4 '13</p></title></series>
  <pubdate>2013</pubdate>
  <issue>11</issue>
  <fpage>11:1</fpage>
  <lpage>-11:10</lpage>
</bibl>

<bibl id="B59">
  <title><p>Network diversity and economic development</p></title>
  <aug>
    <au><snm>Eagle</snm><fnm>N.</fnm></au>
    <au><snm>Macy</snm><fnm>M.</fnm></au>
    <au><snm>Claxton</snm><fnm>R.</fnm></au>
  </aug>
  <source>Science</source>
  <publisher>American Association for the Advancement of Science</publisher>
  <pubdate>2010</pubdate>
  <volume>328</volume>
  <issue>5981</issue>
  <fpage>1029</fpage>
</bibl>

<bibl id="B60">
  <title><p>Mobile Communications Reveal the Regional Economy in C{\^o}te
  d'Ivoire</p></title>
  <aug>
    <au><snm>Mao</snm><fnm>H.</fnm></au>
    <au><snm>Shuai</snm><fnm>X.</fnm></au>
    <au><snm>Ahn</snm><fnm>Y.Y.</fnm></au>
    <au><snm>Bollen</snm><fnm>J.</fnm></au>
  </aug>
  <source>Mobile Phone Data for Development - Analysis of mobile phone datasets
  for the development of Ivory Coast</source>
  <publisher>Orange D4D Challenge</publisher>
  <pubdate>2013</pubdate>
</bibl>

<bibl id="B61">
  <title><p>Poverty on the Cheap: Estimating Poverty Maps Using Aggregated
  Mobile Communication Networks</p></title>
  <aug>
    <au><snm>Smith Clarke</snm><fnm>C.</fnm></au>
    <au><snm>Mashhadi</snm><fnm>A.</fnm></au>
    <au><snm>Capra</snm><fnm>L.</fnm></au>
  </aug>
  <source>Proceedings of the 32nd annual ACM conference on Human factors in
  computing systems</source>
  <pubdate>2014</pubdate>
  <fpage>511</fpage>
  <lpage>-520</lpage>
</bibl>

<bibl id="B62">
  <title><p>Socio-Economic Levels and Human Mobility</p></title>
  <aug>
    <au><snm>Frias Martinez</snm><fnm>V.</fnm></au>
    <au><snm>Virseda</snm><fnm>J.</fnm></au>
    <au><snm>Frias Martinez</snm><fnm>E.</fnm></au>
  </aug>
  <source>Qual Meets Quant Workshop-QMQ</source>
  <pubdate>2010</pubdate>
</bibl>

<bibl id="B63">
  <title><p>Forecasting socioeconomic trends with cell phone
  records</p></title>
  <aug>
    <au><snm>Frias Martinez</snm><fnm>V.</fnm></au>
    <au><snm>Soguero Ruiz</snm><fnm>C.</fnm></au>
    <au><snm>Frias Martinez</snm><fnm>E.</fnm></au>
    <au><snm>Josephidou</snm><fnm>M.</fnm></au>
  </aug>
  <source>Proceedings of the 3rd ACM Symposium on Computing for
  Development</source>
  <pubdate>2013</pubdate>
  <fpage>15</fpage>
</bibl>

<bibl id="B64">
  <title><p>Evaluating socio-economic state of a country analyzing airtime
  credit and mobile phone datasets</p></title>
  <aug>
    <au><snm>Gutierrez</snm><fnm>T.</fnm></au>
    <au><snm>Krings</snm><fnm>G.</fnm></au>
    <au><snm>Blondel</snm><fnm>V.D.</fnm></au>
  </aug>
  <source>arXiv preprint arXiv:1309.4496</source>
  <pubdate>2013</pubdate>
</bibl>

<bibl id="B65">
  <title><p>Temporal Networks</p></title>
  <aug>
    <au><snm>Holme</snm><fnm>P.</fnm></au>
    <au><snm>Saram\"aki</snm><fnm>J.</fnm></au>
  </aug>
  <source>Physics reports</source>
  <pubdate>2012</pubdate>
  <volume>519</volume>
  <issue>3</issue>
  <fpage>97</fpage>
  <lpage>-125</lpage>
</bibl>

<bibl id="B66">
  <title><p>The dynamics of a mobile phone network</p></title>
  <aug>
    <au><snm>Hidalgo</snm><fnm>C.A.</fnm></au>
    <au><snm>Rodriguez Sickert</snm><fnm>C.</fnm></au>
  </aug>
  <source>Physica A: Statistical Mechanics and its Applications</source>
  <publisher>Elsevier</publisher>
  <pubdate>2008</pubdate>
  <volume>387</volume>
  <issue>12</issue>
  <fpage>3017</fpage>
  <lpage>-3024</lpage>
</bibl>

<bibl id="B67">
  <title><p>Predictors of short-term decay of cell phone contacts in a large
  scale communication network</p></title>
  <aug>
    <au><snm>Raeder</snm><fnm>T</fnm></au>
    <au><snm>Lizardo</snm><fnm>O</fnm></au>
    <au><snm>Hachen</snm><fnm>D</fnm></au>
    <au><snm>Chawla</snm><fnm>NV</fnm></au>
  </aug>
  <source>Social Networks</source>
  <publisher>Elsevier</publisher>
  <pubdate>2011</pubdate>
  <volume>33</volume>
  <issue>4</issue>
  <fpage>245</fpage>
  <lpage>-257</lpage>
</bibl>

<bibl id="B68">
  <title><p>Time varying networks and the weakness of strong ties</p></title>
  <aug>
    <au><snm>Karsai</snm><fnm>M</fnm></au>
    <au><snm>Perra</snm><fnm>N</fnm></au>
    <au><snm>Vespignani</snm><fnm>A</fnm></au>
  </aug>
  <source>Scientific reports</source>
  <publisher>Macmillan Publishers Limited. All rights reserved</publisher>
  <pubdate>2014</pubdate>
  <volume>4</volume>
  <url>http://dx.doi.org/10.1038/srep04001</url>
</bibl>

<bibl id="B69">
  <title><p>Limited communication capacity unveils strategies for human
  interaction</p></title>
  <aug>
    <au><snm>Miritello</snm><fnm>G.</fnm></au>
    <au><snm>Rub{\'e}n</snm><fnm>L.</fnm></au>
    <au><snm>Cebrian</snm><fnm>M.</fnm></au>
    <au><snm>Moro</snm><fnm>E.</fnm></au>
  </aug>
  <source>Scientific Reports</source>
  <pubdate>2013</pubdate>
  <volume>3</volume>
</bibl>

<bibl id="B70">
  <title><p>Time as a limited resource: Communication strategy in mobile phone
  networks</p></title>
  <aug>
    <au><snm>Miritello</snm><fnm>G.</fnm></au>
    <au><snm>Moro</snm><fnm>E.</fnm></au>
    <au><snm>Lara</snm><fnm>R.</fnm></au>
    <au><snm>Mart\'inez L\'opez</snm><fnm>R.</fnm></au>
    <au><snm>Belchamber</snm><fnm>J.</fnm></au>
    <au><snm>Roberts</snm><fnm>S.G.B.</fnm></au>
    <au><snm>Dunbar</snm><fnm>R.I.M.</fnm></au>
  </aug>
  <source>Social Networks</source>
  <pubdate>2013</pubdate>
  <volume>35</volume>
  <issue>1</issue>
  <fpage>89</fpage>
  <lpage>-95</lpage>
</bibl>

<bibl id="B71">
  <title><p>The persistence of social signatures in human
  communication</p></title>
  <aug>
    <au><snm>Saram\"aki</snm><fnm>J.</fnm></au>
    <au><snm>Leicht</snm><fnm>E.A.</fnm></au>
    <au><snm>L{\'o}pez</snm><fnm>E.</fnm></au>
    <au><snm>Roberts</snm><fnm>S.G.B.</fnm></au>
    <au><snm>Reed Tsochas</snm><fnm>F.</fnm></au>
    <au><snm>Dunbar</snm><fnm>R.I.M.</fnm></au>
  </aug>
  <source>Proceedings of the National Academy of Sciences</source>
  <pubdate>2014</pubdate>
  <volume>111</volume>
  <issue>3</issue>
  <fpage>942</fpage>
  <lpage>-947</lpage>
</bibl>

<bibl id="B72">
  <title><p>Temporal motifs in time-dependent networks</p></title>
  <aug>
    <au><snm>Kovanen</snm><fnm>L</fnm></au>
    <au><snm>Karsai</snm><fnm>M</fnm></au>
    <au><snm>Kaski</snm><fnm>K</fnm></au>
    <au><snm>Kert{\'e}sz</snm><fnm>J</fnm></au>
    <au><snm>Saram{\"a}ki</snm><fnm>J</fnm></au>
  </aug>
  <source>Journal of Statistical Mechanics: Theory and Experiment</source>
  <publisher>IOP Publishing</publisher>
  <pubdate>2011</pubdate>
  <volume>2011</volume>
  <issue>11</issue>
  <fpage>P11005</fpage>
</bibl>

<bibl id="B73">
  <title><p>Disentangling Social Networks inferred from Call Logs</p></title>
  <aug>
    <au><snm>Cebrian</snm><fnm>M.</fnm></au>
    <au><snm>Pentland</snm><fnm>A.</fnm></au>
    <au><snm>Kirkpatrick</snm><fnm>S.</fnm></au>
  </aug>
  <source>Arxiv preprint arXiv:1008.1357</source>
  <pubdate>2010</pubdate>
</bibl>

<bibl id="B74">
  <title><p>The origin of bursts and heavy tails in human activity</p></title>
  <aug>
    <au><snm>Barab{\'a}si</snm><fnm>AL</fnm></au>
  </aug>
  <source>Nature</source>
  <pubdate>2005</pubdate>
  <volume>435</volume>
  <fpage>207</fpage>
</bibl>

<bibl id="B75">
  <title><p>Small but slow world: How network topology and burstiness slow down
  spreading</p></title>
  <aug>
    <au><snm>Karsai</snm><fnm>M.</fnm></au>
    <au><snm>Kivel{\"a}</snm><fnm>M.</fnm></au>
    <au><snm>Pan</snm><fnm>RK</fnm></au>
    <au><snm>Kaski</snm><fnm>K.</fnm></au>
    <au><snm>Kert{\'e}sz</snm><fnm>J.</fnm></au>
    <au><snm>Barab{\'a}si</snm><fnm>A.L.</fnm></au>
    <au><snm>Saram{\"a}ki</snm><fnm>J.</fnm></au>
  </aug>
  <source>Physical Review E</source>
  <publisher>APS</publisher>
  <pubdate>2011</pubdate>
  <volume>83</volume>
  <issue>2</issue>
  <fpage>025102</fpage>
</bibl>

<bibl id="B76">
  <title><p>Universal features of correlated bursty behaviour</p></title>
  <aug>
    <au><snm>Karsai</snm><fnm>M.</fnm></au>
    <au><snm>Kaski</snm><fnm>K.</fnm></au>
    <au><snm>Barab\'a{}si</snm><fnm>A.L.</fnm></au>
    <au><snm>Kert\'{e}sz</snm><fnm>J.</fnm></au>
  </aug>
  <source>Scientific Reports</source>
  <pubdate>2012</pubdate>
  <volume>2</volume>
</bibl>

<bibl id="B77">
  <title><p>Evidence for a bimodal distribution in human
  communication</p></title>
  <aug>
    <au><snm>Wu</snm><fnm>Y</fnm></au>
    <au><snm>Zhou</snm><fnm>C</fnm></au>
    <au><snm>Xiao</snm><fnm>J</fnm></au>
    <au><snm>Kurths</snm><fnm>J</fnm></au>
    <au><snm>Schellnhuber</snm><fnm>HJ</fnm></au>
  </aug>
  <source>Proceedings of the National Academy of Sciences</source>
  <publisher>National Acad Sciences</publisher>
  <pubdate>2010</pubdate>
  <volume>107</volume>
  <issue>44</issue>
  <fpage>18803</fpage>
  <lpage>-18808</lpage>
</bibl>

<bibl id="B78">
  <title><p>Uncovering individual and collective human dynamics from mobile
  phone records</p></title>
  <aug>
    <au><snm>Candia</snm><fnm>J.</fnm></au>
    <au><snm>Gonz{\'a}lez</snm><fnm>M.C.</fnm></au>
    <au><snm>Wang</snm><fnm>P.</fnm></au>
    <au><snm>Schoenharl</snm><fnm>T.</fnm></au>
    <au><snm>Madey</snm><fnm>G.</fnm></au>
    <au><snm>Barab{\'a}si</snm><fnm>A.L.</fnm></au>
  </aug>
  <source>Journal of Physics A: Mathematical and Theoretical</source>
  <publisher>IOP Publishing</publisher>
  <pubdate>2008</pubdate>
  <volume>41</volume>
  <fpage>224015</fpage>
</bibl>

<bibl id="B79">
  <title><p>Circadian pattern and burstiness in mobile phone
  communication</p></title>
  <aug>
    <au><snm>Jo</snm><fnm>H. H.</fnm></au>
    <au><snm>Karsai</snm><fnm>M.</fnm></au>
    <au><snm>Kert\'{e}sz</snm><fnm>J.</fnm></au>
    <au><snm>Kaski</snm><fnm>K.</fnm></au>
  </aug>
  <source>New Journal of Physics</source>
  <pubdate>2012</pubdate>
  <volume>14</volume>
  <issue>1</issue>
  <fpage>013055</fpage>
</bibl>

<bibl id="B80">
  <title><p>{Understanding individual human mobility patterns}</p></title>
  <aug>
    <au><snm>Gonz{\'a}lez</snm><fnm>M.C.</fnm></au>
    <au><snm>Hidalgo</snm><fnm>C.A.</fnm></au>
    <au><snm>Barab{\'a}si</snm><fnm>A.L.</fnm></au>
  </aug>
  <source>Nature</source>
  <publisher>Nature Publishing Group</publisher>
  <pubdate>2008</pubdate>
  <volume>453</volume>
  <issue>7196</issue>
  <fpage>779</fpage>
  <lpage>-782</lpage>
</bibl>

<bibl id="B81">
  <title><p>Modelling the scaling properties of human mobility</p></title>
  <aug>
    <au><snm>Song</snm><fnm>C.</fnm></au>
    <au><snm>Koren</snm><fnm>T.</fnm></au>
    <au><snm>Wang</snm><fnm>P.</fnm></au>
    <au><snm>Barab{\'a}si</snm><fnm>A.L.</fnm></au>
  </aug>
  <source>Nature Physics</source>
  <publisher>Nature Publishing Group</publisher>
  <pubdate>2010</pubdate>
</bibl>

<bibl id="B82">
  <title><p>Exploring the mobility of mobile phone users</p></title>
  <aug>
    <au><snm>Cs{\'a}ji</snm><fnm>B.</fnm></au>
    <au><snm>Browet</snm><fnm>A.</fnm></au>
    <au><snm>Traag</snm><fnm>V.A.</fnm></au>
    <au><snm>Delvenne</snm><fnm>J. C.</fnm></au>
    <au><snm>Huens</snm><fnm>E.</fnm></au>
    <au><snm>Van Dooren</snm><fnm>P.</fnm></au>
    <au><snm>Smoreda</snm><fnm>Z.</fnm></au>
    <au><snm>Blondel</snm><fnm>V.D.</fnm></au>
  </aug>
  <source>Physica A: Statistical Mechanics and its Applications</source>
  <pubdate>2013</pubdate>
  <volume>392</volume>
  <issue>6</issue>
  <fpage>1459</fpage>
  <lpage>-1473</lpage>
</bibl>

<bibl id="B83">
  <title><p>Mesoscopic structure and social aspects of human
  mobility</p></title>
  <aug>
    <au><snm>Bagrow</snm><fnm>JP</fnm></au>
    <au><snm>Lin</snm><fnm>YR</fnm></au>
  </aug>
  <source>PloS one</source>
  <publisher>Public Library of Science</publisher>
  <pubdate>2012</pubdate>
  <volume>7</volume>
  <issue>5</issue>
  <fpage>e37676</fpage>
</bibl>

<bibl id="B84">
  <title><p>The Differing Tribal and Infrastructural Influences on Mobility in
  Developing and Industrialized Regions</p></title>
  <aug>
    <au><snm>Amini</snm><fnm>A.</fnm></au>
    <au><snm>Kung</snm><fnm>K.</fnm></au>
    <au><snm>Kang</snm><fnm>C.</fnm></au>
    <au><snm>Sobolevsky</snm><fnm>S.</fnm></au>
    <au><snm>Ratti</snm><fnm>C.</fnm></au>
  </aug>
  <source>Mobile Phone Data for Development - Analysis of mobile phone datasets
  for the development of Ivory Coast</source>
  <publisher>Orange D4D Challenge</publisher>
  <pubdate>2013</pubdate>
  <fpage>330</fpage>
  <lpage>-339</lpage>
</bibl>

<bibl id="B85">
  <title><p>Limits of predictability in human mobility</p></title>
  <aug>
    <au><snm>Song</snm><fnm>C.</fnm></au>
    <au><snm>Qu</snm><fnm>Z.</fnm></au>
    <au><snm>Blumm</snm><fnm>N.</fnm></au>
    <au><snm>Barab{\'a}si</snm><fnm>A.L.</fnm></au>
  </aug>
  <source>Science</source>
  <publisher>American Association for the Advancement of Science</publisher>
  <pubdate>2010</pubdate>
  <volume>327</volume>
  <issue>5968</issue>
  <fpage>1018</fpage>
</bibl>

<bibl id="B86">
  <title><p>Human mobility prediction based on individual and collective
  geographical preferences</p></title>
  <aug>
    <au><snm>Calabrese</snm><fnm>F.</fnm></au>
    <au><snm>Di Lorenzo</snm><fnm>G.</fnm></au>
    <au><snm>Ratti</snm><fnm>C.</fnm></au>
  </aug>
  <source>Intelligent Transportation Systems (ITSC), 2010 13th International
  IEEE Conference on</source>
  <pubdate>2010</pubdate>
  <fpage>312</fpage>
  <lpage>-317</lpage>
</bibl>

<bibl id="B87">
  <title><p>The eigenmode analysis of human motion</p></title>
  <aug>
    <au><snm>Park</snm><fnm>J.</fnm></au>
    <au><snm>Lee</snm><fnm>D.S.</fnm></au>
    <au><snm>Gonz{\'a}lez</snm><fnm>M.C.</fnm></au>
  </aug>
  <source>Journal of Statistical Mechanics: Theory and Experiment</source>
  <publisher>IOP Publishing</publisher>
  <pubdate>2010</pubdate>
  <volume>2010</volume>
  <fpage>P11021</fpage>
</bibl>

<bibl id="B88">
  <title><p>A universal model for mobility and migration patterns</p></title>
  <aug>
    <au><snm>Simini</snm><fnm>F</fnm></au>
    <au><snm>Gonzalez</snm><fnm>MC</fnm></au>
    <au><snm>Maritan</snm><fnm>A</fnm></au>
    <au><snm>Barabasi</snm><fnm>AL</fnm></au>
  </aug>
  <source>Nature</source>
  <publisher>Nature Publishing Group, a division of Macmillan Publishers
  Limited. All Rights Reserved.</publisher>
  <pubdate>2012</pubdate>
  <volume>484</volume>
  <issue>7392</issue>
  <fpage>96</fpage>
  <lpage>-100</lpage>
  <url>http://dx.doi.org/10.1038/nature10856</url>
</bibl>

<bibl id="B89">
  <title><p>Inferring human mobility using communication patterns</p></title>
  <aug>
    <au><snm>Palchykov</snm><fnm>V</fnm></au>
    <au><snm>Mitrovic</snm><fnm>M</fnm></au>
    <au><snm>Jo</snm><fnm>HH</fnm></au>
    <au><snm>Saramaki</snm><fnm>J</fnm></au>
    <au><snm>Pan</snm><fnm>RK</fnm></au>
  </aug>
  <source>Scientific reports</source>
  <publisher>Macmillan Publishers Limited. All rights reserved</publisher>
  <pubdate>2014</pubdate>
  <volume>4</volume>
  <url>http://dx.doi.org/10.1038/srep06174</url>
</bibl>

<bibl id="B90">
  <title><p>Ocean of information: fusing aggregate &amp individual dynamics for
  metropolitan analysis</p></title>
  <aug>
    <au><snm>Martino</snm><fnm>M.</fnm></au>
    <au><snm>Calabrese</snm><fnm>F.</fnm></au>
    <au><snm>Di Lorenzo</snm><fnm>G.</fnm></au>
    <au><snm>Andris</snm><fnm>C.</fnm></au>
    <au><snm>Liang</snm><fnm>L.</fnm></au>
    <au><snm>Ratti</snm><fnm>C.</fnm></au>
  </aug>
  <source>Proceedings of the 15th international conference on Intelligent user
  interfaces</source>
  <pubdate>2010</pubdate>
  <fpage>357</fpage>
  <lpage>-360</lpage>
</bibl>

<bibl id="B91">
  <title><p>Eigenbehaviors: Identifying structure in routine</p></title>
  <aug>
    <au><snm>Eagle</snm><fnm>N.</fnm></au>
    <au><snm>Pentland</snm><fnm>A.S.</fnm></au>
  </aug>
  <source>Behavioral Ecology and Sociobiology</source>
  <publisher>Springer</publisher>
  <pubdate>2009</pubdate>
  <volume>63</volume>
  <issue>7</issue>
  <fpage>1057</fpage>
  <lpage>-1066</lpage>
</bibl>

<bibl id="B92">
  <title><p>Mobile Landscapes: using location data from cell phones for urban
  analysis</p></title>
  <aug>
    <au><snm>Ratti</snm><fnm>C.</fnm></au>
    <au><snm>Williams</snm><fnm>S.</fnm></au>
    <au><snm>Frenchman</snm><fnm>D.</fnm></au>
    <au><snm>Pulselli</snm><fnm>RM</fnm></au>
  </aug>
  <source>Environment and Planning B: Planning and Design</source>
  <publisher>PION LTD</publisher>
  <pubdate>2006</pubdate>
  <volume>33</volume>
  <issue>5</issue>
  <fpage>727</fpage>
</bibl>

<bibl id="B93">
  <title><p>Real-time urban monitoring using cell phones: A case study in
  rome</p></title>
  <aug>
    <au><snm>Calabrese</snm><fnm>F.</fnm></au>
    <au><snm>Colonna</snm><fnm>M.</fnm></au>
    <au><snm>Lovisolo</snm><fnm>P.</fnm></au>
    <au><snm>Parata</snm><fnm>D.</fnm></au>
    <au><snm>Ratti</snm><fnm>C.</fnm></au>
  </aug>
  <source>Intelligent Transportation Systems, IEEE Transactions on</source>
  <publisher>IEEE</publisher>
  <pubdate>2011</pubdate>
  <volume>12</volume>
  <issue>1</issue>
  <fpage>141</fpage>
  <lpage>-151</lpage>
</bibl>

<bibl id="B94">
  <title><p>Cellular census: Explorations in urban data collection</p></title>
  <aug>
    <au><snm>Reades</snm><fnm>J.</fnm></au>
    <au><snm>Calabrese</snm><fnm>F.</fnm></au>
    <au><snm>Sevtsuk</snm><fnm>A.</fnm></au>
    <au><snm>Ratti</snm><fnm>C.</fnm></au>
  </aug>
  <source>IEEE Pervasive Computing</source>
  <publisher>IEEE Computer Society</publisher>
  <pubdate>2007</pubdate>
  <fpage>30</fpage>
  <lpage>-38</lpage>
</bibl>

<bibl id="B95">
  <title><p>Eigenplaces: analysing cities using the space- time structure of
  the mobile phone network</p></title>
  <aug>
    <au><snm>Reades</snm><fnm>J.</fnm></au>
    <au><snm>Calabrese</snm><fnm>F.</fnm></au>
    <au><snm>Ratti</snm><fnm>C.</fnm></au>
  </aug>
  <source>Environment and Planning B: Planning and Design</source>
  <pubdate>2009</pubdate>
  <volume>36</volume>
  <issue>5</issue>
  <fpage>824</fpage>
  <lpage>-836</lpage>
</bibl>

<bibl id="B96">
  <title><p>A tale of two cities</p></title>
  <aug>
    <au><snm>Isaacman</snm><fnm>S.</fnm></au>
    <au><snm>Becker</snm><fnm>R.</fnm></au>
    <au><snm>C{\'a}ceres</snm><fnm>R.</fnm></au>
    <au><snm>Kobourov</snm><fnm>S.</fnm></au>
    <au><snm>Rowland</snm><fnm>J.</fnm></au>
    <au><snm>Varshavsky</snm><fnm>A.</fnm></au>
  </aug>
  <source>Proceedings of the Eleventh Workshop on Mobile Computing Systems \&
  Applications</source>
  <pubdate>2010</pubdate>
  <fpage>19</fpage>
  <lpage>-24</lpage>
</bibl>

<bibl id="B97">
  <title><p>From mobile phone data to the spatial structure of
  cities</p></title>
  <aug>
    <au><snm>Louail</snm><fnm>T.</fnm></au>
    <au><snm>Lenormand</snm><fnm>M.</fnm></au>
    <au><snm>Cant{\'u}</snm><fnm>O. G.</fnm></au>
    <au><snm>Picornell</snm><fnm>M.</fnm></au>
    <au><snm>Herranz</snm><fnm>R.</fnm></au>
    <au><snm>Frias Martinez</snm><fnm>E.</fnm></au>
    <au><snm>Ramasco</snm><fnm>J. J.</fnm></au>
    <au><snm>Barthelemy</snm><fnm>M.</fnm></au>
  </aug>
  <source>arXiv preprint arXiv:1401.4540</source>
  <pubdate>2014</pubdate>
</bibl>

<bibl id="B98">
  <title><p>Discovering urban and country dynamics from mobile phone data with
  spatial correlation patterns</p></title>
  <aug>
    <au><snm>Trasarti</snm><fnm>R.</fnm></au>
    <au><snm>Olteanu Raimond</snm><fnm>A. M.</fnm></au>
    <au><snm>Nanni</snm><fnm>M.</fnm></au>
    <au><snm>Couronn{\'e}</snm><fnm>T.</fnm></au>
    <au><snm>Furletti</snm><fnm>B.</fnm></au>
    <au><snm>Giannotti</snm><fnm>F.</fnm></au>
    <au><snm>Smoreda</snm><fnm>Z.</fnm></au>
    <au><snm>Ziemlicki</snm><fnm>C.</fnm></au>
  </aug>
  <source>Telecommunications Policy</source>
  <pubdate>2014</pubdate>
</bibl>

<bibl id="B99">
  <title><p>A review of urban computing for mobile phone traces: current
  methods, challenges and opportunities</p></title>
  <aug>
    <au><snm>Jiang</snm><fnm>S.</fnm></au>
    <au><snm>Fiore</snm><fnm>G. A.</fnm></au>
    <au><snm>Yang</snm><fnm>Y.</fnm></au>
    <au><snm>Ferreira</snm><fnm>J.</fnm></au>
    <au><snm>Frazzoli</snm><fnm>E.</fnm></au>
    <au><snm>Gonz{\'a}lez</snm><fnm>M.C.</fnm></au>
  </aug>
  <source>Proceedings of the 2nd ACM SIGKDD International Workshop on Urban
  Computing</source>
  <pubdate>2013</pubdate>
  <fpage>2</fpage>
</bibl>

<bibl id="B100">
  <title><p>AllAboard: A System for Exploring Urban Mobility and Optimizing
  Public Transport Using Cellphone Data</p></title>
  <aug>
    <au><snm>Berlingerio</snm><fnm>M</fnm></au>
    <au><snm>Calabrese</snm><fnm>F</fnm></au>
    <au><snm>Di Lorenzo</snm><fnm>G</fnm></au>
    <au><snm>Nair</snm><fnm>R</fnm></au>
    <au><snm>Pinelli</snm><fnm>F</fnm></au>
    <au><snm>Sbodio</snm><fnm>M</fnm></au>
  </aug>
  <source>Machine Learning and Knowledge Discovery in Databases</source>
  <publisher>Springer Berlin Heidelberg</publisher>
  <editor>Blockeel, Hendrik and Kersting, Kristian and Nijssen, Siegfried and
  {\v Z}elezn{\'y}, Filip</editor>
  <series><title><p>Lecture Notes in Computer Science</p></title></series>
  <pubdate>2013</pubdate>
  <volume>8190</volume>
  <fpage>663</fpage>
  <lpage>666</lpage>
</bibl>

<bibl id="B101">
  <title><p>MP4-A Project: Mobility Planning For Africa</p></title>
  <aug>
    <au><snm>Nanni</snm><fnm>M.</fnm></au>
    <au><snm>Trasarti</snm><fnm>R.</fnm></au>
    <au><snm>Furletti</snm><fnm>B.</fnm></au>
    <au><snm>Gabrielli</snm><fnm>L.</fnm></au>
    <au><snm>Van Der Mede</snm><fnm>P.</fnm></au>
    <au><snm>De Bruijn</snm><fnm>J.</fnm></au>
    <au><snm>De Romph</snm><fnm>E.</fnm></au>
    <au><snm>Bruil</snm><fnm>G.</fnm></au>
  </aug>
  <source>Mobile Phone Data for Development - Analysis of mobile phone datasets
  for the development of Ivory Coast</source>
  <publisher>Orange D4D Challenge</publisher>
  <pubdate>2013</pubdate>
  <fpage>423</fpage>
  <lpage>-446</lpage>
</bibl>

<bibl id="B102">
  <title><p>Mobility Modeling for Transport Efficiency - Analysis of Travel
  Characteristics Based on Mobile Phone Data</p></title>
  <aug>
    <au><snm>Angelakis</snm><fnm>V.</fnm></au>
    <au><snm>Gundleg{\aa}rd</snm><fnm>D.</fnm></au>
    <au><snm>Rajna</snm><fnm>B.</fnm></au>
    <au><snm>Rydergren</snm><fnm>C.</fnm></au>
    <au><snm>Vrotsou</snm><fnm>K.</fnm></au>
    <au><snm>Carlsson</snm><fnm>R.</fnm></au>
    <au><snm>Forgeat</snm><fnm>J.</fnm></au>
    <au><snm>Hu</snm><fnm>T.H.</fnm></au>
    <au><snm>Liu</snm><fnm>E. L.</fnm></au>
    <au><snm>Moritz</snm><fnm>S.</fnm></au>
    <au><snm>Zhao</snm><fnm>S.</fnm></au>
    <au><snm>Zheng</snm><fnm>Y.</fnm></au>
  </aug>
  <source>Mobile Phone Data for Development - Analysis of mobile phone datasets
  for the development of Ivory Coast</source>
  <publisher>Orange D4D Challenge</publisher>
  <pubdate>2013</pubdate>
  <fpage>412</fpage>
  <lpage>-422</lpage>
</bibl>

<bibl id="B103">
  <title><p>Collective response of human populations to large-scale
  emergencies</p></title>
  <aug>
    <au><snm>Bagrow</snm><fnm>J.P.</fnm></au>
    <au><snm>Wang</snm><fnm>D.</fnm></au>
    <au><snm>Barab{\'a}si</snm><fnm>A.L.</fnm></au>
  </aug>
  <source>PloS one</source>
  <publisher>Public Library of Science</publisher>
  <pubdate>2011</pubdate>
  <volume>6</volume>
  <issue>3</issue>
  <fpage>e17680</fpage>
</bibl>

<bibl id="B104">
  <title><p>Quantifying information flow during emergencies</p></title>
  <aug>
    <au><snm>Gao</snm><fnm>L.</fnm></au>
    <au><snm>Song</snm><fnm>C.</fnm></au>
    <au><snm>Gao</snm><fnm>Z.</fnm></au>
    <au><snm>Barab\'a{}si</snm><fnm>A.L.</fnm></au>
    <au><snm>Bagrow</snm><fnm>J.P.</fnm></au>
    <au><snm>Wang</snm><fnm>D.</fnm></au>
  </aug>
  <source>Scientific Reports</source>
  <pubdate>2014</pubdate>
  <volume>4</volume>
</bibl>

<bibl id="B105">
  <title><p>Understanding Human Mobility Due to Large-Scale Events</p></title>
  <aug>
    <au><snm>Xavier</snm><fnm>F. H. Z.</fnm></au>
    <au><snm>Silveira</snm><fnm>L. M.</fnm></au>
    <au><snm>Almeida</snm><fnm>J. M.</fnm></au>
    <au><snm>Malab</snm><fnm>C. H. S.</fnm></au>
    <au><snm>Ziviani</snm><fnm>A.</fnm></au>
    <au><snm>Marques Neto</snm><fnm>H. T.</fnm></au>
  </aug>
  <source>NetMob 2013 - Third International Conference on the Analysis of
  Mobile Phone Datasets</source>
  <pubdate>2013</pubdate>
</bibl>

<bibl id="B106">
  <title><p>The Social Amplifier - Reaction of Human Communities to
  Emergencies</p></title>
  <aug>
    <au><snm>Altshuler</snm><fnm>Y.</fnm></au>
    <au><snm>Fire</snm><fnm>M.</fnm></au>
    <au><snm>Shmueli</snm><fnm>E.</fnm></au>
    <au><snm>Elovici</snm><fnm>Y.</fnm></au>
    <au><snm>Bruckstein</snm><fnm>A.</fnm></au>
    <au><snm>Pentland</snm><fnm>A.S.</fnm></au>
    <au><snm>Lazer</snm><fnm>D.</fnm></au>
  </aug>
  <source>Journal of Statistical Physics</source>
  <pubdate>2013</pubdate>
  <volume>152</volume>
  <issue>3</issue>
  <fpage>399</fpage>
  <lpage>-418</lpage>
</bibl>

<bibl id="B107">
  <title><p>Predictability of population displacement after the 2010 Haiti
  earthquake</p></title>
  <aug>
    <au><snm>Lu</snm><fnm>X.</fnm></au>
    <au><snm>Bengtsson</snm><fnm>L.</fnm></au>
    <au><snm>Holme</snm><fnm>P.</fnm></au>
  </aug>
  <source>Proceedings of the National Academy of Sciences</source>
  <pubdate>2012</pubdate>
  <volume>109</volume>
  <issue>29</issue>
  <fpage>11576</fpage>
  <lpage>-11581</lpage>
</bibl>

<bibl id="B108">
  <title><p>Interplay between Telecommunications and Face-to-Face Interactions:
  A Study Using Mobile Phone Data</p></title>
  <aug>
    <au><snm>Calabrese</snm><fnm>F.</fnm></au>
    <au><snm>Smoreda</snm><fnm>Z.</fnm></au>
    <au><snm>Blondel</snm><fnm>V.D.</fnm></au>
    <au><snm>Ratti</snm><fnm>C.</fnm></au>
  </aug>
  <source>PloS one</source>
  <publisher>Public Library of Science</publisher>
  <pubdate>2011</pubdate>
  <volume>6</volume>
  <issue>7</issue>
  <fpage>e20814</fpage>
</bibl>

<bibl id="B109">
  <title><p>Human Mobility, Social Ties, and Link Prediction</p></title>
  <aug>
    <au><snm>Wang</snm><fnm>D.</fnm></au>
    <au><snm>Pedreschi</snm><fnm>D.</fnm></au>
    <au><snm>Song</snm><fnm>C.</fnm></au>
    <au><snm>Giannotti</snm><fnm>F.</fnm></au>
    <au><snm>Barab{\'a}si</snm><fnm>A.L.</fnm></au>
  </aug>
  <source>17th ACM SIGKDD Conference on Knowledge Discovery and Data Mining
  (KDD'11)</source>
  <pubdate>2011</pubdate>
</bibl>

<bibl id="B110">
  <title><p>Community computing: Comparisons between rural and urban societies
  using mobile phone data</p></title>
  <aug>
    <au><snm>Eagle</snm><fnm>N.</fnm></au>
    <au><snm>Montjoye</snm><fnm>Y.A.</fnm></au>
    <au><snm>Bettencourt</snm><fnm>L.M.A.</fnm></au>
  </aug>
  <source>2009 International Conference on Computational Science and
  Engineering</source>
  <pubdate>2009</pubdate>
  <fpage>144</fpage>
  <lpage>-150</lpage>
</bibl>

<bibl id="B111">
  <title><p>Directedness of Information Flow in Mobile Phone Communication
  Networks</p></title>
  <aug>
    <au><snm>Peruani</snm><fnm>F.</fnm></au>
    <au><snm>Tabourier</snm><fnm>L.</fnm></au>
  </aug>
  <source>PLoS One</source>
  <pubdate>2011</pubdate>
  <volume>6</volume>
  <issue>12</issue>
  <fpage>e28860</fpage>
</bibl>

<bibl id="B112">
  <title><p>How to detect causality effects on large dynamical communication
  networks: a case study</p></title>
  <aug>
    <au><snm>Tabourier</snm><fnm>L.</fnm></au>
    <au><snm>Stoica</snm><fnm>A.</fnm></au>
    <au><snm>Peruani</snm><fnm>F</fnm></au>
  </aug>
  <source>Communication Systems and Networks (COMSNETS), 2012 Fourth
  International Conference on</source>
  <pubdate>2012</pubdate>
  <fpage>1</fpage>
  <lpage>-7</lpage>
</bibl>

<bibl id="B113">
  <title><p>Dynamical strength of social ties in information
  spreading</p></title>
  <aug>
    <au><snm>Miritello</snm><fnm>G.</fnm></au>
    <au><snm>Moro</snm><fnm>E.</fnm></au>
    <au><snm>Lara</snm><fnm>R.</fnm></au>
  </aug>
  <source>Physical Review E</source>
  <publisher>APS</publisher>
  <pubdate>2011</pubdate>
  <volume>83</volume>
  <issue>4</issue>
  <fpage>045102</fpage>
</bibl>

<bibl id="B114">
  <title><p>Spread of epidemic disease on networks</p></title>
  <aug>
    <au><snm>Newman</snm><fnm>M. E. J.</fnm></au>
  </aug>
  <source>Physical Review E</source>
  <publisher>American Physical Society</publisher>
  <pubdate>2002</pubdate>
  <volume>66</volume>
  <fpage>016128</fpage>
  <url>http://link.aps.org/doi/10.1103/PhysRevE.66.016128</url>
</bibl>

<bibl id="B115">
  <title><p>The Structure and Dynamics of Networks</p></title>
  <aug>
    <au><snm>Newman</snm><fnm>M.E.J.</fnm></au>
    <au><snm>Barabasi</snm><fnm>A. L.</fnm></au>
    <au><snm>Watts</snm><fnm>D. J.</fnm></au>
  </aug>
  <publisher>Princeton: Princeton University Press</publisher>
  <pubdate>2006</pubdate>
</bibl>

<bibl id="B116">
  <title><p>Multiscale Analysis of Spreading in a Large Communication
  Network</p></title>
  <aug>
    <au><snm>Kivel\"{a}</snm><fnm>M.</fnm></au>
    <au><snm>Pan</snm><fnm>RK</fnm></au>
    <au><snm>Kaski</snm><fnm>K.</fnm></au>
    <au><snm>Kert\'{e}sz</snm><fnm>J.</fnm></au>
    <au><snm>Saram\"aki</snm><fnm>J.</fnm></au>
    <au><snm>Karsai</snm><fnm>M.</fnm></au>
  </aug>
  <source>Journal of Statistical Mechanics: Theory and Experiment</source>
  <pubdate>2012</pubdate>
  <volume>2012</volume>
  <issue>03</issue>
  <fpage>P03005</fpage>
</bibl>

<bibl id="B117">
  <title><p>Controllability of complex networks</p></title>
  <aug>
    <au><snm>Liu</snm><fnm>Y.Y.</fnm></au>
    <au><snm>Slotine</snm><fnm>J.J.</fnm></au>
    <au><snm>Barab{\'a}si</snm><fnm>A.L.</fnm></au>
  </aug>
  <source>Nature</source>
  <publisher>Nature Publishing Group</publisher>
  <pubdate>2011</pubdate>
  <volume>473</volume>
  <issue>7346</issue>
  <fpage>167</fpage>
  <lpage>-173</lpage>
</bibl>

<bibl id="B118">
  <title><p>Measuring the Collective Potential of Populations From Dynamic
  Social Interaction Data</p></title>
  <aug>
    <au><snm>Cebrian</snm><fnm>M.</fnm></au>
    <au><snm>Lahiri</snm><fnm>M.</fnm></au>
    <au><snm>Oliver</snm><fnm>N.</fnm></au>
    <au><snm>Pentland</snm><fnm>A.</fnm></au>
  </aug>
  <source>Selected Topics in Signal Processing, IEEE Journal of</source>
  <publisher>IEEE</publisher>
  <pubdate>2010</pubdate>
  <volume>4</volume>
  <issue>4</issue>
  <fpage>677</fpage>
  <lpage>-686</lpage>
</bibl>

<bibl id="B119">
  <title><p>Understanding the spreading patterns of mobile phone
  viruses</p></title>
  <aug>
    <au><snm>Wang</snm><fnm>P.</fnm></au>
    <au><snm>Gonz{\'a}lez</snm><fnm>M.C.</fnm></au>
    <au><snm>Hidalgo</snm><fnm>C.A.</fnm></au>
    <au><snm>Barab{\'a}si</snm><fnm>A.L.</fnm></au>
  </aug>
  <source>Science</source>
  <publisher>American Association for the Advancement of Science</publisher>
  <pubdate>2009</pubdate>
  <volume>324</volume>
  <issue>5930</issue>
  <fpage>1071</fpage>
</bibl>

<bibl id="B120">
  <title><p>New generation of mobile phone viruses and corresponding
  countermeasures</p></title>
  <aug>
    <au><snm>Wang</snm><fnm>P.</fnm></au>
    <au><snm>Gonz{\'a}lez</snm><fnm>M.C.</fnm></au>
    <au><snm>Menezes</snm><fnm>R.</fnm></au>
    <au><snm>Barab{\'a}si</snm><fnm>A.L.</fnm></au>
  </aug>
  <source>Arxiv preprint arXiv:1012.3156</source>
  <pubdate>2010</pubdate>
</bibl>

<bibl id="B121">
  <title><p>Understanding the spread of malicious mobile-phone programs and
  their damage potential</p></title>
  <aug>
    <au><snm>Wang</snm><fnm>P.</fnm></au>
    <au><snm>Gonz{\'a}lez</snm><fnm>M.C.</fnm></au>
    <au><snm>Menezes</snm><fnm>R.</fnm></au>
    <au><snm>Barab\'a{}si</snm><fnm>A.L.</fnm></au>
  </aug>
  <source>International journal of information security</source>
  <pubdate>2013</pubdate>
  <volume>12</volume>
  <issue>5</issue>
  <fpage>383</fpage>
  <lpage>-392</lpage>
</bibl>

<bibl id="B122">
  <title><p>Modeling the economic value of location and preference data of
  mobile users</p></title>
  <aug>
    <au><snm>Baccelli</snm><fnm>F.</fnm></au>
    <au><snm>Bolot</snm><fnm>J.</fnm></au>
  </aug>
  <source>Proc. IEEE Infocom 2011</source>
  <pubdate>2011</pubdate>
</bibl>

<bibl id="B123">
  <title><p>Urban Sensing Using Mobile Phone Network Data: A Survey of
  Research</p></title>
  <aug>
    <au><snm>Calabrese</snm><fnm>F</fnm></au>
    <au><snm>Ferrari</snm><fnm>L</fnm></au>
    <au><snm>Blondel</snm><fnm>VD</fnm></au>
  </aug>
  <source>ACM Computing Surveys (CSUR)</source>
  <publisher>ACM</publisher>
  <pubdate>2014</pubdate>
  <volume>47</volume>
  <issue>2</issue>
  <fpage>25</fpage>
</bibl>

<bibl id="B124">
  <title><p>Identifying important places in people's lives from cellular
  network data</p></title>
  <aug>
    <au><snm>Isaacman</snm><fnm>S.</fnm></au>
    <au><snm>Becker</snm><fnm>R.</fnm></au>
    <au><snm>C{\'a}ceres</snm><fnm>R.</fnm></au>
    <au><snm>Kobourov</snm><fnm>S.</fnm></au>
    <au><snm>Martonosi</snm><fnm>M.</fnm></au>
    <au><snm>Rowland</snm><fnm>J.</fnm></au>
    <au><snm>Varshavsky</snm><fnm>A.</fnm></au>
  </aug>
  <source>Pervasive Computing</source>
  <publisher>Springer</publisher>
  <pubdate>2011</pubdate>
  <fpage>133</fpage>
  <lpage>-151</lpage>
</bibl>

<bibl id="B125">
  <title><p>Mobile phone data from GSM networks for traffic parameter and urban
  spatial pattern assessment: a review of applications and
  opportunities</p></title>
  <aug>
    <au><snm>Steenbruggen</snm><fnm>J.</fnm></au>
    <au><snm>Borzacchiello</snm><fnm>M.T.</fnm></au>
    <au><snm>Nijkamp</snm><fnm>P.</fnm></au>
    <au><snm>Scholten</snm><fnm>H.</fnm></au>
  </aug>
  <source>GeoJournal</source>
  <publisher>Springer</publisher>
  <pubdate>2011</pubdate>
  <fpage>1</fpage>
  <lpage>-21</lpage>
</bibl>

<bibl id="B126">
  <title><p>The path most travelled: Mining road usage patterns from massive
  call data</p></title>
  <aug>
    <au><snm>Toole</snm><fnm>JL</fnm></au>
    <au><snm>Colak</snm><fnm>S</fnm></au>
    <au><snm>Alhasoun</snm><fnm>F</fnm></au>
    <au><snm>Evsukoff</snm><fnm>A</fnm></au>
    <au><snm>Gonzalez</snm><fnm>MC</fnm></au>
  </aug>
  <source>arXiv preprint arXiv:1403.0636</source>
  <pubdate>2014</pubdate>
</bibl>

<bibl id="B127">
  <title><p>Understanding road usage patterns in urban areas</p></title>
  <aug>
    <au><snm>Wang</snm><fnm>P</fnm></au>
    <au><snm>Hunter</snm><fnm>T</fnm></au>
    <au><snm>Bayen</snm><fnm>AM</fnm></au>
    <au><snm>Schechtner</snm><fnm>K</fnm></au>
    <au><snm>Gonz{\'a}lez</snm><fnm>MC</fnm></au>
  </aug>
  <source>Scientific reports</source>
  <publisher>Nature Publishing Group</publisher>
  <pubdate>2012</pubdate>
  <volume>2</volume>
</bibl>

<bibl id="B128">
  <title><p>Crowdsourcing Physical Package Delivery Using the Existing Routine
  Mobility of a Local Population</p></title>
  <aug>
    <au><snm>McInerney</snm><fnm>J.</fnm></au>
    <au><snm>Roger</snm><fnm>A.</fnm></au>
    <au><snm>Jennings</snm><fnm>N. R.</fnm></au>
  </aug>
  <source>Mobile Phone Data for Development - Analysis of mobile phone datasets
  for the development of Ivory Coast</source>
  <publisher>Orange D4D Challenge</publisher>
  <pubdate>2013</pubdate>
  <fpage>447</fpage>
  <lpage>-456</lpage>
</bibl>

<bibl id="B129">
  <title><p>Towards a recomender system for bush taxis</p></title>
  <aug>
    <au><snm>Gambs</snm><fnm>S.</fnm></au>
    <au><snm>Killijian</snm><fnm>M. O.</fnm></au>
  </aug>
  <source>Mobile Phone Data for Development - Analysis of mobile phone datasets
  for the development of Ivory Coast</source>
  <publisher>Orange D4D Challenge</publisher>
  <pubdate>2013</pubdate>
  <fpage>457</fpage>
  <lpage>466</lpage>
</bibl>

<bibl id="B130">
  <title><p>The geography of taste: analyzing cell-phone mobility and social
  events</p></title>
  <aug>
    <au><snm>Calabrese</snm><fnm>F.</fnm></au>
    <au><snm>Pereira</snm><fnm>F.</fnm></au>
    <au><snm>Di Lorenzo</snm><fnm>G.</fnm></au>
    <au><snm>Liu</snm><fnm>L.</fnm></au>
    <au><snm>Ratti</snm><fnm>C.</fnm></au>
  </aug>
  <source>Pervasive Computing</source>
  <publisher>Springer</publisher>
  <pubdate>2010</pubdate>
  <fpage>22</fpage>
  <lpage>-37</lpage>
</bibl>

<bibl id="B131">
  <title><p>Recommending social events from mobile phone location
  data</p></title>
  <aug>
    <au><snm>Quercia</snm><fnm>D.</fnm></au>
    <au><snm>Lathia</snm><fnm>N.</fnm></au>
    <au><snm>Calabrese</snm><fnm>F.</fnm></au>
    <au><snm>Di Lorenzo</snm><fnm>G.</fnm></au>
    <au><snm>Crowcroft</snm><fnm>J.</fnm></au>
  </aug>
  <source>2010 IEEE International Conference on Data Mining</source>
  <pubdate>2010</pubdate>
  <fpage>971</fpage>
  <lpage>-976</lpage>
</bibl>

<bibl id="B132">
  <title><p>Forecasting event attendance with anonymized mobile phone
  data</p></title>
  <aug>
    <au><snm>Cloquet</snm><fnm>C.</fnm></au>
    <au><snm>Blondel</snm><fnm>V.D.</fnm></au>
  </aug>
  <source>submitted to Big Data Research, Elsevier</source>
  <pubdate>2014</pubdate>
</bibl>

<bibl id="B133">
  <title><p>Analyzing the Workload Dynamics of a Mobile Phone Network in Large
  Scale Events</p></title>
  <aug>
    <au><snm>Xavier</snm><fnm>F. H. Z.</fnm></au>
    <au><snm>Silveira</snm><fnm>L. M.</fnm></au>
    <au><snm>Almeida</snm><fnm>J. M.</fnm></au>
    <au><snm>Ziviani</snm><fnm>A.</fnm></au>
    <au><snm>Malab</snm><fnm>C. H. S.</fnm></au>
    <au><snm>Marques Neto</snm><fnm>H. T.</fnm></au>
  </aug>
  <source>Proceedings of the first workshop on Urban networking</source>
  <pubdate>2012</pubdate>
  <fpage>37</fpage>
  <lpage>-42</lpage>
</bibl>

<bibl id="B134">
  <title><p>Monitoring Temporary Populations through Cellular Core Network
  Data</p></title>
  <aug>
    <au><snm>Manfredini</snm><fnm>F.</fnm></au>
    <au><snm>Tagliolato</snm><fnm>P.</fnm></au>
    <au><snm>Di Rosa</snm><fnm>C.</fnm></au>
  </aug>
  <source>Computational Science and Its Applications-ICCSA 2011</source>
  <publisher>Springer</publisher>
  <pubdate>2011</pubdate>
  <fpage>151</fpage>
  <lpage>-161</lpage>
</bibl>

<bibl id="B135">
  <title><p>Analysing repeat visitation on country level with passive mobile
  positioning method: an Estonian case study</p></title>
  <aug>
    <au><snm>Kuusik</snm><fnm>A.</fnm></au>
    <au><snm>Ahas</snm><fnm>R.</fnm></au>
    <au><snm>Tiru</snm><fnm>M.</fnm></au>
  </aug>
  <source>XVII scientific conference on economic policy</source>
  <pubdate>2009</pubdate>
  <fpage>1</fpage>
  <lpage>-3</lpage>
</bibl>

<bibl id="B136">
  <title><p>Quantifying the Impact of Human Mobility on Malaria</p></title>
  <aug>
    <au><snm>Wesolowski</snm><fnm>A.</fnm></au>
    <au><snm>Eagle</snm><fnm>N.</fnm></au>
    <au><snm>Tatem</snm><fnm>A. J.</fnm></au>
    <au><snm>Smith</snm><fnm>D. L.</fnm></au>
    <au><snm>Noor</snm><fnm>A. M.</fnm></au>
    <au><snm>Snow</snm><fnm>R. W.</fnm></au>
    <au><snm>Buckee</snm><fnm>C. O.</fnm></au>
  </aug>
  <source>Science</source>
  <pubdate>2012</pubdate>
  <volume>338</volume>
  <issue>6104</issue>
  <fpage>267</fpage>
  <lpage>-270</lpage>
</bibl>

<bibl id="B137">
  <title><p>On the Use of Human Mobility Proxies for Modeling
  Epidemics</p></title>
  <aug>
    <au><snm>Tizzoni</snm><fnm>M.</fnm></au>
    <au><snm>Bajardi</snm><fnm>P.</fnm></au>
    <au><snm>Decuyper</snm><fnm>A.</fnm></au>
    <au><snm>Kon Kam King</snm><fnm>G.</fnm></au>
    <au><snm>Schneider</snm><fnm>C. M.</fnm></au>
    <au><snm>Blondel</snm><fnm>V.D.</fnm></au>
    <au><snm>Smoreda</snm><fnm>Z.</fnm></au>
    <au><snm>Gonz{\'a}lez</snm><fnm>M.C.</fnm></au>
    <au><snm>Colizza</snm><fnm>V.</fnm></au>
  </aug>
  <source>PLoS computational biology</source>
  <pubdate>2014</pubdate>
  <volume>10</volume>
  <issue>7</issue>
  <fpage>e1003716</fpage>
</bibl>

<bibl id="B138">
  <title><p>An Agent-Based Model of Epidemic Spread Using Human Mobility and
  Social Network Information</p></title>
  <aug>
    <au><snm>Frias Martinez</snm><fnm>E.</fnm></au>
    <au><snm>Williamson</snm><fnm>G.</fnm></au>
    <au><snm>Frias Martinez</snm><fnm>V.</fnm></au>
  </aug>
  <source>Privacy, Security, Risk and Trust (PASSAT) and 2011 IEEE Third
  Inernational Conference on Social Computing (SocialCom), 2011 IEEE Third
  International Conference on</source>
  <pubdate>2011</pubdate>
  <fpage>57</fpage>
  <lpage>64</lpage>
</bibl>

<bibl id="B139">
  <title><p>Data for development: the {D4D} challenge on mobile phone
  data</p></title>
  <aug>
    <au><snm>Blondel</snm><fnm>V.D.</fnm></au>
    <au><snm>Esch</snm><fnm>M.</fnm></au>
    <au><snm>Chan</snm><fnm>C.</fnm></au>
    <au><snm>Cl{\'e}rot</snm><fnm>F.</fnm></au>
    <au><snm>Deville</snm><fnm>P.</fnm></au>
    <au><snm>Huens</snm><fnm>E.</fnm></au>
    <au><snm>Morlot</snm><fnm>F.</fnm></au>
    <au><snm>Smoreda</snm><fnm>Z.</fnm></au>
    <au><snm>Ziemlicki</snm><fnm>C.</fnm></au>
  </aug>
  <source>arXiv preprint arXiv:1210.0137</source>
  <pubdate>2012</pubdate>
</bibl>

<bibl id="B140">
  <title><p>Mitigating Epidemics through Mobile Micro-measures</p></title>
  <aug>
    <au><snm>Kafsi</snm><fnm>M.</fnm></au>
    <au><snm>Kazemi</snm><fnm>E.</fnm></au>
    <au><snm>Maystre</snm><fnm>L.</fnm></au>
    <au><snm>Yartseva</snm><fnm>L.</fnm></au>
    <au><snm>Grossglauser</snm><fnm>M.</fnm></au>
    <au><snm>Thiran</snm><fnm>P.</fnm></au>
  </aug>
  <source>arXiv preprint arXiv:1307.2084</source>
  <pubdate>2013</pubdate>
</bibl>

<bibl id="B141">
  <title><p>Exploiting Cellular Data for Disease Containment and Information
  Campaigns Strategies in Country-Wide Epidemics</p></title>
  <aug>
    <au><snm>Lima</snm><fnm>A.</fnm></au>
    <au><snm>De Domenico</snm><fnm>M.</fnm></au>
    <au><snm>Pejovic</snm><fnm>V.</fnm></au>
    <au><snm>Musolesi</snm><fnm>M.</fnm></au>
  </aug>
  <source>arXiv preprint arXiv:1306.4534</source>
  <pubdate>2013</pubdate>
</bibl>

<bibl id="B142">
  <title><p>Commentary: Containing the Ebola outbreak--the potential and
  challenge of mobile network data</p></title>
  <aug>
    <au><snm>Wesolowski</snm><fnm>A.</fnm></au>
    <au><snm>Buckee</snm><fnm>C. O.</fnm></au>
    <au><snm>Bengtsson</snm><fnm>L.</fnm></au>
    <au><snm>Wetter</snm><fnm>E.</fnm></au>
    <au><snm>Lu</snm><fnm>X.</fnm></au>
    <au><snm>Tatem</snm><fnm>A. J.</fnm></au>
  </aug>
  <source>PLOS Currents Outbreaks</source>
  <pubdate>2014</pubdate>
</bibl>

<bibl id="B143">
  <title><p>Enabling Humanitarian Use of Mobile Phone Data</p></title>
  <aug>
    <au><snm>Montjoye</snm><fnm>Y.A.</fnm></au>
    <au><snm>Kendall</snm><fnm>J.</fnm></au>
    <au><snm>Kerry</snm><fnm>C.F.</fnm></au>
  </aug>
  <source>Issues in Technology Innovation</source>
  <pubdate>2014</pubdate>
  <issue>26</issue>
</bibl>

<bibl id="B144">
  <title><p>Personal Influence, The part played by people in the flow of mass
  communications</p></title>
  <aug>
    <au><snm>Katz</snm><fnm>E.</fnm></au>
    <au><snm>Lazarsfeld</snm><fnm>P. F.</fnm></au>
  </aug>
  <publisher>Transaction Publishers</publisher>
  <pubdate>1970</pubdate>
</bibl>

<bibl id="B145">
  <title><p>Influentials, networks, and public opinion formation</p></title>
  <aug>
    <au><snm>Watts</snm><fnm>D.J.</fnm></au>
    <au><snm>Dodds</snm><fnm>P.S.</fnm></au>
  </aug>
  <source>Journal of consumer research</source>
  <pubdate>2007</pubdate>
  <volume>34</volume>
  <issue>4</issue>
  <fpage>441</fpage>
  <lpage>-458</lpage>
</bibl>

<bibl id="B146">
  <title><p>Network effects in service usage</p></title>
  <aug>
    <au><snm>Szab{\'o}</snm><fnm>G.</fnm></au>
    <au><snm>Barab\'a{}si</snm><fnm>A.L.</fnm></au>
  </aug>
  <source>Arxiv preprint physics/0611177</source>
  <pubdate>2006</pubdate>
</bibl>

<bibl id="B147">
  <title><p>Network-based marketing: Identifying likely adopters via consumer
  networks</p></title>
  <aug>
    <au><snm>Hill</snm><fnm>S.</fnm></au>
    <au><snm>Provost</snm><fnm>F.</fnm></au>
    <au><snm>Volinsky</snm><fnm>C.</fnm></au>
  </aug>
  <source>Statistical Science</source>
  <publisher>Institute of Mathematical Statistics</publisher>
  <pubdate>2006</pubdate>
  <volume>21</volume>
  <issue>2</issue>
  <fpage>256</fpage>
  <lpage>-276</lpage>
</bibl>

<bibl id="B148">
  <title><p>Tracing mobile phone app installations in the" friends and family"
  study</p></title>
  <aug>
    <au><snm>Aharony</snm><fnm>N.</fnm></au>
    <au><snm>Pan</snm><fnm>W.</fnm></au>
    <au><snm>Ip</snm><fnm>C.</fnm></au>
    <au><snm>Pentland</snm><fnm>A.</fnm></au>
  </aug>
  <source>Proceedings of the 2010 Workshop on Information in Networks
  (WIN'10)</source>
  <pubdate>2010</pubdate>
</bibl>

<bibl id="B149">
  <title><p>Dynamic effects of social influence and direct marketing on the
  adoption of high-technology products</p></title>
  <aug>
    <au><snm>Risselada</snm><fnm>H.</fnm></au>
    <au><snm>Verhoef</snm><fnm>P.C.</fnm></au>
    <au><snm>Bijmolt</snm><fnm>T.H.A.</fnm></au>
  </aug>
  <source>Journal of Marketing</source>
  <pubdate>2014</pubdate>
  <volume>78</volume>
  <issue>2</issue>
  <fpage>52</fpage>
  <lpage>-68</lpage>
</bibl>

<bibl id="B150">
  <title><p>Effects of temporal correlations on cascades: Threshold models on
  temporal networks</p></title>
  <aug>
    <au><snm>Backlund</snm><fnm>VP</fnm></au>
    <au><snm>Saram\"aki</snm><fnm>J</fnm></au>
    <au><snm>Pan</snm><fnm>RK</fnm></au>
  </aug>
  <source>Phys. Rev. E</source>
  <publisher>American Physical Society</publisher>
  <pubdate>2014</pubdate>
  <volume>89</volume>
  <fpage>062815</fpage>
  <url>http://link.aps.org/doi/10.1103/PhysRevE.89.062815</url>
</bibl>

<bibl id="B151">
  <title><p>Social leaders in graphs</p></title>
  <aug>
    <au><snm>Blondel</snm><fnm>V.</fnm></au>
    <au><snm>Kerchove</snm><fnm>C.</fnm></au>
    <au><snm>Huens</snm><fnm>E.</fnm></au>
    <au><snm>Van Dooren</snm><fnm>P.</fnm></au>
  </aug>
  <source>Lecture notes in control and information sciences</source>
  <publisher>Citeseer</publisher>
  <pubdate>2006</pubdate>
  <volume>341</volume>
  <fpage>231</fpage>
</bibl>

<bibl id="B152">
  <title><p>Ranking Large Networks: Leadership, Optimization and Distrust (PhD
  thesis)</p></title>
  <aug>
    <au><snm>Kerchove d'Exaerde</snm><fnm>C.</fnm></au>
  </aug>
  <pubdate>2009</pubdate>
</bibl>

<bibl id="B153">
  <title><p>Mobile Phone Data for Development - Analysis of mobile phone
  datasets for the development of Ivory Coast</p></title>
  <publisher>Orange D4D Challenge</publisher>
  <editor>Blondel, V.D. and de Cordes, N. and Decuyper, A. and Deville, P. and
  Raguenez, J. and Smoreda, Z.</editor>
  <pubdate>2013</pubdate>
</bibl>

<bibl id="B154">
  <title><p>The geography and carbon footprint of mobile phone use in C\^ote
  d'Ivoire</p></title>
  <aug>
    <au><snm>Salnikov</snm><fnm>V</fnm></au>
    <au><snm>Schien</snm><fnm>D</fnm></au>
    <au><snm>Youn</snm><fnm>H</fnm></au>
    <au><snm>Lambiotte</snm><fnm>R</fnm></au>
    <au><snm>Gastner</snm><fnm>MT</fnm></au>
  </aug>
  <source>EPJ Data Science</source>
  <publisher>Springer</publisher>
  <pubdate>2014</pubdate>
  <volume>3</volume>
  <issue>1</issue>
  <fpage>1</fpage>
  <lpage>-15</lpage>
</bibl>

<bibl id="B155">
  <title><p>On the relationship between socio-economic factors and cell phone
  usage</p></title>
  <aug>
    <au><snm>Frias Martinez</snm><fnm>V</fnm></au>
    <au><snm>Virseda</snm><fnm>J</fnm></au>
  </aug>
  <source>Proceedings of the Fifth International Conference on Information and
  Communication Technologies and Development</source>
  <pubdate>2012</pubdate>
  <fpage>76</fpage>
  <lpage>-84</lpage>
</bibl>

<bibl id="B156">
  <title><p>Mobile phones and malaria: modeling human and parasite
  travel</p></title>
  <aug>
    <au><snm>Buckee</snm><fnm>CO</fnm></au>
    <au><snm>Wesolowski</snm><fnm>A</fnm></au>
    <au><snm>Eagle</snm><fnm>NN</fnm></au>
    <au><snm>Hansen</snm><fnm>E</fnm></au>
    <au><snm>Snow</snm><fnm>RW</fnm></au>
  </aug>
  <source>Travel medicine and infectious disease</source>
  <publisher>Elsevier</publisher>
  <pubdate>2013</pubdate>
  <volume>11</volume>
  <issue>1</issue>
  <fpage>15</fpage>
  <lpage>-22</lpage>
</bibl>

<bibl id="B157">
  <title><p>Spreading paths in partially observed social networks</p></title>
  <aug>
    <au><snm>Onnela</snm><fnm>J.P.</fnm></au>
    <au><snm>Christakis</snm><fnm>N.A.</fnm></au>
  </aug>
  <source>Physical Review E</source>
  <pubdate>2012</pubdate>
  <volume>85</volume>
  <issue>3</issue>
  <fpage>036106</fpage>
</bibl>

<bibl id="B158">
  <title><p>Are Call Detail Records Biased for Sampling Human
  Mobility?</p></title>
  <aug>
    <au><snm>Ranjan</snm><fnm>G.</fnm></au>
    <au><snm>Zang</snm><fnm>H.</fnm></au>
    <au><snm>Zhang</snm><fnm>Z.L.</fnm></au>
    <au><snm>Bolot</snm><fnm>J.</fnm></au>
  </aug>
  <source>ACM SIGMOBILE Mobile Computing and Communications Review</source>
  <pubdate>2012</pubdate>
  <volume>16</volume>
  <issue>3</issue>
  <fpage>33</fpage>
  <lpage>-44</lpage>
</bibl>

<bibl id="B159">
  <title><p>Measuring large-scale social networks with high
  resolution</p></title>
  <aug>
    <au><snm>Stopczynski</snm><fnm>A</fnm></au>
    <au><snm>Sekara</snm><fnm>V</fnm></au>
    <au><snm>Sapiezynski</snm><fnm>P</fnm></au>
    <au><snm>Cuttone</snm><fnm>A</fnm></au>
    <au><snm>Madsen</snm><fnm>MM</fnm></au>
    <au><snm>Larsen</snm><fnm>JE</fnm></au>
    <au><snm>Lehmann</snm><fnm>S</fnm></au>
  </aug>
  <source>PloS one</source>
  <publisher>Public Library of Science</publisher>
  <pubdate>2014</pubdate>
  <volume>9</volume>
  <issue>4</issue>
  <fpage>e95978</fpage>
</bibl>

<bibl id="B160">
  <title><p>Netflix spilled your Brokeback Mountain secret, lawsuit
  claims</p></title>
  <aug>
    <au><snm>Singel</snm><fnm>R.</fnm></au>
  </aug>
  <source>Threat Level (blog), Wired</source>
  <pubdate>2009</pubdate>
</bibl>

<bibl id="B161">
  <title><p>The'Re-Identification'of Governor William Weld's Medical
  Information: A Critical Re-Examination of Health Data Identification Risks
  and Privacy Protections, Then and Now</p></title>
  <aug>
    <au><snm>Barth Jones</snm><fnm>D.C.</fnm></au>
  </aug>
  <source>Then and Now</source>
  <pubdate>2012</pubdate>
</bibl>

<bibl id="B162">
  <title><p>Commission proposes a comprehensive reform of data protection rules
  to increase users' control of their data and to cut costs for
  businesses</p></title>
  <aug>
    <au><snm>Commission</snm><fnm>E</fnm></au>
  </aug>
  <source>Reference: IP/12/46</source>
  <pubdate>2012</pubdate>
</bibl>

<bibl id="B163">
  <title><p>95/46/EC of the European Parliament and of the Council of 24
  October 1995 on the protection of individuals with regard to the processing
  of personal data and on the free movement of such data</p></title>
  <aug>
    <au><snm>Directive</snm><fnm>EU</fnm></au>
  </aug>
  <source>Official Journal of the EC</source>
  <pubdate>1995</pubdate>
  <volume>23</volume>
  <issue>6</issue>
</bibl>

<bibl id="B164">
  <title><p>Anonymization of location data does not work: a large-scale
  measurement study</p></title>
  <aug>
    <au><snm>Zang</snm><fnm>H.</fnm></au>
    <au><snm>Bolot</snm><fnm>J.</fnm></au>
  </aug>
  <source>submitted to ACM Mobicom</source>
  <pubdate>2011</pubdate>
  <volume>11</volume>
</bibl>

<bibl id="B165">
  <title><p>Unique in the Crowd: The privacy bounds of human
  mobility</p></title>
  <aug>
    <au><snm>Montjoye</snm><fnm>Y.A.</fnm></au>
    <au><snm>Hidalgo</snm><fnm>C.A.</fnm></au>
    <au><snm>Verleysen</snm><fnm>M.</fnm></au>
    <au><snm>Blondel</snm><fnm>V.D.</fnm></au>
  </aug>
  <source>Scientific Reports</source>
  <pubdate>2013</pubdate>
  <volume>3</volume>
</bibl>

<bibl id="B166">
  <title><p>Wherefore art thou r3579x?: anonymized social networks, hidden
  patterns, and structural steganography</p></title>
  <aug>
    <au><snm>Backstrom</snm><fnm>L.</fnm></au>
    <au><snm>Dwork</snm><fnm>C.</fnm></au>
    <au><snm>Kleinberg</snm><fnm>J.</fnm></au>
  </aug>
  <source>Proceedings of the 16th international conference on World Wide
  Web</source>
  <pubdate>2007</pubdate>
  <fpage>181</fpage>
  <lpage>-190</lpage>
</bibl>

<bibl id="B167">
  <title><p>De-anonymizing social networks</p></title>
  <aug>
    <au><snm>Narayanan</snm><fnm>A.</fnm></au>
    <au><snm>Shmatikov</snm><fnm>V.</fnm></au>
  </aug>
  <source>2009 30th IEEE Symposium on Security and Privacy</source>
  <pubdate>2009</pubdate>
  <fpage>173</fpage>
  <lpage>-187</lpage>
</bibl>

<bibl id="B168">
  <title><p>Not So Unique in the Crowd: a Simple and Effective Algorithm for
  Anonymizing Location Data</p></title>
  <aug>
    <au><snm>Song</snm><fnm>Y</fnm></au>
    <au><snm>Dahlmeier</snm><fnm>D</fnm></au>
    <au><snm>Bressan</snm><fnm>S</fnm></au>
  </aug>
  <source>Proceeding of the 1st International Workshop on Privacy-Preserving
  IR: When Information Retrieval Meets Privacy and Security (PIR 2014)</source>
  <pubdate>2014</pubdate>
  <fpage>19</fpage>
</bibl>

<bibl id="B169">
  <title><p>{D4D-Senegal}: The Second Mobile Phone Data for Development
  Challenge</p></title>
  <aug>
    <au><snm>Montjoye</snm><fnm>Y.A.</fnm></au>
    <au><snm>Smoreda</snm><fnm>Z.</fnm></au>
    <au><snm>Trinquart</snm><fnm>R.</fnm></au>
    <au><snm>Ziemlicki</snm><fnm>C.</fnm></au>
    <au><snm>Blondel</snm><fnm>V.D.</fnm></au>
  </aug>
  <source>arXiv preprint arXiv:1407.4885</source>
  <pubdate>2014</pubdate>
</bibl>

<bibl id="B170">
  <title><p>\textit{k}-anonymity: a model for protecting privacy</p></title>
  <aug>
    <au><snm>Sweeney</snm><fnm>L.</fnm></au>
  </aug>
  <source>International Journal of Uncertainty, Fuzziness and Knowledge-Based
  Systems</source>
  <pubdate>2002</pubdate>
  <volume>10</volume>
  <issue>05</issue>
  <fpage>557</fpage>
  <lpage>-570</lpage>
</bibl>

<bibl id="B171">
  <title><p>Human mobility modeling at metropolitan scales</p></title>
  <aug>
    <au><snm>Isaacman</snm><fnm>S.</fnm></au>
    <au><snm>Becker</snm><fnm>R.</fnm></au>
    <au><snm>C{\'a}ceres</snm><fnm>R.</fnm></au>
    <au><snm>Martonosi</snm><fnm>M.</fnm></au>
    <au><snm>Rowland</snm><fnm>J.</fnm></au>
    <au><snm>Varshavsky</snm><fnm>A.</fnm></au>
    <au><snm>Willinger</snm><fnm>W.</fnm></au>
  </aug>
  <source>Proceedings of the 10th international conference on Mobile systems,
  applications, and services</source>
  <pubdate>2012</pubdate>
  <fpage>239</fpage>
  <lpage>-252</lpage>
</bibl>

<bibl id="B172">
  <title><p>DP-WHERE: Differentially Private Modeling of Human
  Mobility</p></title>
  <aug>
    <au><snm>Mir</snm><fnm>D. J.</fnm></au>
    <au><snm>Isaacman</snm><fnm>S.</fnm></au>
    <au><snm>C{\'a}ceres</snm><fnm>R.</fnm></au>
    <au><snm>Martonosi</snm><fnm>M.</fnm></au>
    <au><snm>Wright</snm><fnm>R.N.</fnm></au>
  </aug>
  <source>Big Data, 2013 IEEE International Conference on</source>
  <pubdate>2013</pubdate>
  <fpage>580</fpage>
  <lpage>-588</lpage>
</bibl>

<bibl id="B173">
  <title><p>Reality Mining and Personal Privacy</p></title>
  <aug>
    <au><snm>Madan</snm><fnm>A.</fnm></au>
    <au><snm>Waber</snm><fnm>B.N.</fnm></au>
    <au><snm>Ding</snm><fnm>M.</fnm></au>
    <au><snm>Kominers</snm><fnm>P.</fnm></au>
    <au><snm>Pentland</snm><fnm>A.S.</fnm></au>
  </aug>
  <pubdate>2009</pubdate>
</bibl>

<bibl id="B174">
  <title><p>Making sense from Snowden</p></title>
  <aug>
    <au><snm>Landau</snm><fnm>S.</fnm></au>
  </aug>
  <source>IEEE Security \& Privacy Magazine</source>
  <pubdate>2013</pubdate>
  <issue>3</issue>
  <fpage>5463</fpage>
</bibl>

<bibl id="B175">
  <title><p>Reality mining of mobile communications: Toward a new deal on
  data</p></title>
  <aug>
    <au><snm>Pentland</snm><fnm>A.</fnm></au>
  </aug>
  <source>The Global Information Technology Report 2008--2009</source>
  <pubdate>2009</pubdate>
  <fpage>1981</fpage>
</bibl>

<bibl id="B176">
  <title><p>Engineering a common good: Fair use of aggregated, anonymized
  behavioral data</p></title>
  <aug>
    <au><snm>Eagle</snm><fnm>N</fnm></au>
  </aug>
  <source>First International Forum on the Application and Management of
  Personal Electronic Information</source>
  <pubdate>2009</pubdate>
</bibl>

<bibl id="B177">
  <title><p>Applying data mining to telecom churn management</p></title>
  <aug>
    <au><snm>Hung</snm><fnm>S.Y.</fnm></au>
    <au><snm>Yen</snm><fnm>D.C.</fnm></au>
    <au><snm>Wang</snm><fnm>H.Y.</fnm></au>
  </aug>
  <source>Expert Systems with Applications</source>
  <publisher>Elsevier</publisher>
  <pubdate>2006</pubdate>
  <volume>31</volume>
  <issue>3</issue>
  <fpage>515</fpage>
  <lpage>-524</lpage>
</bibl>

<bibl id="B178">
  <title><p>Social ties and their relevance to churn in mobile telecom
  networks</p></title>
  <aug>
    <au><snm>Dasgupta</snm><fnm>K.</fnm></au>
    <au><snm>Singh</snm><fnm>R.</fnm></au>
    <au><snm>Viswanathan</snm><fnm>B.</fnm></au>
    <au><snm>Chakraborty</snm><fnm>D.</fnm></au>
    <au><snm>Mukherjea</snm><fnm>S.</fnm></au>
    <au><snm>Nanavati</snm><fnm>A.A.</fnm></au>
    <au><snm>Joshi</snm><fnm>A.</fnm></au>
  </aug>
  <source>Proceedings of the 11th international conference on Extending
  database technology: Advances in database technology</source>
  <pubdate>2008</pubdate>
  <fpage>668</fpage>
  <lpage>-677</lpage>
</bibl>

<bibl id="B179">
  <title><p>Predicting customer churn in mobile networks through analysis of
  social groups</p></title>
  <aug>
    <au><snm>Richter</snm><fnm>Y.</fnm></au>
    <au><snm>Yom Tov</snm><fnm>E.</fnm></au>
    <au><snm>Slonim</snm><fnm>N.</fnm></au>
  </aug>
  <source>Proceedings of the 2010 SIAM International Conference on Data Mining
  (SDM 2010)</source>
  <pubdate>2010</pubdate>
</bibl>

<bibl id="B180">
  <title><p>Estimating the effect of word of mouth on churn and cross-buying in
  the mobile phone market with Markov logic networks</p></title>
  <aug>
    <au><snm>Dierkes</snm><fnm>T.</fnm></au>
    <au><snm>Bichler</snm><fnm>M.</fnm></au>
    <au><snm>Krishnan</snm><fnm>R.</fnm></au>
  </aug>
  <source>Decision Support Systems</source>
  <publisher>Elsevier</publisher>
  <pubdate>2011</pubdate>
</bibl>

<bibl id="B181">
  <title><p>Dynamic pricing in cellular networks, a mobility model with a
  provider-oriented approach</p></title>
  <aug>
    <au><snm>Fitkov Norris</snm><fnm>ED</fnm></au>
    <au><snm>Khanifar</snm><fnm>A.</fnm></au>
  </aug>
  <source>3G Mobile Communication Technologies, 2001. Second International
  Conference on (Conf. Publ. No. 477)</source>
  <pubdate>2001</pubdate>
  <fpage>63</fpage>
  <lpage>-67</lpage>
</bibl>

<bibl id="B182">
  <title><p>An Empirical Analysis of Mobile Voice Service and Sms: A Structural
  Model</p></title>
  <aug>
    <au><snm>Kim</snm><fnm>Y.</fnm></au>
    <au><snm>Telang</snm><fnm>R.</fnm></au>
    <au><snm>Vogt</snm><fnm>W.B.</fnm></au>
    <au><snm>Krishnan</snm><fnm>R.</fnm></au>
  </aug>
  <source>Management Science</source>
  <publisher>INFORMS</publisher>
  <pubdate>2010</pubdate>
  <volume>56</volume>
  <issue>2</issue>
  <fpage>234</fpage>
  <lpage>-252</lpage>
</bibl>

<bibl id="B183">
  <title><p>You're so predictable</p></title>
  <aug>
    <au><snm>Barab{\'a}si</snm><fnm>A.L.</fnm></au>
  </aug>
  <source>Physics World</source>
  <pubdate>2010</pubdate>
  <fpage>22</fpage>
  <lpage>-26</lpage>
</bibl>

<bibl id="B184">
  <title><p>Ranking stability and super-stable nodes in complex
  networks</p></title>
  <aug>
    <au><snm>Ghoshal</snm><fnm>G.</fnm></au>
    <au><snm>Barab{\'a}si</snm><fnm>A.L.</fnm></au>
  </aug>
  <source>Nature Communications</source>
  <publisher>Nature Publishing Group</publisher>
  <pubdate>2011</pubdate>
  <volume>2</volume>
  <fpage>394</fpage>
</bibl>

</refgrp>
} 

\end{backmatter}
\end{document}